\documentclass[12pt]{article}
\pdfoutput=1
\usepackage{graphics,amssymb,float}
\usepackage{graphicx}
\usepackage{rotating}
\usepackage{amssymb,amsmath,amsbsy}
\usepackage{mathrsfs}
\usepackage{dsfont}
\usepackage{color}
\usepackage{cite}
\usepackage{enumerate}
\usepackage{caption}
\usepackage{scalerel}

\let\Re\relax
\let\Im\relax

\DeclareMathOperator{\Re}{Re}
\DeclareMathOperator{\Im}{Im}

\renewcommand\star{\scriptscriptstyle{\bigstar}}

\renewcommand{\theequation}{\arabic{section}.\arabic{equation}}

\newenvironment{Eqnarray}
         {\arraycolsep 0.14em\begin{eqnarray}}{\end{eqnarray}}
\def\phm{\phantom{-}}
\def\vev#1{\langle #1 \rangle}
\def\beq{\begin{equation}}
\def\eeq{\end{equation}}
\def\beqa{\begin{Eqnarray}}
\def\eeqa{\end{Eqnarray}}
\def\beq{\begin{equation}}
\def\eeq{\end{equation}}
\def\ifmath#1{\relax\ifmmode #1\else $#1$\fi}
\def\ls#1{\ifmath{_{\lower1.5pt\hbox{$\scriptstyle #1$}}}}
\def\lss#1{\ifmath{^{\,\lower2.5pt\hbox{$\scriptstyle #1$}}}}

\def\half{\tfrac{1}{2}}

\def\quarter{\tfrac{1}{4}}

\def\lsim{\mathrel{\raise.3ex\hbox{$<$\kern-.75em\lower1ex\hbox{$\sim$}}}}
\def\gsim{\mathrel{\raise.3ex\hbox{$>$\kern-.75em\lower1ex\hbox{$\sim$}}}}
\def\ifmath#1{\relax\ifmmode #1\else $#1$\fi}

\def\eq#1{Eq.~(\ref{#1})}
\def\Eq#1{Eq.~(\ref{#1})}

\def\eqs#1#2{Eqs.~(\ref{#1}) and (\ref{#2})}
\def\eqss#1#2#3{Eqs.~(\ref{#1}), (\ref{#2}) and (\ref{#3})}
\def\eqst#1#2{Eqs.~(\ref{#1})--(\ref{#2})}

\def\nn{\nonumber}
\def\vev#1{\langle #1 \rangle}
\def\ddel{\!\!\mathrel{\raise1.7ex\hbox{$\leftrightarrow$\kern-.85em
\lower1.7ex\hbox{$\partial$}}}\!}
\def\lsup#1{^{\lower 6pt\hbox{$\scriptstyle#1$}}}
\def\llsup#1{^{\lower 2.5pt\hbox{$\scriptstyle#1$}}}
\def\abar{{\bar a}}
\def\bbar{{\bar b}}
\def\cbar{{\bar c}}
\def\dbar{{\bar d}}
\def\ebar{{\bar e}}
\def\fbar{{\bar f}}
\def\gbar{{\bar g}}
\def\hbarr{{\bar h}}
\def\kbar{{\bar k}}
\def\iso{\mathchoice{\cong}{\cong}{\isoS}{\cong}}
\def\isoS{\vbox{\baselineskip 0pt  \lineskip 0.5pt
    \ialign{$ \mathsurround=0pt  \scriptstyle \hfil ## \hfil $\crcr
        \sim \crcr = \crcr}}}

\allowdisplaybreaks 

\begin{document}
\begin{titlepage}
\begin{center}

\vspace*{-1.5cm}
\begin{flushright}
SCIPP-18/04\\
\end{flushright}

\vspace*{1.5cm}
{\Large\bf 
Symmetries and Mass Degeneracies
in the Scalar Sector} 

\vspace*{0.8cm}

\renewcommand{\thefootnote}{\fnsymbol{footnote}}

{\large
Howard~E.~Haber$^{a}$\footnote[1]{Email: haber@scipp.ucsc.edu},
O.M.~Ogreid$^{b}$\footnote[2]{Email: omo@hvl.no},
P.~Osland$^{c}$\footnote[3]{Email: Per.Osland@uib.no} and
M. N. Rebelo$^{d}$\footnote[4]{Email: rebelo@tecnico.ulisboa.pt}
}

\renewcommand{\thefootnote}{\arabic{footnote}}

\vspace*{0.8cm}
{\normalsize \it
$^a\,$Santa Cruz Institute for Particle Physics, University of California, 1156 High Street, \\ Santa Cruz, CA 95064, USA\\[2mm]
$^b\,$Western Norway University of Applied Sciences, Postboks 7030, N-5020 Bergen, Norway\\[2mm]
$^c\,$Department of Physics and Technology, University of Bergen,
Postboks 7803, N-5020 Bergen, Norway\\[2mm]
$^d\,$Departamento de F\'{\i}sica and Centro de F\'{\i}sica Te\'{o}rica de Part\'{\i}culas (CFTP), Instituto
Superior T\'{e}cnico, Av. Rovisco Pais, P-1049-001 Lisboa, Portugal}
\vspace{0.8cm}

\begin{abstract}
We explore some aspects of models with two and three SU(2) scalar doublets that lead to mass degeneracies among some of the physical scalars. In Higgs sectors with two scalar doublets, the exact degeneracy of scalar masses, without an artificial fine-tuning of the scalar potential parameters, is possible
only in the case of the inert doublet model (IDM),  where the scalar potential respects a global U(1)  symmetry that is not broken by the vacuum.
In the case of three doublets, we introduce and analyze the 
replicated inert doublet model, which possesses two inert doublets of scalars. We then generalize this model to obtain a scalar potential, first proposed 
by Ivanov and Silva, with a CP4 symmetry that guarantees the existence of pairwise degenerate scalar states among two pairs of neutral scalars and two pairs of charged scalars.
Here, CP4 is a generalized CP symmetry with the property that $({\rm CP}4)^n$ is the identity operator only for integer $n$ values that are multiples of 4. 
The form of the CP4-symmetric scalar potential is simplest when expressed in the Higgs basis, where the neutral scalar field vacuum expectation value resides entirely in one of the scalar doublet fields.   The symmetries of the model permit
a term in the scalar potential with a complex coefficient that cannot be removed by any redefinition of the scalar fields within the class of Higgs bases
(in which case, we say that no real Higgs basis exists).
A striking feature of the CP4-symmetric model is 
that it preserves CP even in the absence of a real Higgs basis, as illustrated by the cancellation of 
the contributions to the CP violating form factors of the effective $ZZZ$ and $ZWW$ vertices. 
\end{abstract}

\end{center}
\end{titlepage}

\section{Introduction}\label{sec:intro}
\setcounter{equation}{0}

After the initial discovery of the Higgs boson in 2012\cite{Aad:2012tfa,Chatrchyan:2012xdj}, certain anomalies in the Higgs data (which have since disappeared)
motivated the exploration of the possibility that the 125 GeV Higgs signal was comprised of two nearly mass-degenerate scalar states\cite{Gunion:2012gc,Gunion:2012he,Ferreira:2012nv,Drozd:2012vf,Grossman:2013pt,Munir:2013wka,Efrati:2013ini,David:2014jla,Han:2015pwa}.   Although the present Higgs data is consistent with the Standard Model\cite{Khachatryan:2016vau,Sirunyan:2018koj,ATLAS:2018doi}, one cannot yet rule out the presence of a mass degenerate scalar state at 125 GeV\cite{Bian:2017gxg}. 

In this work, we consider the implications of a mass degeneracy among two (or more) scalar states of an extended Higgs sector.   Such a mass degeneracy can be either accidental or the result of a symmetry.  A trivial example of such a phenomenon arises in any doublet extended Higgs model.  All such models possess a mass degenerate state, namely the charged Higgs boson,~$H^\pm$.  Indeed, $H^+$ and $H^-$ are mass-degenerate due to the U(1)$_{\rm EM}$ gauge symmetry.  Moreover, the $H^+$ and $H^-$ are distinguishable by their electric charge, which can be experimentally probed using photons.  
Suppose that this probe were unavailable (or equivalently, suppose one could turn off electromagnetism).  In this case, would it be possible for an experiment to reveal the existence of a mass-degenerate scalar?
In this very simple example, one could not physically distinguish (on an event by event basis) between the two degenerate states that comprise the charged Higgs scalar.  Nevertheless, there would in principle be observables that are sensitive to the number of mass-degenerate scalar states present.   For example, in the CP-conserving two Higgs doublet model, the decay rate for the decay of a heavy CP-even neutral scalar, $H\to H^+ H^-$ (if kinematically allowed) is proportional to the number of degrees of freedom in the final state.  If we express the charged Higgs field as a linear combination of real scalar fields, $H^\pm =(\phi_1\pm i\phi_2)/\sqrt{2}$, then the decay rate for $H\to H^+ H^-$ is the (incoherent) sum of the decay rates for $H\to \phi_1\phi_1$ and $\phi_2\phi_2$.  These two rates are identical, and the sum yields a multiplicity factor of 2.  This multiplicity factor provides the experimental signal for mass-degenerate scalars.\footnote{Note that the decay rate for $Z\to H^+ H^-$ is equal to the decay rate for $Z\to \phi_1\phi_2$.  In this case, the off-diagonal nature of the $Z\phi_1\phi_2$ coupling implies that no multiplicity factor is present.  Nevertheless, one can still infer the existence of mass-degenerate states, since the decays $Z\to\phi_1\phi_1, \phi_2\phi_2$ are forbidden by Bose statistics.}

Apart from the trivial mass degeneracy of $H^\pm$, we would like to explore in this paper the possibility of exactly mass-degenerate neutral scalars and/or mass-degenerate charged Higgs pairs in extended Higgs sectors.   In each case, the critical questions to ask are: (i) 
is the origin of the exact mass degeneracy natural? and (ii) how can the mass degenerate scalars be distinguished experimentally?
Exact mass degeneracies are natural if they are a consequence of an unbroken symmetry.  In particular, accidental mass degeneracies require an artificial fine-tuning of 
the scalar potential parameters, and in this sense we shall call them unnatural (and in our view not especially interesting).  If mass-degenerate states are present, it is of interest to determine how to probe them experimentally.  In some cases, one can identify the presence of mass degenerate states on an event by event basis.  In other cases, the only signal of the mass degeneracy is a measurable multiplicity factor that can be determined when averaging over initial state degeneracies and summing over final state degeneracies.\footnote{For example, a quark of a given flavor is a mass degenerate state due to its three possible colors.   Although the color of a quark cannot be identified experimentally, the presence of the color degree of freedom can be experimentally verified by the color multiplicity factor (most famously exhibited in the observed cross section for $e^+e^-$ annihilation into quark-antiquark pairs.)}
Our focus in this paper is extended multi-doublet Higgs sectors with mass-degenerate scalar states.

It is particularly instructive to discuss mass degeneracy in the scalar sector starting from
the so-called Higgs basis. This corresponds to a subset of all possible scalar field parameterizations in which
only one Higgs doublet, denoted by $H_1$, acquires a non-zero positive vacuum expectation value (vev), 
while all the other scalar fields of the Higgs basis ($H_2$, $H_3,\ldots H_n$) have zero vev~\cite{Donoghue:1978cj,Georgi:1978ri,Botella:1994cs,Branco:1999fs,Davidson:2005cw}. 
The neutral and charged Goldstone bosons reside entirely in $H_1$, as this is the only doublet that possesses a non-zero vev,
together with a neutral Higgs field that, in the absence of mixing
with the neutral fields of the other Higgs doublets, behaves like the Standard Model (SM) Higgs boson.
In this sense, this subset of scalar bases may be viewed as Standard Model aligned bases.  That is, the Higgs basis is actually a family of basis choices, since one is always free to perform an arbitrary U($n-1$) transformation among $H_2$, $H_3,\ldots H_n$ while preserving the vev of $H_1$ \cite{Ogreid:2017alh}.  There is no loss of
generality in choosing any particular scalar basis as a starting point. In order to identify
the physical neutral scalars one must diagonalize the corresponding scalar squared-mass matrix.  As expected, the end result is independent of the initial choice of basis for the scalar fields.

In section \ref{thdmdegenerate}, we study under what circumstances there is an exact mass degeneracy in the familiar two-Higgs-doublet model (2HDM) \cite{Gunion:1989we,Branco:2011iw}. 
In this model there are three physical
neutral fields and one charged  field, so we only consider potential mass degeneracies among the neutral fields.\footnote{One can also examine mass degeneracies between the charged Higgs boson and one of the neutral Higgs bosons.  For example, in a custodial symmetric 2HDM, the CP-odd Higgs scalar is degenerate with the charged Higgs boson~\cite{Gerard:2007kn,Haber:2010bw,Grzadkowski:2010dj,Nishi:2011gc}.   
However, custodial symmetry is not an exact symmetry of the full electroweak Lagrangian.  Thus, given a custodial symmetric 2HDM scalar potential,
any potential mass degeneracies between neutral and charged scalars is at best approximate. 
We do not consider such mass degeneracies further in this work.}
We begin our 2HDM analysis by studying possible mass degeneracies among the neutral scalar states of the 
inert doublet model (IDM)~\cite{Ma:2006km,Barbieri:2006dq}.  The scalar potential of this model possesses a
discrete $\mathbb{Z}_2$ symmetry that is unbroken by the vacuum.  In this case, the CP-even neutral component of $H_1$ in the Higgs basis is a mass eigenstate whose tree-level couplings are precisely those of the SM Higgs boson.  The real and imaginary parts of the neutral component of $H_2$ are odd under the discrete $\mathbb{Z}_2$ symmetry, and have opposite signs under CP.  We will denote these two neutral states by $H$ and $A$, although there is no way to identify which of these two states is CP-even and which is CP-odd.\footnote{Equivalently, one can propose two different definitions of CP (called, say, CPa and CPb), such that $H$ is CPa-even and $A$ is CPa-odd, and vice versa for CPb.  Either definition can be consistently used to define the CP symmetry of the bosonic sector of the IDM. \label{cpab}}
It is possible that $h$ is degenerate in mass with either $H$ or $A$, but such mass degeneracies are accidental in nature since neither case can arise due to a symmetry.  Moreover, these mass-degenerate states are physically distinguishable, since $h$ is even whereas $H$ and $A$ are odd under the $\mathbb{Z}_2$ symmetry.   In contrast, an exact mass degeneracy of $H$ and $A$ can arise if the $\mathbb{Z}_2$ symmetry of the scalar potential is promoted to a continuous U(1) symmetry.   In our terminology, this mass degeneracy of $H$ and $A$  is natural.  Nevertheless, the two mass-degenerate states can still be physically distinguished due to the coupling of these states to $W^\pm H^\mp$.  

One can now extend the above analysis to an arbitrary 2HDM.  
One can show that with one exception,  all 2HDM mass degeneracies are accidental. 
The one exceptional case of a natural mass degeneracy is precisely the case of $m_H=m_A$ in the IDM.   
This conclusion can also be obtained by considering all possible symmetries of the 2HDM scalar potential.   Among these symmetries, we can identify those that can potentially guarantee the mass degeneracy of scalar states.   By examining the consequences of these symmetries, we again confirm that 
the only possible neutral scalar mass degeneracy in the 2HDM arises in the IDM as previously noted.

In section \ref{threedmdegenerate} we consider possible mass-degeneracies in the three Higgs doublet model. 
Using the previous 2HDM analysis of mass degeneracies of the IDM, we construct a three Higgs doublet model (3HDM) generalization of the IDM, which we call the replicated inert doublet model (RIDM).  In this model, two of the three Higgs doublets are inert, and  
four mass-degenerate scalar pairs exist (two involving the charged scalar states from the inert doublets and two involving the neutral scalar states from the inert doublets).  We can explicitly identify the symmetries that are responsible for these mass degeneracies.  We then investigate the possibility of adding new terms to the scalar potential that partially break these symmetries while preserving the mass degeneracies.   In this way, we arrive at a model first proposed by Ivanov and Silva~\cite{Ivanov:2015mwl}. The Ivanov and Silva scalar potential
possesses a discrete subgroup of the continuous symmetries that govern the RIDM, that maintains the mass degeneracies of the RIDM.   This discrete subgroup is the generalized CP symmetry, CP4, which has the property that
$({\rm CP}4)^n$ is the identity operator only for integer $n$ values that are multiples of 4.  The CP4 symmetry is distinguished from the conventional CP symmetry (denoted henceforth by CP2), which has the property that $(\rm{CP}2)^2$ is the identity operator. Some properties of specialized 3HDMs have also been analyzed recently in Ref.\cite{Kopke:2018vyw}.

One of the most notable properties of the Ivanov-Silva (IS) model is that one can write down the most general CP4-invariant scalar potential with three Higgs doublets, which has the feature that at least one of the coefficients of the quartic terms of the scalar potential must be complex (with a nonvanishing imaginary part).   
Indeed, as demonstrated explicitly in Appendix A,
one cannot redefine the scalar fields within the family of Higgs bases
such that all the coefficients of the scalar potential are real.  
In this case, we say that no real Higgs basis exists.
This means that CP2 is not a symmetry of the IS scalar potential and vacuum.

In section~\ref{sect:Z-to-QQQQ}, we identify the existence of a physical observable that is present if 
no CP2 symmetry exists that commutes with the CP4 symmetry of the IS model.\footnote{However, this leaves open the possibility of the existence of a CP2 symmetry that does not commute with the CP4 symmetry~\cite{Ivanov:2018ime}; in this case, a real Higgs basis exists and a conventional CP symmetry can be defined.}
As an example, we focus on $Z$ decay into four inert neutral scalars (with some details relegated to Appendix~B).
Nevertheless, the CP4 invariance guarantees that all CP-violating observables involving the Higgs/gauge boson sector of the theory must be absent.  
For example, we provide an instructive analysis in section~\ref{sect:ZZZ-ZWW}  that shows how the CP4 symmetry of the IS model with no real Higgs basis ensures the cancellation of contributions to the CP-violating form factors of the effective $ZZZ$ and $ZW^+ W^-$ vertices up to three-loop order.   
Finally, we state our conclusions in section~\ref{conclude}.  

\section{2HDM mass degeneracies}
\label{thdmdegenerate}
\setcounter{equation}{0}

Consider the 2HDM, consisting of two hypercharge-one, doublet scalar fields, $\Phi_1$ and $\Phi_2$.
The most general gauge-invariant renormalizable scalar potential is
\beqa
\mathcal{V}&=& m_{11}^2 \Phi_1^\dagger \Phi_1+ m_{22}^2 \Phi_2^\dagger \Phi_2 -[m_{12}^2
\Phi_1^\dagger \Phi_2+{\rm h.c.}]
+\half \lambda_1(\Phi_1^\dagger \Phi_1)^2+\half \lambda_2(\Phi_2^\dagger \Phi_2)^2
+\lambda_3(\Phi_1^\dagger \Phi_1)(\Phi_2^\dagger \Phi_2)
\nn\\
&&\quad
+\lambda_4( \Phi_1^\dagger \Phi_2)(\Phi_2^\dagger \Phi_1)
 +\left\{\half \lambda_5 (\Phi_1^\dagger \Phi_2)^2 +\big[\lambda_6 (\Phi_1^\dagger
\Phi_1) +\lambda_7 (\Phi_2^\dagger \Phi_2)\big] \Phi_1^\dagger \Phi_2+{\rm
h.c.}\right\}\,.\label{genpot}
\eeqa
We shall assume that the minimum of the scalar potential is electric charge conserving, in which case
only the neutral scalar fields possess a nonzero vacuum expectation value (vev), $\vev{\Phi_i^0}=v_i/\sqrt{2}$, where the $v_i$ are potentially
complex.  The Fermi constant, $G_F$ fixes the value of 
\beq
v^2\equiv |v_1|^2+|v_2|^2=(\sqrt{2}G_F)^{-1}\simeq (246~{\rm GeV})^2\,.
\eeq

Employing  a new scalar field basis consisting of two orthonormal linear combinations of $\Phi_1$ and $\Phi_2$ does not modify the physical predictions of the model.
One convenient choice is the Higgs basis, in which the redefined doublet fields (denoted below
by $H_1$ and $H_2$) have the property that $H_1$ has a non-zero vev whereas $H_2$ has a zero
vev~\cite{Donoghue:1978cj,Georgi:1978ri,Botella:1994cs}. In particular, we define new Higgs doublet fields:
\beq \label{higgsbasispot}
H_1=\begin{pmatrix}H_1^+\\ H_1^0\end{pmatrix}\equiv \frac{1}{v}(v_1^* \Phi_1+v_2^*\Phi_2)\,,
\qquad\quad 
H_2=\begin{pmatrix} H_2^+\\ H_2^0\end{pmatrix}\equiv\frac{1}{v}(-v_2 \Phi_1+v_1\Phi_2)\,.
\eeq
It follows that $\vev{H_1^0}=v$ and $\vev{H_2^0}=0$.
The Higgs basis is uniquely defined
up to an overall rephasing, $H_2\to e^{i\chi} H_2$ (which does not alter the fact that
$\vev{H_2^0}=0$).  In the Higgs basis, the scalar potential of Eq.~(\ref{genpot})  is
denoted as~\cite{Branco:1999fs,Davidson:2005cw}:
\beqa \mathcal{V}&=& Y_1 H_1^\dagger H_1+ Y_2 H_2^\dagger H_2 +[Y_3
H_1^\dagger H_2+{\rm h.c.}]
+\half Z_1(H_1^\dagger H_1)^2+\half Z_2(H_2^\dagger H_2)^2
+Z_3(H_1^\dagger H_1)(H_2^\dagger H_2)
\nn\\
&&\quad
+Z_4( H_1^\dagger H_2)(H_2^\dagger H_1)
+\left\{\half Z_5 (H_1^\dagger H_2)^2 +\big[Z_6 (H_1^\dagger
H_1) +Z_7 (H_2^\dagger H_2)\big] H_1^\dagger H_2+{\rm
h.c.}\right\}\,,\label{hbasispot}
\eeqa
where $Y_1$, $Y_2$ and $Z_1,\ldots,Z_4$ are real parameters,
whereas $Y_3$, $Z_5$, $Z_6$ and $Z_7$ are potentially complex parameters.
Imposing the scalar potential minimum conditions yields,
\beq \label{YZ}
Y_1=-\half Z_1 v^2\,,\qquad\quad Y_3=-\half Z_6 v^2\,.
\eeq

\subsection{Mass degeneracies of the inert doublet model (IDM)}
\label{sec:idm}

We wish to study the consequences of a 2HDM in which two or three of the neutral Higgs scalars are degenerate in mass.
For simplicity, we shall first specialize to the inert 2HDM (the so-called IDM)\cite{Ma:2006km,Barbieri:2006dq} in which there is an exact discrete $\mathbb{Z}_2$ 
symmetry that is preserved by the vacuum, under which all particles of the SM and one of the two Higgs doublet fields 
(which contains the observed Higgs boson) are even and the second
Higgs doublet field is odd under the multiplicative discrete symmetry. In particular, the discrete symmetry of the IDM is manifest in the Higgs basis,
where we identify $H_1$ as even and $H_2$ as odd under the $\mathbb{Z}_2$ symmetry.   
It then follows that $Y_3=Z_6=Z_7=0$.

The IDM scalar potential is CP-conserving since one can eliminate the phase of $Z_5$ (the only remaining potentially complex scalar potential parameter) by appropriately rephasing the Higgs basis field $H_2$.   The Higgs basis fields are 
\beq
H_1=\begin{pmatrix} G^+ \\ \frac{1}{\sqrt{2}}\bigl[v+h+iG^0\bigr] \end{pmatrix}\,,\qquad\quad H_2=\begin{pmatrix} H^+ \\ \frac{1}{\sqrt{2}}\bigl[H+iA\bigr]\end{pmatrix}\,,
\eeq
where $G^\pm$ and $G^0$ are the Goldstone bosons that provide the longitudinal degrees of freedom of the massive $W^\pm$ and $Z^0$ gauge bosons.
The physical mass spectrum of the IDM is given by,
\beqa
&&m_h^2=Z_1 v^2\,,\qquad\qquad\qquad\qquad\quad m^2_{H^\pm}=Y_2+\half Z_3 v^2\,,\label{mass1}\\
&&m_{A}^2=m_{H^\pm}^2+\half(Z_4- Z_5)v^2\,,\qquad\,\, m_H^2=m_A^2+Z_5 v^2\,.\label{mass2}
\eeqa
\clearpage

For completeness, we exhibit the Higgs couplings of the IDM in the unitary gauge below (where the Goldstone fields are set to zero).  First, the interactions of the Higgs bosons and the gauge bosons are governed by,\footnote{The photon field $A_\mu$ should not be confused with the scalar field $A$.}
\beqa
\mathscr{L}_{VVH}&=&\left(gm_W W_\mu^+W^{\mu\,-}+\frac{g}{2c_W} m_Z
Z_\mu Z^\mu\right)h
\,,\label{VVH}\\[8pt]
\mathscr{L}_{VVHH}&=&\left[\quarter g^2 W_\mu^+W^{\mu\,-}
+\frac{g^2}{8c_W^2}Z_\mu Z^\mu\right](h^2+H^2+A^2) \nonumber \\
&& +\left[\half g^2 W_\mu^+ W^{\mu\,-}+
e^2A_\mu A^\mu+\frac{g^2}{c_W^2}\left(\half
-s_W^2\right)^2Z_\mu Z^\mu +\frac{2ge}{c_W}\left(\half
-s_W^2\right)A_\mu Z^\mu\right]H^+H^-
\nn \\
&& +\biggl\{ \left(\half eg A^\mu W_\mu^+
-\frac{g^2s_W^2}{2c_W}Z^\mu W_\mu^+\right)
H^-(H+iA) +{\rm h.c.}\biggr\}
\,,\label{VVHH} \\[8pt]
\mathscr{L}_{VHH}&=&\frac{g}{2c_W}\,
Z^\mu A\,\ddel_\mu H -\half g\biggl[iW_\mu^+
H^-\ddel\lsup{\,\mu} (H+iA)
+{\rm h.c.}\biggr] \nn \\
&& +\left[ieA^\mu+\frac{ig}{c_W}\left(\half -s_W^2\right)
Z^\mu\right]H^+\ddel_\mu H^-, \label{VHH}
\eeqa
where $s_W\equiv \sin\theta_W$, $c_W\equiv\cos\theta_W$.
The trilinear and quadrilinear Higgs self-interactions are governed by
\beqa
\mathscr{L}_{3h}&=&-\half v\bigl[Z_1 h^3+(Z_3+Z_4)h(H^2+A^2)+Z_5h(H^2-A^2)\bigr]-vZ_3hH^+H^-\,, \label{hhh} \\[6pt]
\mathscr{L}_{4h}&=&
-\tfrac{1}{8}\bigl[Z_1 h^4+Z_2(H^2+A^2)^2+2(Z_3+Z_4)h^2(H^2+A^2)+2Z_5 h^2(H^2-A^2)\bigr]\nn \\[6pt]
&&-\half H^+ H^-\bigl[Z_2(H^2+A^2+H^+ H^-)+Z_3 h^2\bigr].   \label{hhhh}
\eeqa

The tree-level couplings of $h$ to SM particles obtained above are precisely those of the SM Higgs boson, corresponding to the exact Higgs alignment limit~\cite{Craig:2013hca,Haber:2013mia,Asner:2013psa,Carena:2013ooa,Carena:2014nza,Dev:2014yca,Bernon:2015qea,Bernon:2015wef,Pilaftsis:2016erj}
(as expected in light of $Z_6=0$).  Moreover,
an examination of the above couplings implies that $h$ is CP-even (to be identified with the SM Higgs boson) and $H$ and $A$ have opposite CP-quantum numbers (one is odd and the other is even) based on the $ZAH$ coupling.%
\footnote{Under the rephasing, $H_2\to iH_2$, we note that $Z_5\to -Z_5$, $H\to -A$ and $A\to H$.  One can check that the masses of $H$ and $A$ and their couplings are invariant under this rephasing.}    Note that the CP is not uniquely defined by the IDM interactions, since two candidate definitions of CP exist (called CPa and CPb in footnote~\ref{cpab}), where $H$ is CPa-even and $A$ is CPa-odd, and vice versa for CPb.
Either definition of CP can be used consistently in exploring the phenomenology of the IDM.

Finally, we note that under the $\mathbb{Z}_2$ symmetry of the IDM,
the quarks and leptons can be chosen to be even.   Consequently,  the tree-level couplings of $h$ to fermion pairs are identical to those of the SM Higgs boson, whereas $H$, $A$ and $H^\pm$ do not couple to the SM fermions.

We now examine the possibility of mass degeneracies in the IDM.  First, consider the case of $m_h=m_H$ or $m_h=m_A$.   In this case, it is possible to physically distinguish between $h$ and its mass-degenerate partner due to their opposite $\mathbb{Z}_2$ quantum numbers.  For example, since all SM bosons and fermions are even under the $\mathbb{Z}_2$ symmetry, it follows that
the gluon-gluon (via a top quark loop), $WW$ and $ZZ$ fusion processes can only produce $h$ whereas Drell-Yan production (via virtual $s$-channel $Z$ exchange) can only produce $H$ in association with $A$.   Hence, despite the mass degeneracy, the two mass-degenerate scalars are physically distinguishable.  Note that the mass degeneracy of $h$ and its scalar partner is not radiatively stable.  For example, if $h$ and $H$ are mass degenerate states, then the one-loop contributions to the $hh$ two-point function (such as $ZZ$ and $WW$ intermediate states) differ from the corresponding contributions to the $HH$ two-point function (e.g. the $AZ$ intermediate state).  Indeed, the tree-level condition for the mass degeneracy of $h$ and $H$, 
\beq \label{cond}
Z_1v^2=Y_2+\half Z_{345}v^2,
\eeq
where $Z_{345}\equiv Z_3+Z_4+Z_5$, is unnatural; i.e., \eq{cond} is not the result of some symmetry.\footnote{The scenario where $m_h=m_H=m_A$ is a special case of the $h$--$H$ mass degeneracy.  In the triply mass-degenerate scenario, $h$ is also distinguished from $\phi^\pm$ by its U(1) charge, which is zero. For example, there is no coupling of $ZW^\pm H^\mp h$ in contrast to the $Z W^\pm H^\mp\phi^\pm$ couplings exhibited in \eq{lint}.}

Second, consider the case of $m_H=m_A$, which corresponds to $Z_5=0$.  In this case, the IDM scalar potential possesses a continuous U(1) symmetry, which is not spontaneously broken by the vacuum.\footnote{This case has also been noted in Ref.~\cite{Pilaftsis:2016erj}.}  
It is this symmetry that is responsible for the mass degenerate states $H$ and $A$.  One can now define eigenstates of U(1) charge,
\beq \label{PQcharge}
\phi^\pm=\frac{1}{\sqrt{2}}\bigl[H\pm iA\bigr]\,.
\eeq
The relevant interaction terms of $\phi^{\pm}$ are
\beqa
\mathscr{L}_{\rm int}&=&\left[\half g^2 W_\mu^+W^{\mu\,-}
+\frac{g^2}{4c_W^2}Z_\mu Z^\mu\right]\phi^+\phi^- +\frac{ig}{2c_W}Z^\mu\phi^-\ddel_\mu\phi^+ -\frac{g}{\sqrt2}\biggl[iW_\mu^+
H^-\ddel\lsup{\,\mu} \phi^+ +{\rm h.c.}\biggr]  \nn \\[6pt]
&& +\frac{eg}{\sqrt2} \left(A^\mu W_\mu^+H^-\phi^+ + A^\mu W_\mu^-H^+\phi^-\right)
-\frac{g^2s_W^2}{\sqrt2c_W}\left(Z^\mu W_\mu^+H^-\phi^+ + Z^\mu W_\mu^-H^+\phi^-\right)
\nn \\[6pt]
&&- v(Z_3+Z_4)h\phi^+\phi^--\half\bigl[Z_2(\phi^+\phi^-)^2+(Z_3+Z_4)h^2\phi^+\phi^-\bigr]-Z_2 H^+ H^-\phi^+\phi^-\,.\label{lint}
\eeqa
Although $\phi^\pm$ are mass degenerate states, they can be distinguished.  For example, Drell-Yan production via a virtual $s$-channel $W^+$ exchange can produce $H^+$ in association with $\phi^-$, whereas virtual $s$-channel $W^-$ exchange can produce $H^-$ in association with $\phi^+$.  Thus, the sign of the charged Higgs boson reveals the U(1)-charge of the produced neutral scalar. The origin of this correlation lies in the fact that, by construction, $H^+$ and $\phi^+$ both reside in $H_2$, whereas $H^-$ and $\phi^-$ both reside in $H_2^\dagger$.

\subsection{2HDM mass degeneracies beyond the IDM}
\label{sec:idmbeyond}

Although the IDM is a rather special case among all possible 2HDMs,
the conclusions concerning mass degeneracies are robust.   Allowing for the most general
2HDM scalar potential, the Higgs sector is CP-violating if the relative phases among $Z_5$, $Z_6$ and $Z_7$ cannot be removed by rephasing the Higgs basis field $H_2$.  
It is convenient to introduce three invariant quantities\cite{Lavoura:1994fv,Botella:1994cs,Davidson:2005cw}, whose imaginary parts are given by,\footnote{Basis-invariant expressions for the $J_i$ are given in Ref.\cite{Davidson:2005cw}.}
\beq \label{imJ}
\Im J_1=\Im(Z_6^* Z_7)\,,\qquad \Im J_2=\Im(Z_5^* Z_6^2)\,,\qquad \Im J_3=\Im\bigl[Z_5^*(Z_6+Z_7)^2\bigr]\,.
\eeq
The Higgs sector of the 2HDM is CP-violating unless $\Im J_1=\Im J_2=\Im J_3=0$.
The origin of the CP-violation can either be explicit or 
spontaneous~\cite{Lee:1973iz,Branco:1985aq}.\footnote{To
test for explicit CP-violation, one must employ invariant quantities that are independent of the scalar field vacuum expectation values~\cite{Branco:2005em,Gunion:2005ja,Grzadkowski:2016szj}.}  

Note that the neutral scalar squared-mass matrix does not involve the Higgs basis parameter~$Z_7$.  In particular, $Z_7$ only enters in the Higgs boson cubic and quartic self-couplings.
Hence, if $\Im J_2=\Im(Z_5^* Z_6^2)=0$, then the neutral Higgs mass eigenstates behave like eigenstates of CP in their tree-level interactions with the gauge bosons and fermions  (independently of the values of $\Im J_1$ and $\Im J_3$).
Moreover, the neutral scalar squared-mass matrix breaks up into a block diagonal form consisting of a $2\times 2$ block (whose diagonalization yields the two CP-even neutral scalars) and a $1\times 1$ block (which yields the CP-odd neutral scalar).

Consider the possibility of mass degeneracies among the neutral scalars of the most general 2HDM.  
We now recall a remarkable tree-level relation of the CP-violating 2HDM\cite{Mendez:1991gp,Lavoura:1994fv,Haber:2006ue},
\begin{align} \label{remarkable}
\Im J_2&={\rm Im}(Z_5^* Z_6^2)=\frac{2 s_{13}c_{13}^2 s_{12}c_{12}}{v^6}(m_2^2-m_1^2)(m_3^2-m_1^2)(m_3^2-m_2^2)\,,
\end{align}
where the $m_i$ ($i=1,2,3$) are the masses of the three neutral Higgs bosons of the 2HDM, $s_{12}\equiv\sin\theta_{12}$, $c_{12}\equiv \cos\theta_{12}$, etc., and
$\theta_{12}$ and $\theta_{13}$ are invariant mixing angles that are associated with the diagonalization of the neutral Higgs squared-mass matrix in the Higgs basis.\footnote{Details on the definition of the mixing angles and their relations to the Higgs basis scalar potential parameters can be found in Ref.\cite{Haber:2006ue}.  However, we will not need any of these details for the present argument.} 
In Ref.~\cite{Grzadkowski:2014ada}, the three CP-odd invariants $\Im J_i$ have been expressed  in terms of the neutral scalar masses and the couplings of the neutral Higgs mass eigenstates to charged pairs, $e_i$ $(H_iW^+W^-)$ and $q_i$ $(H_iH^+H^-)$.    
In particular,\footnote{In Ref.~\cite{Lavoura:1994fv}, the invariant defined in \eq{Eq:Im_J2} is called $J_1$.}
\begin{equation} \label{Eq:Im_J2}
\Im J_2=\frac{2e_1 e_2 e_3}{v^9}(m_1^2-m_2^2)(m_2^2-m_3^2)(m_3^2-m_1^2).
\end{equation}

If any two of the three neutral Higgs bosons are mass-degenerate, then either \eq {remarkable} or (\ref{Eq:Im_J2}) implies that $\Im J_2=0$, and the corresponding neutral scalar mass-eigenstates will behave as states of definite CP in their interactions with gauge bosons and fermions.
Nevertheless, if $\Im J_1\neq 0$ and/or $\Im J_3\neq 0$ (which would imply that $\Im Z_7\neq 0$ in the Higgs basis where $Z_5$ and $Z_6$ are simultaneously real), then CP-violating Higgs self-couplings must be present.
Moreover, radiative corrections will generate a non-zero $\Im J_2$ and yield neutral Higgs states of indefinite CP.   That is, if CP-violation in the scalar sector is present, the tree-level relation $\Im J_2=0$ can only be realized via an artificial fine-tuning of the parameters.  Nevertheless, one can consider the implications of a tree-level mass degeneracy among the neutral Higgs scalars of the 2HDM.
The above discussion illustrates the power of using scalar basis invariant conditions to analyze the CP properties of 
multi-Higgs models~\cite{Branco:2005em,Gunion:2005ja,Branco:2015gna,Varzielas:2016zjc,deMedeirosVarzielas:2017ote}.

In light of \eq{imJ}, if a tree-level mass degeneracy among the neutral Higgs scalars is present, then it is possible to rephase the Higgs basis field $H_2$ such that $Z_5$ and $Z_6$ are simultaneously real. 
Thus, in the analysis that follows, we shall analyze the most general 2HDM scalar potential assuming that $Z_5$ and $Z_6$ are real parameters.
The squared-masses of the charged Higgs boson, $H^\pm$, and the CP-odd Higgs boson, $A$ are given by,\footnote{In the notation employed in \eqst{ma}{dhhhh}, $h$ and $H$ [$A$] refer to the neutral scalars that behave as CP-even [odd] mass eigenstates in their tree-level interactions with the gauge bosons and fermions.  Indeed, CP-violating interactions are present in \eqs{dhhh}{dhhhh} if $\Im Z_7\neq 0$.}
\beq \label{ma}
m_{H^\pm}^2=Y_2+\half Z_3 v^2\,,\qquad\quad m_A^2=m_{H^\pm}^2+\half(Z_4-Z_5)v^2\,.
\eeq
The squared-masses of the CP-even Higgs bosons, $h$ and $H$ are the eigenvalues of the $2\times 2$ matrix,
\beq
\mathcal{M}_H^2=\begin{pmatrix} Z_1 v^2 & \quad Z_6 v^2 \\  Z_6 v^2 & \quad m_A^2+Z_5 v^2\end{pmatrix}.
\eeq
That is,
\beq \label{mhH}
m^2_{H,h}=\frac12\biggl\{m_A^2+(Z_1+Z_5)v^2\pm\sqrt{[m_A^2-(Z_1-Z_5)v^2]^2+4Z_6^2 v^4\,}\biggr\}\,.
\eeq
Mass degenerate states arise if one of the following two quantities is zero,
\beq \label{twocond}
Z_5(m_A^2-Z_1 v^2)+Z_6^2 v^2=0\quad \text{or}\quad \bigl[m_A^2-(Z_1-Z_5)v^2\bigr]^2+4Z_6^2 v^4=0\,,
\eeq
where $m_A^2$ is given by \eq{ma}.

The case of $m_h=m_H$ arises when the second condition given in \eq{twocond} is satisfied.  It then follows that $m_A^2=(Z_1-Z_5)v^2$ and $Z_6=0$, and the latter then yields the IDM mass spectrum.   
As in the case of the IDM, the mass degeneracy of $h$ and $H$ requires a fine tuning of the parameters shown in \eq{cond}.  In principle, it is possible that $Z_7\neq 0$, but in this case, $Z_6=0$ is not a natural condition since the $\mathbb{Z}_2$ symmetry of the IDM is not present.  Nevertheless, even if one accepts the two fine tuned conditions needed in this scenario, the arguments presented above \eq{cond} still apply.  Namely, $Z_6=0$ corresponds to the exact alignment limit (at tree-level), in which case the tree-level interactions of the Higgs bosons and gauge bosons are still the same as those of the IDM [cf.~\eqst{VVH}{VHH}], whereas the tree-level trilinear and quadrilinear Higgs self-interactions given in \eqs{hhh}{hhhh} are modified by the addition of the following
terms, 
\beqa
\delta\mathscr{L}_{3h}&=&-\tfrac14 v \bigl[Z_7(H+iA)+Z_7^*(H-iA)\bigr](HH+AA+2H^+ H^-)\,,\label{dhhh} \\[5pt]
\delta\mathscr{L}_{4h}&=&-\tfrac14\bigl[Z_7(H+iA)+Z_7^*(H-iA)\bigr](HH+AA+2H^+ H^-)h\,,\label{dhhhh}
\eeqa
after rephasing the Higgs basis field $H_2$ such that $Z_5$ is real.

The cases $m_h=m_A$ or $m_H=m_A$ arise when the first condition given in \eq{twocond} is satisfied.  This condition also requires a fine-tuning of the parameters.
Moreover,  approximate Higgs alignment (as suggested by the LHC Higgs data) is not achieved unless $m_A^2\gg Z_1 v^2$ or $|Z_6|\ll 1$.
Nevertheless, the physical distinction of the mass degenerate states is due to the CP quantum numbers of the neutral scalar states (which are preserved in the tree-level Higgs interactions with gauge bosons and with fermions).  
One can therefore distinguish between the corresponding production mechanisms of the degenerate scalars that are mediated by gauge boson fusion or Drell-Yan production via $s$-channel gauge boson exchange.  

Finally, we consider the triply mass-degenerate case of $m_h=m_H=m_A$.  In this case, both conditions given in \eq{twocond} must be satisfied, which yields
$Z_5=Z_6=0$ and $m_A^2=Z_1 v^2$.   This leaves $Z_7$ as the only potentially complex parameter of the scalar potential in the Higgs basis.  Thus, one is free to rephase the Higgs basis field $H_2$ such that $Z_7$ is real, and we conclude that the Higgs scalar potential and vacuum must be CP-conserving.
However, as long as $Z_7\neq 0$, the triply mass-degenerate case is unnatural,  since the $\mathbb{Z}_2$ symmetry of the IDM is not present.

\subsection{Natural 2HDM mass degeneracies: a symmetry based approach}
\label{sec:naturaldegen}

In sections \ref{sec:idm} and \ref{sec:idmbeyond}, we derived the conditions that yield mass degeneracies among the neutral scalars of the 2HDM by brute force.   Namely, we obtained explicit expressions for the neutral scalar masses and then derived the corresponding relations among Higgs basis parameters for which mass degeneracies were present.   We then checked whether any of these relations were a consequence of a symmetry, and if yes we concluded that the corresponding mass degeneracy was natural.
In this section, we will obtain the same result by considering all possible symmetries of the 2HDM scalar potential.  Since the complete list of such symmetries is known~\cite{Deshpande:1977rw,Ivanov:2007de,Ferreira:2009wh,Ferreira:2010yh,Battye:2011jj,Pilaftsis:2011ed}, we can be sure that our catalog of natural mass degeneracies of the 2HDM is complete.

We shall make use of the classification of symmetries presented in Ref.~\cite{Ferreira:2009wh}, which identifies three possible Higgs family symmetries, $\mathbb{Z}_2$, U(1) and SO(3), and three classes of generalized CP-symmetries, denoted by GCP1, GCP2 and GCP3, respectively, as summarized 
in Table~\ref{tab}.\footnote{In  Ref.~\cite{Ferreira:2009wh} the three classes of generalized CP transformations are denoted by CP1, CP2 and CP3 respectively. This nomenclature for the generalized CP-symmetries is awkward, in light of the notation that will be employed in section~\ref{threedmdegenerate}.  To avoid confusion, we have appended the letter G (for ``general'') in denoting the three classes of generalized CP transformations of the 2HDM.} 
In the GCP transformation laws of Table~\ref{tab}, we have introduced the conjugation symbol ${}^{\star}$, which when applied to an SU(2) multiplet of scalar fields is defined by $\Phi^{\star}\equiv \bigl[\Phi^\dagger\bigr]\llsup{T}$,
where the dagger refers both to hermitian conjugation of the quantum field operator when acting on the Hilbert space, and to complex conjugate transpose when acting on an SU(2) multiplet of fields.  
\vskip -0.01in
\begin{table}[hb!]
\begin{tabular}{|cccc|} 
\hline
\rule{0pt}{\normalbaselineskip} 
symmetry & \hspace{0.5in} transformation law  & \hspace{0.7in} \phantom{transformation law}  & \\[2pt]
\hline 
\rule{0pt}{\normalbaselineskip} 
$\mathbb{Z}_2$ & $\Phi_1\to \Phi_1$ &  $\Phi_2\to -\Phi_2$ &\\
U(1)  & $\Phi_1\to  \Phi_1$ &  $\Phi_2\to e^{2i\theta}\Phi_2$ &\\
SO(3) & $\Phi_a\to U_{ab}\Phi_b$ & $U\in {\rm U}(2)/{\rm U}(1)_{\rm Y}$& (\text{for $a$, $b=1,2$} )\\
GCP1 & $\Phi_1\to\Phi_1^{\star}$ &  $\Phi_2\to\Phi_2^{\star}$ & \\
GCP2 & $\Phi_1\to\Phi_2^{\star}$ &  $\Phi_2\to -\Phi_1^{\star}$ & \\
GCP3 & $\Phi_1 \rightarrow \Phi_1^{\star} \cos\theta+\Phi_2^{\star} \sin\theta$ &
$\Phi_2 \rightarrow -\Phi_1^{\star}\sin\theta+\Phi_2^{\star}\cos\theta$ & \text{(for $0<\theta<\half\pi$)}  \\[2pt]
\hline\hline 
\rule{0pt}{\normalbaselineskip} 
$\Pi_2$& $\Phi_1\to\Phi_2$ & $\Phi_2\to \Phi_1$ & \\[2pt]
\hline 
\end{tabular}
\caption{\small Possible symmetries of the 2HDM scalar potential that are respected by the SU(2)$\times$U(1)$_{\rm Y}$ gauge kinetic terms of the scalar fields.  The corresponding symmetry transformation laws are given in a basis where the symmetry is manifest.  Note that a scalar potential that is invariant under the mirror discrete
symmetry, $\Pi_2$, is also invariant under the $\mathbb{Z}_2$ in another scalar field basis~\cite{Davidson:2005cw}. \label{tab}}
\end{table}

\begin{table}[t!]
\begin{tabular}{|ccccccccccccc|}
\hline
\rule{0pt}{\normalbaselineskip} 
symmetry & $m_{11}^2$ &  $m_{22}^2$ & $m_{12}^2$ & $\lambda_1$ &
 $\lambda_2$ & $\lambda_3$ & $\lambda_4$ &
$\Re\lambda_5$ & $\Im\lambda_5$ & $\lambda_6$ & $\lambda_7$ & \\[2pt]
\hline
\rule{0pt}{\normalbaselineskip} 
$\mathbb{Z}_2$ & - & -   & 0 & - 
   &  -  &  -  &  -  &  - &  -
   & 0 & \phm 0  & \\
U(1) & -  & -    & 0  & - 
 & -  & - &  -  & 0 &
0 & 0 & \phm 0 & \\
SO(3)  & - &   $ m_{11}^2$ & 0 & - 
   & $\lambda_1$ &- & $\lambda_1 - \lambda_3$ & 0 &
0 & 0 & \phm 0  & \\
GCP1   & -  & -  & real & - 
 & -  &  - &  -  &
- & 0 & real &\phm real  & \\
GCP2   & -  & $m_{11}^2$ & 0  & - 
  & $\lambda_1$ &  - &  -  & - &-
   &  -  & $-\lambda_6$ & \\
GCP3  & -   & $m_{11}^2$ & 0 & - 
   & $\lambda_1$ &  -  &  - &
$\lambda_1 - \lambda_{3}-\lambda_4$ & 0 & 0 & \phm 0 & \\[2pt]
\hline\hline
\rule{0pt}{\normalbaselineskip} 
$\Pi_2$  & -  & $ m_{11}^2$ & real & - &
    $ \lambda_1$ &  - &   -&  
- & 0 & - & $\lambda_6^*$ &
\\
$\mathbb{Z}_2\oplus\Pi_2$ & -  & $ m_{11}^2$ & $0$ & - &
    $ \lambda_1$ &  - &   - &  
- & 0 & 0 &\phm 0 &
\\
U(1)$\oplus\Pi_2$ & -  & $ m_{11}^2$ & $0$ & -  &  $\lambda_1$  & - & - & 0 & 0 & 0 & \phm 0 &
\\[2pt]
\hline
\end{tabular}
\caption{\small Impact of the symmetries defined in Table~\ref{tab} on the coefficients of the 2HDM scalar potential [cf.~\eq{genpot}] in a 
basis where the symmetry is manifest.  A short dash indicates the absence of a constraint.  A scalar potential that is simultaneously invariant under $\mathbb{Z}_2$ and $\Pi_2$ is
also invariant under GCP2 in another scalar field basis~\cite{Davidson:2005cw,Ferreira:2009wh}.  Likewise, a scalar potential that is simultaneously invariant under U(1) and $\Pi_2$ is also invariant under GCP3 in another scalar field basis~\cite{Ferreira:2009wh}.  The symbol $\oplus$ is being used above to indicate that two symmetries are enforced simultaneously within the same scalar field basis.
\label{tabconstraints}}
\end{table}

 \noindent
 We shall not consider the seven additional accidental symmetries of the 2HDM scalar potential identified in Refs.~\cite{Battye:2011jj,Pilaftsis:2011ed}, that utilized mixed Higgs family and generalized CP transformations that leave the SU(2) gauge kinetic terms of the scalar fields invariant.   An example of such a symmetry is the well known custodial symmetry that is respected by the 2HDM scalar potential when $m_{H^\pm}=m_A$~\cite{Gerard:2007kn,Haber:2010bw,Grzadkowski:2010dj,Nishi:2011gc}.  However, this class of symmetries is violated by the U(1)$_{\rm Y}$ gauge kinetic term of the scalar potential (as well as by the Yukawa couplings that are responsible for mass differences between up and down-type fermions).   Hence, any exact mass degeneracies arising from these seven accidental symmetries will be spoiled, in the absence of an artificial fine tuning of the Higgs scalar potential parameters.\footnote{In cases of accidental symmetries, i.e. symmetries of the scalar potential that are not respected by the full theory,
the would-be mass degeneracies are only approximate, with calculable mass splittings.  The possibility of such approximate mass degeneracies, although technically natural, 
is not the subject of this paper.}

Possible natural mass degeneracy of the 2HDM must be the consequence of one of the symmetries listed in Table~\ref{tab}.   Starting from a generic scalar potential given by \eq{genpot}, if the scalar potential respects one of the symmetries listed in Table~\ref{tab}, then a scalar basis is picked out in which the symmetry is manifest.  In this basis, the coefficients of the scalar potential are constrained according to Table~\ref{tabconstraints}.\footnote{It can be shown that for each of the symmetries listed in Table~\ref{tabconstraints},
a scalar field basis exists in which all scalar potential parameters and the neutral scalar field vacuum expectation values are simultaneously real, in which case CP (as defined by GCP1 in Table~\ref{tab}) is conserved by the scalar sector Lagrangian and vacuum.}
It is straightforward to check that the possible discrete symmetries of the 2HDM, namely $\mathbb{Z}_2$, GCP1, GCP2 (or equivalently, $\mathbb{Z}_2\oplus\Pi_2$), do not yield scalar potentials that lead to scalar mass degeneracies.  Thus, we henceforth focus on U(1), SO(3) and GCP3 (and the related U(1)$\oplus \Pi_2$ symmetry).

Given a 2HDM scalar potential with a Peccei-Quinn [U(1)$_{\rm PQ}$] symmetry~\cite{Peccei:1977ur} (or equivalently the U(1) transformation specified in 
Table~\ref{tab}\footnote{In Ref.~\cite{Peccei:1977ur}, a U(1)$_{\rm PQ}$ transformation of the 2HDM scalar fields is given by $\Phi_1\to e^{-i\theta}\Phi_1$ and $\Phi_2\to e^{i\theta}\Phi_2$.
The U(1) transformation specified in Table~\ref{tab} corresponds to combining
the U(1)$_{\rm PQ}$ transformation with a hypercharge U(1)$_{\rm Y}$ transformation, $\Phi_i\to e^{i\theta}\Phi_i$ (for $i=1,2$).}) 
that is spontaneously broken by the vacuum, the scalar sector will contain a massless CP-odd (Goldstone) scalar~\cite{Weinberg:1977ma,Wilczek:1977pj}.  In such cases, no mass degeneracy is present (without further constraints on the scalar potential parameters).   However, if the U(1) symmetry is manifestly realized in the Higgs basis, then the U(1) symmetry is unbroken by the vacuum,
resulting in a mass degeneracy between the two neutral scalars residing in the Higgs basis field $H_2$.   Indeed, this has already been shown in Section \ref{sec:idm} [see text above \eq{PQcharge}], in the case of the mass degeneracy, $m_H=m_A$, of the IDM with $Z_5=0$. 

In the case of a 2HDM scalar potential with an SO(3) symmetry, the form of the scalar potential is invariant with respect to all possible changes of the scalar basis.  Hence, it follows that the scalar potential parameters in the Higgs basis satisfy $Y_1=Y_2$, $Z_1=Z_2=Z_3+Z_4$ and $Y_3=Z_5=Z_6=Z_7=0$.   Using \eqss{YZ}{mass1}{mass2}, it follows that $m_H=m_A=0$.   The presence of two massless (Goldstone) scalars is a consequence of the spontaneous breaking of the SO(3) global symmetry by the vacuum.  This scalar mass degeneracy is a special case of the mass degeneracy in the case of the IDM with $Z_5=0$, in which additional constraints among the Higgs basis parameters result in the pair of massless scalar states.

Finally, let us consider the case of a 2HDM scalar potential with a GCP3 symmetry.   Suppose that the GCP3 symmetry is manifestly realized in a basis where
\beq \label{phivevs}
\vev{\Phi_1^0}=\frac{v_1}{\sqrt{2}}\,,\qquad\quad
\vev{\Phi_2^0}=\frac{v_2}{\sqrt{2}}e^{i\xi}\,,
\eeq
where $v_1$ and $v_2$ are positive.
We define $\tan\beta\equiv v_2/v_1$ (so that $0<\beta<\half\pi$).   Then, in light of the constraints on the GCP3 scalar potential parameters given in Table~\ref{tabconstraints}, the scalar potential minimum conditions yield (e.g., see eqs.~(3)--(5) of Ref.~\cite{Haber:2015pua}),
\beqa
m_{11}^2 & = & -\half v^2\bigl[\lambda_1-2\lambda_5\sin^2\beta\sin^2\xi\bigr]\,, \\
m_{22}^2 & = & -\half v^2\bigl[\lambda_1-2\lambda_5\cos^2\beta\sin^2\xi\bigr]\,, \\
m_{12}^2\sin\xi&=& v^2\lambda_5 \sin\beta\cos\beta\sin\xi\cos\xi\,.
\eeqa
We can assume that $\lambda_5\neq 0$, since otherwise we would be dealing with an SO(3)-symmetric scalar potential.
Setting $m_{11}^2=m_{22}^2$ and $m_{12}^2=0$ then yields two conditions,
\beq
\sin^2 \xi\cos 2\beta=0\,,\qquad\quad  \sin\xi\cos\xi\sin2\beta=0\,.
\eeq
Hence, there are two classes of vacua,
\begin{enumerate}[\hspace{1cm} A.]
\item $\sin\xi=0$ and $\beta$ arbitrary ($0<\beta<\half\pi$)\,,
\item
$\cos\xi=0$ and $\cos 2\beta=0$.
\end{enumerate}

We now can calculate the parameters of the GCP3 scalar potential in the Higgs basis (e.g., see eqs.~(11)--(20) of Ref.~\cite{Haber:2015pua}) in the two Cases A and B defined above,
\begin{enumerate}[\hspace{1cm} A.]
\item
$Y_1=Y_2, \quad Y_3=Z_6=Z_7=0, \quad Z_1=Z_2=\lambda_1, \qquad\quad Z_3+Z_4=\lambda_1-\lambda_5,\quad Z_5=\lambda_5$\,,
\item
$Y_1=Y_2, \quad Y_3=Z_6=Z_7=0, \quad Z_1=Z_2=\lambda_1-\lambda_5, \quad Z_3+Z_4=\lambda_1+\lambda_5,\quad Z_5=0$\,.
\end{enumerate}
\clearpage

\noindent
In particular, $Z_1=Z_2=Z_3+Z_4+Z_5$ in Case A, and $Z_1=Z_2\neq Z_3+Z_4$ (and $Z_5=0$) in Case~B.
We can now make use of \eqs{mass1}{mass2} [along with \eq{YZ} to eliminate $Y_2$ by virtue of $Y_1=Y_2$] to compute the neutral scalar mass spectrum in the 
two cases,\footnote{Note that the positivity of the squared masses are consistent with the conditions on the 2HDM scalar potential parameters first obtained in Ref.~\cite{Deshpande:1977rw}.}  
\begin{enumerate}[\hspace{1cm} A.]
\item
$m_h^2=Z_1 v^2,\quad m_H^2=0,\quad m_A^2=(Z_3+Z_4-Z_1)v^2,\quad m_{H^\pm}^2=\half(Z_3-Z_1)v^2$\,,
\item
$m_h^2=Z_1 v^2,\quad m^2_H=m^2_A=\half(Z_3+Z_4-Z_1) v^2,\qquad\,\,\, m_{H^\pm}^2=\half(Z_3-Z_1)v^2$\,.
\end{enumerate}
Note that Cases A and B correspond to degenerate vacua, since in both cases the value of the scalar potential (in the Higgs basis) at its minimum is $V_{\rm min}=-\tfrac18 Z_1 v^4=-\tfrac18  v^2m_h^2$.

In light of Table~\ref{tabconstraints}, we can identify case A as corresponding to realizing a GCP3 symmetry in the Higgs basis,\footnote{This means that we can relax the restrictions of $\beta\neq 0, \half\pi$ in defining Case A.   Note that for $\beta=\half\pi$ one can simply interchange the definitions of $\Phi_1$ and $\Phi_2$ to recover the Higgs basis result (corresponding to $\beta=0$).} 
and case B corresponding to realizing a U(1)$\oplus\Pi_2$ symmetry in the Higgs basis.\footnote{This result is not surprising given that U(1)$\oplus\Pi_2$ is equivalent to GCP3 in another scalar field basis.}
In particular, case A exhibits a massless Goldstone boson corresponding to the spontaneous breaking of GCP3 by the vacuum.   In contrast,
in case B, the GCP3 symmetry possesses a continuous U(1) subgroup, which is unbroken by the vacuum, that protects the mass degeneracy, $m_H=m_A$.  Indeed, this case is again a special case of the IDM with $Z_5=0$, where the additional constraints, $Y_1=Y_2$ and $Z_1=Z_2$ are imposed.   

As a check, it is instructive to evaluate the consequences of a 2HDM scalar potential with a U(1)$\oplus\Pi_2$ symmetry that is manifestly realized in the $\Phi_1$--$\Phi_2$ basis.   
In the following, we employ primed coefficients, $\lambda_i^\prime$ to distinguish this case from the one above where the GCP3 symmetry is manifestly realized in the $\Phi_1$--$\Phi_2$ basis.
Using the results of Table~\ref{tabconstraints}, the scalar potential minimum conditions yield,
\beqa
m_{11}^2&=& -\half v^2\bigl[\lambda^\prime_1\cos^2\beta+(\lambda^\prime_3+\lambda^\prime_4)\sin^2\beta\bigr]\,,\\
m_{22}^2&=& -\half v^2\bigl[\lambda^\prime_1\sin^2\beta+(\lambda^\prime_3+\lambda^\prime_4)\cos^2\beta\bigr]\,,
\eeqa
under the assumption that $0<\beta<\half\pi$. 
We can assume that $\lambda^\prime_1\neq \lambda^\prime_3+\lambda^\prime_4$, since otherwise we would be dealing with an SO(3)-symmetric scalar potential. Setting $m_{11}^2=m_{22}^2$ then yields $\cos 2\beta=0$, with $\xi$ arbitrary.  Without loss of generality, one can set $\xi=0$, since the scalar potential is unchanged under a rephasing of $\Phi_2$.  As before, we can now compute the scalar potential parameters in the Higgs basis,
\beqa
&&Y_1=Y_2,\qquad Y_3=Z_6=Z_7=0,\qquad Z_1=Z_2=\half(\lambda^\prime_1+\lambda^\prime_3+\lambda^\prime_4),\nonumber \\
&& Z_3+Z_4=\lambda^\prime_1,\qquad Z_5=\half(\lambda^\prime_1-\lambda^\prime_3-\lambda^\prime_4)\,. \label{PQ2}
\eeqa
In particular, note that $Z_1=Z_2=Z_3+Z_4-Z_5$.
Using \eqss{YZ}{mass1}{mass2}, we obtain 
\beq
m_h^2=Z_1 v^2\,,\quad m_H^2=(Z_3+Z_4-Z_1)v^2\,,\quad m_A^2=0\,,\quad m_{H^\pm}^2=\half(Z_3-Z_1)v^2\,,
\eeq
which is the same mass spectrum as Case A of the GCP3-symmetric scalar potential with $H$ and~$A$ interchanged.   This result can be understood by noting that \eq{PQ2}
takes the standard form of the GCP3-symmetric scalar potential in the Higgs basis after rephasing the Higgs basis field $H_2\to iH_2$, which interchanges $H$ and $A$ and transforms $Z_5\to -Z_5$.  
\clearpage

Finally, the case of $\beta=0$ or $\beta=\half\pi$ must be treated separately and corresponds to a manifest realization of the U(1)$\oplus\Pi_2$ symmetry in the Higgs basis.  This vacuum is degenerate with the one considered above, since in both cases, $V_{\rm min}=-\tfrac18 Z_1 v^4=-\tfrac18  v^2m_h^2$.  Indeed, this latter case corresponds to Case B of the GCP3-symmetric scalar potential treated above, where the neutral scalar mass spectrum exhibits a mass degeneracy, $m_H=m_A$.

In summary, massless scalar (Goldstone boson) states $A$, $H$ or ($A$, $H$) exist in the 2HDM with a scalar potential that exhibits, respectively, a U(1), GCP3 or SO(3) symmetry manifestly realized in a generic $\Phi_1$--$\Phi_2$ basis, which agrees with the results of Table~2 of Ref.~\cite{Pilaftsis:2011ed}.  Nevertheless, in the special cases where U(1) or U(1)$\oplus\Pi_2$  are manifestly realized in the Higgs basis (the latter corresponding to the Case B solution of the GCP3-symmetric scalar potential), the corresponding U(1) subgroups of theses symmetries are \text{not} spontaneously broken by the vacuum, and the neutral scalar mass spectrum exhibits a mass degeneracy,
$m_H=m_A$.   In the case of the SO(3)-symmetric scalar potential, this mass degeneracy is realized by a pair of 
massless Goldstone boson states.  

Thus, we conclude that mass-degenerate neutral scalars can arise naturally in the 2HDM \textit{only} in the case of the IDM with $Z_5=0$.
All other cases of mass-degenerate scalars require an artificial fine-tuning of the scalar potential parameters,
in agreement with the analysis of section~\ref{sec:idmbeyond}.  Furthermore, this conclusion is unaffected by the interactions of the scalars with the vector bosons.  Indeed, the
Higgs boson--gauge boson interactions, $\mathscr{L}_{\rm int}$, given by \eq{lint} show that the global U(1) symmetry responsible for the mass degeneracy of $H$ and $A$ is an exact symmetry of $\mathscr{L}_{\rm int}$.  Finally, as previously noted, 
the Higgs basis field $H_2$ of the IDM is odd whereas
all other scalar, fermion and vector fields are even under the discrete $\mathbb{Z}_2$ symmetry.   
This can be achieved by employing Type-I Yukawa couplings~\cite{Hall:1981bc} where fermions couple only to the Higgs basis field $H_1$.   In this case, the global U(1) symmetry of the IDM scalar potential with $Z_5=0$ will also be respected by the Yukawa interactions. 
However, a GCP3 [or equivalently U(1)$\oplus\Pi_2$] or SO(3) symmetry of the IDM scalar potential will be explicitly broken by the Yukawa interactions.   Hence, 
the U(1)-symmetric IDM is the only 2HDM for which an exact mass degeneracy of $H$ and $A$ can be preserved.

\section{3HDM mass degeneracies and the Ivanov Silva model}
\label{threedmdegenerate}
\setcounter{equation}{0}

In extended Higgs sectors with more than two scalar doublets, it is now possible to have mass-degenerate charged Higgs pairs as well as mass-degenerate neutral scalars \cite{Olaussen:2010aq}.  In this section, we explore new phenomena associated with mass degenerate scalars that arises for the first time in the three-Higgs doublet model (3HDM).
 
As a warmup exercise, we return to the IDM and add a second inert doublet and consider possible mass degeneracies among the scalar fields of the two inert doublets.  We then perturb the resulting model to obtain a version of the 3HDM that is equivalent to a model first introduced by Ivanov and Silva\cite{Ivanov:2015mwl}.

\subsection{The replicated inert doublet model (RIDM)}
\label{sec:ridm}

The IDM introduced in section~\ref{sec:idm} can be generalized by introducing additional inert scalar doublets.   In this section, we consider a 3HDM that consists of two inert hypercharge-one electroweak doublets, in which the inert doublets contain mass-degenerate scalar states.  The resulting models shall be called the replicated inert doublet model (RIDM).  As in the case of the IDM, we work in the Higgs basis in which the first Higgs doublet field $H_1$ contains the SM Higgs boson.  The RIDM consists of $H_1$, with $\vev{H_1^0}=v/\sqrt{2},$ and two inert doublet fields $H_2$ and $H_3$, with $\vev{H_2}=\vev{H_3}=0$, and a scalar potential given by,
\beqa 
\mathcal{V}_{\rm RIDM}&=& Y_1 H_1^\dagger H_1+ Y_2 \left(H_2^\dagger H_2 +H_3^\dagger H_3\right)
+\half Z_1(H_1^\dagger H_1)^2+\half Z_2(H_2^\dagger H_2+H_3^\dagger H_3)^2\nonumber\\
&&
+Z_3(H_1^\dagger H_1)\left(H_2^\dagger H_2+H_3^\dagger H_3\right)
+Z_4\left[( H_1^\dagger H_2)(H_2^\dagger H_1)+( H_1^\dagger H_3)(H_3^\dagger H_1)\right]\nonumber\\
&&
+\half Z_5\left\{ (H_1^\dagger H_2)^2 +(H_2^\dagger H_1)^2+(H_1^\dagger H_3)^2 +(H_3^\dagger H_1)^2\right\}\,.\label{ridmpot}
\eeqa
Without loss of generality, we have chosen $Z_5$ real and non-negative in \eq{ridmpot}, which is always possible by an appropriate rephasing of the scalar fields $H_2$ and $H_3$.  Hence, if follows that the bosonic sector of the RIDM is CP-conserving.

The charged and neutral components of the Higgs basis doublet fields of the RIDM are also mass eigenstate fields,
\beq \label{Hridm}
H_1=\begin{pmatrix} G^+ \\ \frac{1}{\sqrt{2}}\bigl[v+h_{\rm SM}+iG^0\bigr] \end{pmatrix},\quad
H_2=\begin{pmatrix} H^+ \\ \frac{1}{\sqrt{2}}\bigl[H+iA\bigr]\end{pmatrix},\quad
H_3=\begin{pmatrix} h^+ \\ \frac{1}{\sqrt{2}}\bigl[h+ia\bigr]\end{pmatrix},
\eeq
with a minor change of notation from the IDM.  The corresponding squared masses of the 
neutral and charged scalars are given by, 
\beqa
m^2_{H^\pm}&=&m^2_{h^\pm}=Y_2+\half Z_3 v^2\,,\qquad m^2_H=m^2_h=Y_2+\half(Z_3+Z_4+Z_5)v^2\,,\nn \\
m^2_A&=&m^2_a=Y_2+\half(Z_3+Z_4-Z_5)v^2\,.\label{ridmmasses}
\eeqa
By assumption, $Z_5\geq 0$, in which case $m_H=m_h\geq m_A=m_a$.\footnote{In particular, note that if $Z_5=0$ then there is an enhanced mass degeneracy in which $m_H=m_h=m_A=m_a$.  
\label{fn:enhanced}}   
Thus, the RIDM possesses four mass-degenerate scalar pairs: $(H^\pm, h^\pm)$, $(H,h)$ and $(A,a)$.   These mass degeneracies can be understood as a consequence of a continuous global Higgs flavor symmetry (where Higgs flavor corresponds to the multiplicity of Higgs doublets).

In order to explicitly exhibit the relevant symmetries, it is convenient to focus on the neutral scalar states of the doublet fields $H_2$ and $H_3$, denoted henceforth by the complex fields, 
\beq \label{Hzero}
H^0\equiv \frac{H+iA}{\sqrt{2}}\,,\qquad\quad h^0\equiv \frac{h+ia}{\sqrt{2}}\,,
\eeq
respectively.   Let us first focus on the kinetic energy terms and the terms in \eq{ridmpot} in the \textit{absence} of the term proportional to $Z_5$.   Then, one can check that the neutral complex scalar fields $H^0$ and $h^0$ appear only in the combination $H^{0\,\dagger}H^0+h^{0\,\dagger}h^0=\half(H^2+h^2+A^2+a^2)$.   Thus, excluding $Z_5$, the scalar Lagrangian possesses an O(4) global symmetry, that is responsible for four mass-degenerate neutral scalar states.

It is instructive to see how this symmetry arises when employing the complex basis $\varphi_i=\{H^0,h^0\}$ (for $i=1,2$).  Noting that $\varphi^{\dagger\,i}\varphi_i=H^{0\,\dagger}H^0+h^{0\,\dagger}h^0$ (the sum over the repeated index $i$ is implicit), it is clear that
the scalar Lagrangian (in the absence of $Z_5$) is invariant under a U(2) global symmetry, $\varphi_i\to U_i{}^j\varphi_j$, with $U\in{\rm U(2)}$.  However, the corresponding symmetry group is in fact larger than U(2).  Working in the complex basis, it is straightforward to verify that the quantity
$\varphi^{\dagger\,i}\varphi_i$ is invariant with respect to 
\beq \label{global}
\varphi_i\to U_i{}^j\varphi_j+(V^*)^i{}_j\varphi^{\dagger\,j}\,,
\eeq
where $U$ and $V$ are complex $2\times 2$ matrices (which are not in general unitary), provided that the following two conditions are satisfied:
\beqa
&&(i)~~~~(U^\dagger U+V^\dagger V)_i{}^j=\delta_i{}^j\,,
\label{ucond1}\\
&&(ii)~~~V^T U~\hbox{\rm is an antisymmetric matrix}\,.\label{ucond2}
\eeqa

One can now check that \eq{global} corresponds to an O(4) symmetry transformation.  More explicitly, the $4\times 4$ matrix,
\beq \label{qortho}
\mathcal{Q}=\left(\begin{array}{cc}\Re(U+V) &\,\, -\Im(U+V) \\
\Im(U-V) & \phantom{-}\,\, \Re(U-V) \end{array}\right)
\eeq
is an orthogonal matrix if and only if $U$ and $V$ satisfy \eqs{ucond1}{ucond2}.  Indeed, one can check that in light of \eqs{ucond1}{ucond2}, the global symmetry specified by \eq{global} is governed by 6 continuous parameters as expected for an O(4) transformation. 
Two special cases of \eq{global} are noteworthy.  First, if $V=0$, then $U$ is unitary and we recover the U(2) global symmetry mentioned previously.  Second, if $U=0$ then \eq{global} corresponds to a generalized CP transformation [cf.~\eq{gcptrans}].\footnote{The unified treatment of Higgs family transformations and generalized CP transformations has been advocated previously in Ref.~\cite{Battye:2011jj}.  A related discussion emphasizing the promotion of the U(2) basis transformation to an enlarged group of O(4) transformations appears in Ref.~\cite{Aranda:2016qmp}.}   
Both symmetries are present in the scalar Lagrangian if the $Z_5$ coupling is neglected,
and either one would be sufficient to guarantee the mass degeneracy of 
$H$, $h$, $A$ and $a$.

In the absence of the $Z_5$ coupling, the full O(4) global symmetry is respected by the pure scalar Lagrangian.  However, when we include the coupling of the scalar doublets to the gauge bosons, one must replace the ordinary derivative, $\partial_\mu$, with the SU(2)$\times$U(1) gauge covariant derivative, $D_\mu$, in the scalar kinetic energy term.  The resulting coupling of the scalars to the vector bosons partially breaks the O(4) symmetry.  Employing the complex basis, it is easy to check that the symmetry transformation specified by \eq{global} is unbroken if and only if either $U=0$ or $V=0$, namely the two special cases just highlighted above.\footnote{This result is not surprising given that \eq{global} transforms the scalar field into a linear combination of two fields of opposite hypercharge unless either $U=0$ or $V=0$.}   
That is, the kinetic energy term $(D^\mu\varphi)^{i\,\dagger} (D_\mu\varphi)_i$ is invariant under a U(2) symmetry (corresponding to $V=0$) and under the generalized CP symmetry (corresponding to $U=0$).   Mathematically, the unbroken global symmetry that remains is the semi-direct product U(2)$\rtimes\mathbb{Z}_2$.\footnote{This symmetry is a generalization of the U(1) symmetry (and the associated CP symmetry) of the IDM with $Z_5=0$ treated in section~\ref{sec:idm}, and provides the motivation for our choice of the RIDM scalar potential given in \eq{ridmpot}.}  

We now examine the consequence of including the term of \eq{ridmpot} proportional to $Z_5$.  Focusing again on the neutral complex scalar fields $H^0$ and $h^0$ [cf.~\eq{Hzero}], we see that a new combination of fields arises, $\varphi_i\varphi_i+{\rm h.c.}=(H^0)^2+(H^{0\,\dagger})^2+(h^0)^2+(h^{0\,\dagger})^2$.   This term is invariant with respect to \eq{global} provided that the conditions specified in \eqs{ucond1}{ucond2} are replaced by
\enlargethispage{1.25\baselineskip}
\beqa
&&(i^\prime)~~~~(U^T U+V^T V)_i{}^j=\delta_i{}^j\,,
\label{ucond1p}\\
&&(ii^\prime)~~~V^\dagger U~\hbox{\rm is an antihermitian matrix}\,.\label{ucond2p}
\eeqa
The conditions specified by \eqs{ucond1p}{ucond2p} are compatible with those of \eqs{ucond1}{ucond2} if $U$ and $V$ are real matrices.   

Consequently, $\mathcal{Q}$ specified in \eq{qortho} is now a block diagonal orthogonal $4\times 4 $ matrix,
\beq \label{ohtwo}
\mathcal{Q}=\left(\begin{array}{cc}U+V &\,\, 0 \\
0 & \phantom{-}\,\, U-V \end{array}\right)\,,
\eeq
where $(U\pm V)^T(U\pm V)=\mathds{1}_{2\times 2}$ as a consequence of \eqs{ucond2}{ucond1p}.  That is, the scalar Lagrangian is invariant under a global O(2)$\times$O(2) symmetry, which explains the presence of the mass-degenerate scalars $(H,h)$ and $(A,a)$, respectively.
The breaking of the four-fold mass degeneracy to the two mass-degenerate pairs is due to the scalar potential term proportional to $Z_5$, as is evident from \eq{ridmmasses}.  Finally, after promoting the derivative to the gauge covariant derivative in the scalar kinetic energy term, the remaining symmetry is O(2)$\rtimes\mathbb{Z}_2$.

For completeness, we note that the degeneracy of the charged Higgs scalars $(H^\pm,h^\pm)$ is governed by the full O(4) symmetry, which is broken down to U(2)$\rtimes\mathbb{Z}_2$ after promoting the derivatives of the scalar kinetic energy term to gauge covariant derivatives.  This is easily seen by noting that 
in the unitary gauge (in which the Goldstone fields do not explicitly appear), the physical charged scalar fields do not appear in the scalar potential term proportional to $Z_5$.  Finally, if $Z_4=Z_5=0$, we can make use of the vertical SU(2) global symmetry (which when gauged corresponds to the SU(2) electroweak gauge group) to conclude that all eight charged and neutral inert scalars are mass-degenerate.

Next, we examine all the bosonic couplings of the RIDM in the unitary gauge (where the Goldstone fields are set to zero).   The Higgs boson interactions with the gauge bosons and the Higgs boson self couplings of the RIDM are listed below.
\beqa
\mathscr{L}_{VVH}&=&\left(gm_W W_\mu^+W^{\mu\,-}+\frac{g}{2c_W} m_Z
Z_\mu Z^\mu\right)h_{\rm SM}
\,,\\[8pt]
\mathscr{L}_{VVHH}&=&\left[\quarter g^2 W_\mu^+W^{\mu\,-}
+\frac{g^2}{8c_W^2}Z_\mu Z^\mu\right](h_{\rm SM}^2+H^2+h^2+A^2+a^2) \nonumber \\
&& +\left[\half g^2 W_\mu^+ W^{\mu\,-}+
e^2A_\mu A^\mu+\frac{g^2}{c_W^2}\left(\half
-s_W^2\right)^2Z_\mu Z^\mu +\frac{2ge}{c_W}\left(\half
-s_W^2\right)A_\mu Z^\mu\right](H^+H^- +h^+ h^-)
\nn \\
&& +\biggl\{ \left(\half eg A^\mu W_\mu^+
-\frac{g^2s_W^2}{2c_W}Z^\mu W_\mu^+\right)
\bigl[H^-(H+iA)+h^-(h+ia)\bigr] +{\rm h.c.}\biggr\}
\,,\label{lvvhh}  \\
\mathscr{L}_{VHH}&=&\frac{g}{2c_W}\,
Z^\mu( A\,\ddel_\mu H+a\,\ddel_\mu h) -\half g\bigg\{iW_\mu^+\bigl[
H^-\ddel\lsup{\,\mu} (H+iA)+h^-\ddel\lsup{\,\mu} (h+ia)\bigr]
+{\rm h.c.}\biggr\} \nn \\
&& +\left[ieA^\mu+\frac{ig}{c_W}\left(\half -s_W^2\right)
Z^\mu\right](H^+\ddel_\mu H^-+h^+\ddel_\mu h^-), \label{lvhh}
\eeqa
\clearpage
\beqa
\mathscr{L}_{3h}&=&-\half v\bigl[Z_1 h_{\rm SM}^3+(Z_3+Z_4)h_{\rm SM}(H^2+A^2+h^2+a^2)+Z_5h_{\rm SM}(H^2-A^2+h^2-a^2)\bigr]\nn \\[6pt]
&& -vZ_3h_{\rm SM}(H^+H^-+h^+ h^-)\,, \label{l3h}  \\
\mathscr{L}_{4h}&=&-\tfrac18\biggl[ Z_1 h_{\rm SM}^4+ Z_2\bigl(H^2+A^2+h^2+a^2\bigr)^2 +2(Z_3+Z_4)h_{\rm SM}^2(H^2+h^2+A^2+a^2)\nn \\
&&\qquad +2Z_5 h_{\rm SM}^2(H^2+h^2-A^2-a^2)\biggr]-\half Z_3 h_{\rm SM}^2(H^+ H^- +h^+ h^-)
\nn \\[4pt]
&&\qquad -\half Z_2 (H^+ H^- + h^+ h^-)(H^2+A^2+h^2+a^2+H^+ H^- + h^+ h^-)
\,.\label{l4h}
\eeqa
In the RIDM, there is no experimental measurement that can physically distinguish the degenerate scalars,
$(H^\pm,h^\pm)$, $(H,h)$ and $(A,a)$.   However, a multiplicity factor will appear after summing over final mass-degenerate states, e.g., $Z\to HA$, $ha$ 
doubles the rate into a pair of neutral scalars.

\subsection{An alternative basis choice for the RIDM}
\label{alt-basis}

So far, our discussion has employed the $\{H_1,H_2,H_3\}$ basis of doublet scalar fields.   This is one choice among a family of Higgs bases defined such that $\vev{H_1^0}=v/\sqrt{2}$ and $\vev{H_2^0}=\vev{H_3^0}=0$.   Indeed, the Higgs basis is unique only up to an arbitrary U(2) transformation of the doublet fields $H_2$ and $H_3$.   In the following, we shall denote the $\{H_1,H_2,H_3\}$ basis as the $H23$-basis, since the scalar potential of \eq{ridmpot} provides a simple 3HDM extension of the inert 2HDM.

It will prove useful to consider another choice of scalar field basis that is related to the $H23$-basis as follows,\footnote{Further details are provided in Appendix~\ref{app:basis}.}
\beqa
&&\mathcal{R}\equiv \frac{1}{\sqrt{2}}\bigl(H_2+iH_3\bigr)=\begin{pmatrix} R^\dagger \\ \frac{1}{\sqrt{2}}\bigl(P+iQ^\dagger\bigr)\end{pmatrix}\,,\nn \\[15pt]
&&\mathcal{S}\equiv \frac{1}{\sqrt{2}}\bigl(H_2-iH_3\bigr)=\begin{pmatrix} S^\dagger \\ \frac{1}{\sqrt{2}}\bigl(P^\dagger+iQ\bigr)\end{pmatrix}\,.
\eeqa
This defines the $\{H_1,\mathcal{R},\mathcal{S}\}$ basis of doublet scalar field, henceforth denoted as the $RS$-basis.  Note that
since the real neutral fields $(H,h)$ and $(A,a)$ are mass-degenerate pairs, respectively, 
one can combine the mass-degenerate real fields into complex fields,
\beq \label{PQdef}
P\equiv \frac{H+ih}{\sqrt{2}}\,,\qquad Q\equiv \frac{A-ia}{\sqrt{2}}\,,
\eeq
where $M_P\geq M_Q$ (in our convention where $Z_5\geq 0$).\footnote{The relative minus sign in the definition of the imaginary parts of $P$ and $Q$ has been introduced for later convenience.}  The corresponding conjugate fields are
\beq \label{antiPQdef}
P^\dagger\equiv \frac{H-ih}{\sqrt{2}}\,,\qquad Q^\dagger\equiv \frac{A+ia}{\sqrt{2}}\,,
\eeq
Likewise, since $H^\pm$ and $h^\pm$ are mass-degenerate charged fields, one is free to define,
\begin{alignat}{2}
R&=\frac{H^- - ih^-}{\sqrt2}, &\quad S&=\frac{H^- + ih^-}{\sqrt2}, \label{RSdef}\\
R^\dagger&=\frac{H^+ + ih^+}{\sqrt2}, &\quad S^\dagger&=\frac{H^+ - ih^+}{\sqrt2},\label{RSdagdef}
\end{alignat}
where $R$ and $S$ are negatively charged mass-degenerate scalars and 
the corresponding conjugate fields, $R^\dagger$ and $S^\dagger$, are positively charged mass-degenerate scalars.

In the $RS$-basis, the scalar potential is given by
\beqa 
\!\!\!\!\!\!
\mathcal{V}_{\rm RIDM-RS}&=& Y_1 H_1^\dagger H_1+ Y_2 \bigl(\mathcal{R}^\dagger \mathcal{R} +\mathcal{S}^\dagger \mathcal{S}\bigr)
+\half Z_1(H_1^\dagger H_1)^2+\half {\bar Z}_2(\mathcal{R}^\dagger\mathcal{R}+\mathcal{S}^\dagger \mathcal{S})^2
+Z_3(H_1^\dagger H_1)(\mathcal{R}^\dagger \mathcal{R}+\mathcal{S}^\dagger \mathcal{S}) \nn \\
&& +Z_4\bigl[( H_1^\dagger \mathcal{R})(\mathcal{R}^\dagger H_1)+( H_1^\dagger \mathcal{S})(\mathcal{S}^\dagger H_1)\bigr]
+ \bar{Z}^\prime_5\bigl[(H_1^\dagger \mathcal{R})(H_1^\dagger \mathcal{S}) +(\mathcal{R}^\dagger H_1)(\mathcal{S}^\dagger H_1)\bigr]\,,\label{ridmpotRS}
\eeqa
where $\bar{Z}_2=Z_2$ and $\bar{Z}_5^\prime=Z_5$.\footnote{The reason for introducing the notation $\bar{Z}_2$ and $\bar{Z}^\prime_5$ in \eq{ridmpotRS} is clarified in section~\ref{sec:beyondridm}.}
One can then rewrite the RIDM 
couplings given in \eqst{lvvhh}{l4h} in terms of the neutral scalar fields  $P$ and $Q$ and the charged scalar fields $R$ and $S$ (and the corresponding conjugated fields), 
\beqa
&&\hspace{-0.2in}\mathscr{L}_{VVHH}=\left[\tfrac14 g^2 W_\mu^+W^{\mu\,-}
+\frac{g^2}{8c_W^2}Z_\mu Z^\mu\right]\bigl(h^2_{\rm SM}+2|P|^2+2|Q|^2\bigr) \nonumber \\
&& +\left[\half g^2 W_\mu^+ W^{\mu\,-}+
e^2A_\mu A^\mu+\frac{g^2}{c_W^2}\left(\half
-s_W^2\right)^2Z_\mu Z^\mu +\frac{2ge}{c_W}\left(\half-
s_W^2\right)A_\mu Z^\mu\right](R^\dagger R+S^\dagger S)
\nn \\
&&
+
\biggl\{ \left(\half eg A^\mu W_\mu^+
-\frac{g^2s_W^2}{2c_W}Z^\mu W_\mu^+\right)
\bigl[R(P+iQ^\dagger)+S(P^\dagger+iQ)\bigr] +{\rm h.c.}\biggr\},
\\
&&\hspace{-0.2in}\mathscr{L}_{VHH}=\frac{g}{2c_W}\,
Z^\mu( Q\,\ddel_\mu P+Q^\dagger\,\ddel_\mu P^\dagger)  +\left[ieA^\mu+\frac{ig}{c_W}\left(\half -s_W^2\right)
Z^\mu\right]
(R^\dagger\ddel_\mu R+S^\dagger\ddel_\mu S) \nn \\
&& \qquad\qquad 
-\half g\biggl\{iW_\mu^+\bigl[R\,\ddel^{\,\mu} (P+iQ^\dagger) +S\,\ddel^{\,\mu} (P^\dagger+iQ)\bigl]
+{\rm h.c.}\biggr\}
\label{ISVHH} \\[8pt]
&&\hspace{-0.2in}\mathscr{L}_{3h}=-\half vZ_1 h_{\rm SM}^3-v\bigl[(Z_3+Z_4)h_{\rm SM}(|P|^2+|Q|^2)+\bar{Z}^\prime_5h_{\rm SM}(|P|^2-|Q|^2)\bigr]
-vZ_3h_{\rm SM}(R^\dagger R+S^\dagger S)\,,  \nn \\[4pt]
&& \phantom{line} .\label{lagpq3} \\
&&\hspace{-0.2in} \mathscr{L}_{4h}=-\tfrac18 Z_1 h_{\rm SM}^4 
-\half \bar{Z}_2\bigl(|P|^2+|Q|^2\bigr)\bigl(|P|^2+|Q|^2+2R^\dagger R + 2S^\dagger S\bigr)-\half(Z_3+Z_4) h_{\rm SM}^2(|P|^2+|Q|^2) \nn \\[6pt]
&& \qquad\qquad -\half \bar{Z}^\prime_5 h_{\rm SM}^2 (|P|^2-|Q|^2)-\half\bigl[\bar{Z}_2(R^\dagger R+S^\dagger S) +Z_3 h^2_{\rm SM}\bigr] (R^\dagger R+S^\dagger S)\,.\label{lagpq4}
\eeqa

\subsection{Mass degeneracies beyond the RIDM}
\label{sec:beyondridm}

In this section, we add additional terms to the RIDM scalar potential while preserving the mass degeneracies of the model.  Naively, one can add to the RIDM scalar potential any gauge invariant quartic term involving the doublet fields $H_2$ and $H_3$ without upsetting the mass degeneracies of \eq{ridmmasses}.  However, the resulting tree-level mass degeneracies will be unnatural
unless they are a consequence of a symmetry.  

The simplest possible modification of the RIDM is to remove the $(H^\dagger_2 H_2)(H^\dagger_3 H_3)$ term entirely from the scalar potential.  That is, we can define a RIDM$^\prime$ scalar potential as,
\beq \label{ridmpotprime}
\mathcal{V}_{\rm RIDM'}= \mathcal{V}_{\rm RIDM}-Z_2(H^\dagger_2 H_2)(H^\dagger_3 H_3)\,.
\eeq
Note that the term in $\mathcal{V}_{\rm RIDM'}$ that is proportional to $Z_2$ is now given by $\half\bigl[(H_2^\dagger H_2)^2+(H_3^\dagger H_3)^2\bigr]$.
Indeed, one can argue that \eq{ridmpotprime} provides the simplest 3HDM generalization of the IDM.  
In the case of the RIDM$^\prime$, the tree-level mass degeneracies are no longer a consequence of a continuous symmetry, which is now explicitly broken by the presence of the explicit term in \eq{ridmpotprime} that is proportional to $(H^\dagger_2 H_2)(H^\dagger_3 H_3)$.   Indeed, this term is invariant only under a discrete subgroup of O(2)$\times$O(2) [which is the symmetry group of the RIDM scalar Lagrangian as discussed in section~\ref{sec:ridm}].  
In the notation of \eqs{global}{ohtwo}, consider the following two discrete subgroups of the O(2)$\times$O(2) symmetry group, 
\beqa
&& (1)~~~U=g\,,\qquad\quad V=0\,, \label{Rsymm1} \\
&& (2)~~~U=0\,,\qquad\quad V=g\,,\label{Rsymm2}
\eeqa
where $g$ is a $2\times 2$ matrix that acts on the Higgs basis fields $H_2$ and $H_3$ regarded as a two dimensional vector.   Then, the term  $(H^\dagger_2 H_2)(H^\dagger_3 H_3)$ is invariant under the two discrete subgroups above if $g\in D_4\iso \{\mathds{1},-\mathds{1}, R,-R,S,-S,Z,-Z\}$, where $\mathds{1}$ is the $2\times 2$ identity matrix and
\beq \label{RSZ}
R=\begin{pmatrix} 1 &\,\,\,\phm 0 \\ 0 &\,\,\, \phm 1\end{pmatrix}\,,\qquad\quad
S=\begin{pmatrix} 1 &\,\,\, \phm 0 \\ 0 &\,\,\,-1\end{pmatrix}\,,\qquad\quad
Z=\begin{pmatrix} 0 &\,\,\, -1 \\ 1 &\,\,\, \phm 0\end{pmatrix}\,.
\eeq
We recognize $D_4$ as the dihedral group of order eight, which is the symmetry group of the square~\cite{Ishimori:2012zz}.  Both discrete subgroups [\eqs{Rsymm1}{Rsymm2}] are isomorphic to $D_4$.  Following the discussion below \eq{ohtwo}, we conclude that the RIDM$^\prime$ scalar Lagrangian is invariant under a discrete
$D_4\times D_4$ symmetry, which is responsible for the presence of the mass-degenerate scalars $(H,h)$ and $(A,a)$, respectively.  Finally, after promoting the derivative to the gauge covariant derivative in the scalar kinetic energy term, the remaining symmetry is $D_4\rtimes\mathbb{Z}_2$.  

A comprehensive treatment of natural scalar mass degeneracies in the 3HDM would require a complete classification of 3HDM scalar potential symmetries, along the lines of the 2HDM analysis given in section~\ref{sec:naturaldegen}.\footnote{A complete catalog of all possible finite symmetry groups of the 3HDM is known (as well as some additional partial results); however the complete classification of all possible symmetries of the 3HDM remains an open problem~\cite{Ivanov:2012ry,Ivanov:2012fp}.}
In this paper, we shall ask a less ambitious question:
can one break the discrete symmetry identified above further while still naturally maintaining the mass-degenerate states of the RIDM.   The answer turns out to be affirmative.
This investigation led us to a particular 3HDM originally introduced by Ivanov and Silva~\cite{Ivanov:2015mwl} for other reasons that will be reviewed below.  

The Ivanov-Silva (IS)  model 
was constructed to exhibit a number of curious properties~\cite{Ivanov:2015mwl,Aranda:2016qmp}, which appear to rely on the existence of degenerate states in the scalar spectrum.
In particular, the IS scalar potential does not respect the conventional CP symmetry, $H_i\to H_i^{\star}$, where the latter  satisfies $({\rm CP})^2=\mathds{1}$, but instead respects a generalized CP symmetry of the form $H_i\to X_{ij}H_j^{\star}$ for some unitary matrix $X$.
In particular, the generalized CP symmetry of the IS scalar potential, denoted by CP4, is of order 4, signifying that
$({\rm CP4})^4=\mathds{1}$ and $({\rm CP4})^2\neq\mathds{1}$.  Moreover, no Higgs basis of scalar fields exists in which all the parameters of the IS scalar potential are simultaneously real.  As noted in section~\ref{sec:naturaldegen}, this property is in stark contrast with the 2HDM in which the existence of any generalized CP symmetry implies that the 2HDM scalar potential automatically respects the conventional CP symmetry, i.e.~a basis of scalar fields exists such that the corresponding 2HDM scalar potential parameters are real~\cite{Ferreira:2009wh,Ferreira:2010yh}.

In Appendix A, we demonstrate that starting from the IS scalar potential, one can perform a basis change in order to obtain a more convenient form of the scalar potential.  By making an appropriate U(2) transformation to define the Higgs basis fields, $H_2$ and $H_3$, we find that the IS scalar potential takes on the following form in the $H23$-basis,
\beq \label{Vis}
\mathcal{V}_{\rm IS}= \mathcal{V}_{\rm RIDM}+Z_3^\prime (H_2^\dagger H_2)(H_3^\dagger H_3)+Z_4^\prime(H_2^\dagger H_3)(H_3^\dagger H_2) +\bigl[Z_8 (H_2^\dagger H_3)^2+Z_9(H_2^\dagger H_3)(H_2^\dagger H_2-H_3^\dagger H_3)+{\rm h.c.}\bigr]\,,
\eeq
where $\mathcal{V}_{\rm RIDM}$ is given in \eq{ridmpot}.  In general, $Z_8$ and $Z_9$ are complex parameters.\footnote{As shown in Appendix~\ref{app:real}, one can perform an SO(2) rotation to redefine the fields $H_2$ and $H_3$ 
to remove the complex phase from either $Z_8$ or $Z_9$.}

We shall continue to use \eq{Hridm} to express the Higgs basis fields in terms of mass-eigenstate fields. 
Since none of the extra terms in \eq{Vis} involve the Higgs basis field $H_1$, the tree-level mass relations of \eq{ridmmasses} are not modified.  
We now argue that the mass-degeneracies of $(H^\pm, h^\pm)$, $(H,h)$ and $(A,a)$ are stable due to the presence of a symmetry.  The O(2)$\times$O(2) symmetry of the RIDM (prior to gauging the scalar kinetic energy terms) that is responsible for the mass degeneracies among the neutral Higgs mass eigenstates is broken by the new terms beyond $\mathcal{V}_{\rm RIDM}$ contained
in \eq{Vis}.   Indeed, after the extra terms are included, no unbroken continuous subgroup of O(2)$\times$O(2) remains.

In the notation of \eqs{global}{ohtwo}, consider the following two discrete subgroups of~the
O(2)$\times$O(2) symmetry group, 
\beqa
&& (1)~~~U=Z\,,\qquad\quad V=0\,, \label{symm1} \\
&& (2)~~~U=0\,,\qquad\quad V=Z\,,\label{symm2}
\eeqa
where $Z$ is given by \eq{RSZ}.
The $2\times 2$ matrix $Z$ acts on the Higgs basis fields $H_2$ and $H_3$.
Both discrete subgroups [\eqs{symm1}{symm2}] are isomorphic to 
$\mathbb{Z}_4=\bigl\{\mathds{1},-\mathds{1},Z,-Z\bigr\}$.  Note that $Z^2=-\mathds{1}$, where $\mathds{1}$ is the $2\times 2$ identity matrix.\footnote{In this case, gauging the scalar kinetic energy terms does not reduce the symmetry group further.}

Consider first the discrete symmetry defined in \eq{symm1}.
The fields $H_2$ and $H_3$ are odd under $-\mathds{1}$, which simply identifies the two inert doublets.
The elements $Z$ (and $-Z$) act non-trivially on the inert doublets.  However, \eq{Vis} is invariant with respect to 
\beq \label{Z}
\begin{pmatrix}  H_2\\ H_3\end{pmatrix}\rightarrow \begin{pmatrix} 0 & -1\\ 1 & \phm 0\end{pmatrix}\begin{pmatrix}  H_2\\ H_3\end{pmatrix}\,,
\eeq
if and only if $Z_8$ and $Z_9$ are both real.  In the model of IS where there is an unremovable complex phase in the scalar potential, only the subgroup $\mathbb{Z}_2=\{\mathds{1},-\mathds{1}\}$ of $\mathbb{Z}_4$ survives.  In particular, the residual symmetry in this case is not sufficient to explain the mass degeneracies of the IS model.

The discrete symmetry defined in \eq{symm2} is a generalized CP symmetry.  In particular, the IS scalar potential is invariant under 
\beq \label{newcp4}
H_i\to X_{ij}H_j^{\star}\,, \qquad \quad \text{where $X=\begin{pmatrix} 1 &\phm 0 & \phm 0 \\ 0 & \phm 0 & -1 \\ 0 & \phm 1 & \phm 0
\end{pmatrix}$,}
\eeq
This symmetry, which is also isomorphic to $\mathbb{Z}_4$, is the CP4 symmetry advertised above.   Moreover, this discrete symmetry is sufficient to explain the mass degeneracies of the IS model (in the case of an unremovable complex phase in the IS scalar potential).

It is instructive to consider the Higgs couplings of the IS model.   Only the quartic Higgs couplings of the RIDM are modified as follows,
\beqa
\delta\mathscr{L}_{4h}&=&-\tfrac14(Z_3^\prime+Z_4^\prime)\bigl[(H^2+A^2)(h^2+a^2)+4H^+ H^- h^+ h^-\bigr]-\half Z_3^\prime\bigl[(H^2+A^2)h^+ h^-+(h^2+a^2)H^+ H^-\bigr]\nn \\
&& -\half Z'_4\bigl[\bigl(Hh+Aa+i(Ha-hA)\bigr)H^+ h^- + \bigl(Hh+Aa-i(Ha-hA)\bigr)h^+ H^-\bigr] \nn \\
&& -\tfrac14 Z_8\bigl[Hh+Aa+i(Ha-hA)+2h^+ H^-\bigr]^2-\tfrac14 Z^*_8\bigl[Hh+Aa-i(Ha-hA)+2H^+ h^-\bigr]^2 \nn \\
&& -\tfrac14 Z_9\bigl(H^2+A^2-h^2-a^2+2H^+ H^- -2h^+ h^-\bigr)\bigl[Hh+Aa+i(Ha-hA)+2h^+ H^-\bigr]\nn \\
&& -\tfrac14 Z^*_9\bigl(H^2+A^2-h^2-a^2+2H^+ H^- -2h^+ h^-\bigr)\bigl[Hh+Aa-i(Ha-hA)+2H^+ h^-\bigr]\,.\label{deltalag4}
\eeqa

It is convenient to re-express the neutral scalar fields appearing in \eq{deltalag4} in terms of the complex neutral fields $P$ and $Q$ and their conjugates introduced in \eqs{PQdef}{antiPQdef}, and the charged fields $R$ and $S$ and their conjugates defined in \eqs{RSdef}{RSdagdef}. 
Note that the fields $P$, $Q$ and the corresponding conjugate fields $P^\dagger$ and $Q^\dagger$ are each eigenstates of CP4.\footnote{This means that each of the four states, $P$, $Q$, $P^\dagger$ and $Q^\dagger$, are CP4-self conjugate (they are their own antiparticles).   Moreover, $P$ and the corresponding conjugate state $P^\dagger$ are mass-degenerate, but are otherwise unrelated fields (and similarly for $Q$ and $Q^\dagger$).}
In particular, under a CP4 transformation, $P\to iP$, $Q\to iQ$, $P^\dagger\to -iP^\dagger$, and $Q^\dagger\to -iQ^\dagger$. 
Likewise, under a CP4 transformation, $R\to -iS^\dagger$, $R^\dagger\to iS$, $S\to iR^\dagger$ and $S^\dagger\to -iR$.  Note that these transformation properties are consistent with the requirement that $({\rm CP}4)^4=\mathds{1}$.  

We can evaluate the four-scalar interaction Lagrangian directly in the $RS$-basis.   We first must rewrite \eq{Vis} in the RS-basis,
\beq \label{VisRS}
\mathcal{V}_{\rm IS-RS}= \mathcal{V}_{\rm RIDM-RS}+\bar{Z}_3^\prime (\mathcal{R}^\dagger \mathcal{R})(\mathcal{S}^\dagger \mathcal{S})+\bar{Z}_4^\prime(\mathcal{R}^\dagger \mathcal{S})(\mathcal{S}^\dagger \mathcal{R}) +\bigl[\bar{Z}_8 (\mathcal{R}^\dagger \mathcal{S})^2+\bar{Z}_9(\mathcal{R}^\dagger \mathcal{S})(\mathcal{R}^\dagger \mathcal{R}-\mathcal{S}^\dagger \mathcal{S})+{\rm h.c.}\bigr]\,,
\eeq
where $\mathcal{V}_{\rm RIDM-RS}$ is given by \eq{ridmpotRS}.  The relations between the unbarred and barred parameters are derived in Appendix~\ref{app:basis},
\beqa
\bar{Z}_2&=&Z_2+\half(Z_3^\prime+Z_4^\prime-2\Re Z_8)\,,\qquad\qquad\quad\,\,\,\,
\bar{Z}_3^\prime=-Z_4^\prime+2\Re Z_8\,,\label{zeetwonew}\\
\bar{Z}_4^\prime&=&\half(Z_4^\prime-Z_3^\prime+2\Re Z_8)\,,\qquad\qquad
\qquad\qquad 
\bar{Z}_5^\prime=Z_5\,,\\
\bar{Z}_8&=&-\tfrac14(Z_3^\prime+Z_4^\prime+2\Re Z_8)+i\Re Z_9\,,\qquad\quad
\bar{Z}_9= \Im Z_9+i\Im Z_8\,. \label{z89}
\eeqa
The quartic interactions given in \eq{lagpq4} are then modified by employing the new definition of $\bar{Z_2}$ given in \eq{zeetwonew}
and adding the following terms,
\clearpage

 \beqa
 \delta\mathscr{L}_{4h}&=&
 -\tfrac14 \bar{Z}_3^\prime\bigl[|P|^2+|Q|^2+2|R|^2-i(PQ-P^\dagger Q^\dagger)\bigr]\bigl[|P|^2+|Q|^2+2|S|^2+i(PQ-P^\dagger Q^\dagger)\bigr] \nn \\[5pt]
  && -\tfrac14\bar{Z}_4^\prime\bigl(P^2+Q^{\dagger\,2}+2R^\dagger S\bigr)\bigl(P^{\dagger\,2}+Q^2+2S^\dagger R\bigr)  \nn \\[5pt]
&& -\tfrac14 \bar{Z}_8(P^{\dagger\,2}+Q^2+2S^\dagger R)^2-\tfrac14 \bar{Z}_8^*(P^2+Q^{\dagger\,2}+2R^\dagger S)^2 \nn \\[5pt]
 && +\half\bigl[i(PQ-P^\dagger Q^\dagger)-R^\dagger R+S^\dagger S\bigr]\bigl[\bar{Z}_9(P^{\dagger\,2}+Q^2+2S^\dagger R)+\bar{Z}_9^*(P^2+Q^{\dagger\,2}+2R^\dagger S)\bigr]\,.  \nn \\
 &&\phantom{line}
 \label{l4hRS}
 \eeqa
 
We now consider the possible effects of the Yukawa interactions.
It is remarkable that it is possible to construct a CP4-invariant Yukawa interaction Lagrangian where the fermions transform nontrivially under a CP4 transformation~\cite{Aranda:2016qmp,Ferreira:2017tvy,Ivanov:2017bdx}.  In such a model, the mass degeneracies identified above that are a consequence of the CP4 symmetry are of course maintained.  Alternatively,
if the fermions couple exclusively to the Higgs basis field $H_1$ (as in the case of the IDM), then the Yukawa interactions are invariant with respect to the 
$\mathbb{Z}_2$ discrete symmetry defined below \eq{Z},\footnote{Note that this $\mathbb{Z}_2$ symmetry is isomorphic to $(\rm{CP}4)^2$, which remains an exact symmetry of the model.} 
under which the inert doublet
fields, $H_2$ and $H_3$, are odd and all other fields of the model ($H_1$, gauge bosons and fermions) are even.  However, the CP4 symmetry is no longer a symmetry of the complete model.  That is, if we define the CP4 transformation to be the conventional CP transformation when acting on the fermions and gauge fields, then the CP4 symmetry of the model will be violated by the presence of the unremovable CP-violating phase in the CKM mixing matrix.   Nevertheless, it is not clear whether this violation is sufficient to remove the scalar mass degeneracies of the IS model that were protected by the (now accidental) CP4 symmetry of the scalar potential.   This is an open question that we hope to revisit in a future work.
 
Finally, it is instructive to note that the scalar mass degeneracies of the CP4-invariant 3HDM is just the simplest example of a larger class of multi-Higgs models with degenerate scalars that are a consequence of a generalized CP symmetry.  In Ref.~\cite{Ivanov:2018qni}, Ivanov and Laletin demonstrate how to construct $N$ Higgs doublet models with a generalized CP symmetry of order $2k$ (denoted by CP$2k$) with positive integer $k$.   Nontrivial cases arise only for $2k=2^p$ with integer $p\geq 1$.   The simplest nontrivial models of this type (CP8 and CP16) require at least $N=5$ Higgs doublets.  Such models necessarily have mass-degenerate neutral scalars and mass-degenerate charged Higgs pairs.   A further exploration of models of this type is beyond the scope of this work.
 
\section{An observable distinction between CP2 and CP4}
\label{sect:Z-to-QQQQ}
\setcounter{equation}{0}

The distinction between the IS scalar potential in the $H23$-basis with $Z_8$ and $Z_9$ real or complex is physical.\footnote{In making this assertion, we have implicitly assumed that $Z_5\neq 0$.   The case of $Z_5=0$, which is special due to the enhanced mass degeneracy noted in footnote~\ref{fn:enhanced}, will be treated at the end of this section.\label{fnz5}}
To demonstrate this assertion, we focus on the neutral scalar self-interactions in $\delta\mathscr{L}_{4h}$ that are linear in the fields $P$ or $Q$ (or their complex conjugates),
 \beq \label{lin}
 \delta\mathscr{L}_{4h}\ni \half \Im Z_8\bigl[(PQ-P^\dagger Q^\dagger)(P^2-Q^2-P^{\dagger\,2}+Q^{\dagger\,2})\bigr]
 +\half i\Im Z_9\bigl[(PQ-P^\dagger Q^\dagger)(P^2+Q^2+P^{\dagger\,2}+Q^{\dagger\,2})\bigr]\,,
 \eeq
 where we have used \eq{z89} to re-express $\bar{Z}_9$ [which appears in \eq{l4hRS}] in terms of the $H23$-basis parameters, $\Im Z_8$ and $\Im Z_9$.
Self-interaction terms of this type are absent if $Z_8$ and $Z_9$ are both real.  Hence, the presence of these terms signals a CP4-symmetric IS scalar potential that does not respect the conventional CP symmetry, $H_i\to H_i^{\star}$.
Here we provide two specific examples.  First, \eq{ISVHH} shows the existence of a $ZPQ$ interaction, which would permit the decay $Z\to PQ, P^*Q^*$, if kinematically available.  Since $M_Q\leq M_P$, let us further suppose that $M_Q<\tfrac14 m_Z<M_P$.   In this case, the $P$ and $P^*$ would be virtual.   One possible decay of the virtual $P$ or $P^*$ makes use of the existence of the four-scalar interaction given in \eq{lin}.  If this interaction is present, the decay $Z\to QQQQ^*$, $Q^* Q^* Q^* Q$ is allowed and provides unambiguous evidence that either $Z_8$ and/or $Z_9$ possesses a nonzero imaginary part.
A second example makes use of the $W^+H^-P$, $W^+h^-P$, $W^+H^-Q$, and $W^+h^-Q$ interactions of \eq{ISVHH}.   In this case, we can consider the decay of a charged $W$ into a charged Higgs boson and $P$ (or $P^*$).   We can now make use of \eq{lin} to decay the virtual $P$ or $P^*$ into $QQQ$, $QQQ^*$, $QQ^*Q^*$, $Q^* Q^* Q^*$.   
Note that in each of the two cases above, there are multiple four-scalar final states involving mass-degenerate scalars.   In computing the experimentally observed rates, one must compute the squared amplitude for each of the possible final states, and then multiply the final result by a multiplicity factor that counts the number of possible final states.

In contrast, suppose that \eq{Z} were a symmetry of the IS scalar potential.  In this case, the corresponding transformation properties of the scalar fields are,
 $P\to iP$, $Q\to -iQ$, $P^\dagger\to -iP^\dagger$, $Q^\dagger\to iQ^\dagger$,  $H^{\pm}\to -h^\pm$, and $h^\pm\to H^\pm$.  One would then immediately conclude that $Z_8=Z_8^*$ and $Z_9=Z_9^*$, as expected.  In particular, \eq{lin} is \textit{not} invariant under \eq{Z}, and thus the four scalar decay modes listed above would necessarily be absent.

 As an exercise, we have evaluated the decay rate for $Z\to QQQQ^*$, $QQ^* Q^* Q^*$, in an approximation where $M_Q=0$ and $M_P\gg m_Z$.  The computation is presented in Appendix~\ref{app:QQQQ}.  
 The end result is
\beq \label{zqqqq}
\frac{\Gamma(Z\to QQQQ^*,QQ^* Q^* Q^*)}{\Gamma(Z\to \nu\bar{\nu})}=\frac{(\Im Z_8)^2+(\Im Z_9)^2}{3\cdot  5\cdot 2^{8}\, \pi^4 }\left(\frac{m_Z}{M_P}\right)^4\,.
\eeq
This result implies that the quantity $(\Im Z_8)^2+(\Im Z_9)^2$ must be a physical quantity,
and hence invariant with respect to scalar basis changes that are consistent with the form of the IS scalar potential 
given by \eq{Vis} in the $H23$-basis.  

However, the family of Higgs bases is larger than the set of scalar field bases in which the IS scalar potential has the form of \eq{Vis}, as discussed in Appendix~\ref{app:real}.   In special cases, it is possible that there exists a real Higgs basis even if $(\Im Z_8)^2+(\Im Z_9)^2\neq 0$.  In such cases, one can transform the fields ($H_2, H_3)\to (\bar{H}_2,\bar{H}_3)$, where $\bar{H}_i\to\bar{H_i}^{\star}$ is a symmetry of the Lagrangian; i.e., the model exhibits a CP2 symmetry.\footnote{The notation CP2 derives from the property, $(\text{CP2})^2=\mathds{1}$.}   
In the IS model, the existence of a nonzero decay rate for $Z\to QQQQ^*$, $QQ^* Q^* Q^*$ implies that no CP2 symmetry that commutes with the CP4 symmetry is present.\footnote{We say that the CP2 symmetry commutes with CP4 if the application of these two symmetry transformations on the scalar fields does not depend on the order in which the transformations are applied.  For further details see Appendix B of Ref.~\cite{Ivanov:2018ime}.}  However, this leaves open the possibility of a CP2 symmetry that does not commute with CP4.  In Appendix~\ref{app:rbasis}, we provide two examples in which 
$(\Im Z_8)^2+(\Im Z_9)^2\neq 0$ in the $H23$ basis, but nevertheless a real Higgs basis exists: (i)~$\Im Z_8\neq 0$ and $Z_9=0$ and (ii) $\Im Z_8=0$, $\Re Z_9=0$ and $\Im Z_9\neq 0$.   In both these examples, the CP2 symmetry that exists does not commute with the CP4 symmetry, even though the decay rate for $Z\to QQQQ^*$, $QQ^* Q^* Q^*$ is nonzero.   In light of the results of Appendix~\ref{app:inv}, this is a generic feature of a noncommuting CP2 symmetry in the IS model.   Equivalently, the nonexistence or existence of the decay $Z\to QQQQ^*$, $QQ^* Q^* Q^*$ is a physical distinction between the 3HDM with a CP4 symmetric IS scalar potential that either preserves or does not preserve a commuting CP2 symmetry.

Finally, we return to the special case of $Z_5=0$ (cf.~footnote~\ref{fnz5}).
In Appendix~\ref{app:inv}, we have demonstrated explicitly that if $Z_5\neq 0$, then there exists a ratio of two basis-invariant quantities, which when evaluated in the $H23$-basis yields $(\Im Z_8)^2+(\Im Z_9)^2$. 
Moreover, if $Z_5=0$, then it is possible to change the basis of scalar fields of the IS model, in which the form of the IS potential is still given by \eq{Vis} but $\Im Z_8=\Im Z_9=0$.
This result appears to be in contradiction to the result of \eq{zqqqq}.    The resolution of this apparent paradox can be obtained by noting that if $Z_5=0$, then $M_P=M_Q$.   
Since \eq{zqqqq} was derived under the assumption that $M_Q=0$ and $M_P\gg m_Z$, \eq{zqqqq} no longer applies if $Z_5=0$.   But, more importantly, if $Z_5=0$ (so that $M_P=M_Q$), then the decay $Z\to QQQQ^*,QQ^* Q^* Q^*$ is no longer an experimental observable, since one must also include four scalar decays involving $P$ and $P^*$.   The possible four-body final states involve all possible combinations of $P$ and $Q$ scalars, such that either one or three of the final state scalars are complex-conjugated.   Some of the vertices that contribute to these final states are present even if $\Im Z_8=\Im Z_9=0$.   For example, there is a four-scalar $|P|^2 |Q|^2$ interaction that contributes to $Z\to QQPP^*$.   One must compute the squared amplitude for each possible final state and then add the amplitudes incoherently to obtain the final experimentally observable decay rate.  
This decay rate will involve a complicated combination of the IS potential coefficients, which will correspond to the appropriate invariant quantity in the case of $Z_5=0$.  Thus, the possibility of finding a new basis for the IS potential in which $\Im Z_8=\Im Z_9=0$ when $Z_5=0$ is no longer paradoxical.

\section{The $ZZZ$ and $ZWW$ vertices}
\label{sect:ZZZ-ZWW}
\setcounter{equation}{0}

In the CP-violating 2HDM, CP violation may manifest itself at loop level in the effective $ZZZ$ and $ZWW$ vertices. 
In that model, one finds that CP-violating form factors can be described in terms of the three invariants introduced in \eq{imJ}~\cite{Grzadkowski:2016lpv}.
In section~\ref{sect:Z-to-QQQQ}, we noted the existence of a physical observable that could distinguish between the CP4-conserving IS models in which a CP2 symmetry that commutes with CP4 is either present or absent.   However, from a spacetime viewpoint, this physical observable was CP-even.   This raises the question as to whether any observable can exist in a CP4-invariant theory that is CP2-odd.  The answer to this question is no.   
For example, there is no way to distinguish between CP2 and CP4
on the level of the form factors themselves.   Thus, if the theory respects at least one generalized CP symmetry, then 
all CP-violating form factors must be absent.   

It is instructive to check the cancellation of contributions to the CP-violating form factors of the effective $ZZZ$ and $ZWW$ vertices in a CP4-conserving, CP2-violating theory (neglecting any effects from the Higgs-fermion Yukawa interactions).
The general $ZZZ$ vertex function (with all $Z$ bosons off-shell) can be expressed in terms of 14 different Lorentz structures \cite{Hagiwara:1986vm,Nieves:1996ff,Gounaris:1999kf,Gounaris:2000dn,Baur:2000ae}, all preserving parity.
Some of these vanish when one or more $Z$ are on-shell.
Let us characterize them by momenta and 
Lorentz indices ($p_1,\mu$), ($p_2,\alpha$) and ($p_3,\beta$), and let $Z_1$ be off-shell while $Z_2$ and $Z_3$ are on-shell. Furthermore, we assume that $Z_1$ couples to a pair of leptons such as $e^+e^-$, and terms proportional to the lepton mass will be neglected. 
Denoting $\ell\equiv p_2-p_3 \equiv 2p_2-p_1$,
the $ZZZ$ vertex structure reduces to the form~\cite{Gounaris:1999kf} 
\begin{equation}
-i\Gamma_{ZZZ}^{\alpha\beta\mu}
=\frac{p_1^2-M_Z^2}{M_Z^2}\left[f_4^Z(p_1^{\alpha}g^{\mu\beta}+p_1^{\beta}g^{\mu\alpha})
+f_5^Z\epsilon^{\mu\alpha\beta\rho}\ell_\rho\right]\,.
\end{equation}
The dimensionless form factor $f_4^Z$ violates CP while $f_5^Z$ conserves CP.
%

\begin{figure}[t!]
\refstepcounter{figure}
\label{Fig:ZZZ}
\addtocounter{figure}{-1}
\begin{center}
\includegraphics[scale=0.8]{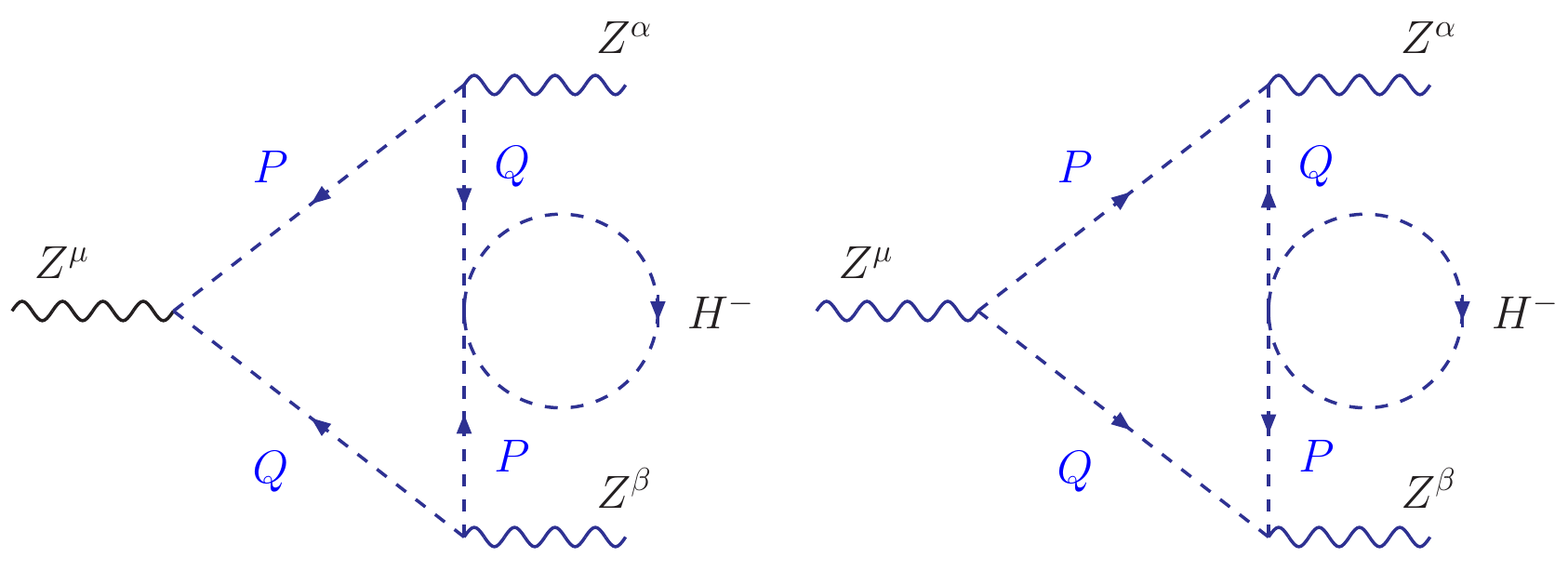}
\end{center}
\vspace*{-4mm}
\caption{\small A typical pair of Feynman diagrams for $Z\to ZZ$ at two-loop order.}
\end{figure}

For example, consider the case of the 2HDM.
At the one-loop level, CP violating effects yield a non-zero contribution to the $ZZZ$ vertex function,~$f_4$, that is
proportional to $\Im J_2$ of Eq.~(\ref{Eq:Im_J2}) \cite{Grzadkowski:2016lpv}. Thus, only one of the three invariants of \eq{imJ} contributes. Indeed, in light of \eq{remarkable}, it follows that a non-zero $\Im J_2$ requires all three neutral Higgs bosons to be non-degenerate in mass, and the $Z$ boson couples to all three non-diagonal neutral Higgs pairs.

In order to understand how the IS model conserves CP (while not respecting CP2), it is instructive to see how the CP-violating effects cancel at loop level in the effective $ZZZ$ (and $ZWW$) vertices. In order to do this we have employed the software package \texttt{FeynArts}\cite{Hahn:2000kx} and written a \texttt{FeynArts} model file containing all the bosonic couplings of the IS-model.  We have automated the construction of the diagrams contributing to the effective $ZZZ$-vertex and evaluated their amplitude (the loop integrals are kept unevaluated in symbolic form). We are only interested in those contributions to each diagram that contain $\Im Z_8$ and/or $\Im Z_9$, since such contributions could be a signal of CP violation.

At the one-loop level there are no diagrams containing $\Im Z_8$ and/or $\Im Z_9$. Such contributions can only arise from a four-point scalar vertex. This means that this four-point vertex must be ``internal"; i.e., none of the external $Z$-fields can be part of this vertex. None of the $ZZZ$ one-loop topologies can accommodate this. 
Diagrams containing $\Im Z_8$ and/or $\Im Z_9$ first appear at two-loop order. But even if there are individual diagrams with this type of contribution, the sum of the contributions is zero when we add the amplitudes for all the individual diagrams within each topology. 
A pair of cancelling diagrams are shown in Fig.~\ref{Fig:ZZZ}.
The same happens for diagrams at three-loop order. Repeating this exercise for the $ZWW$ vertex we find the same result. Hence, there are no contributions at one-, two- or three-loop order containing $\Im Z_8$ and/or $\Im Z_9$ after adding the amplitudes for all the individual diagrams within each topology.  The arguments presented at the beginning of this section imply that this cancellation persists to all orders in perturbation theory.

 \section{Conclusions}
 \label{conclude}
 \setcounter{equation}{0}
In this work we discussed the interplay between symmetries and natural mass degeneracies
in the scalar sector.   Some cases of scalar mass degeneracy are accidental, i.e.~they are not the 
result of an exact symmetry and  therefore can be implemented only by an artificial fine tuning of the scalar potential parameters.  
The Higgs basis \cite{Donoghue:1978cj,Georgi:1978ri,Botella:1994cs}, in which the neutral scalar field vacuum expectation value resides entirely in one of the scalar doublet fields, 
is especially suitable for our study.     
We began by examining the two Higgs doublet model (2HDM), with particular attention given to
the special case of the inert doublet model (IDM), which possesses an unbroken $\mathbb{Z}_2$ symmetry under which one ``inert'' scalar doublet is odd, and all other fields of the model are even.
In all cases in which the 2HDM exhibited scalar mass degeneracies (whether natural or accidental), the mass degenerate states can be experimentally distinguished from each other.  Moreover, with one exception, we found that all 2HDM mass degeneracies are accidental. 
The one exceptional case of 2HDM scalars that can be naturally degenerate in mass are 
the two neutral scalar states of the inert doublet of the IDM.  This result was also confirmed by examining all possible symmetries of the 2HDM scalar potential and analyzing which of these symmetries can guarantee the presence of mass degenerate scalar states.

For models with three Higgs doublets, the analysis of the general case 
becomes significantly more elaborate.  We focused first on a 3HDM generalization of the IDM with mass degenerate scalars, which we denoted as the replicated IDM (RIDM), 
where the two doublets $H_2$ and $H_3$ are invariant 
under two separate unbroken  $\mathbb{Z}_2$ symmetries and the model is CP conserving. 
In this framework $H_2$ and $H_3$ are composed of mass eigenstate fields, 
that do not mix with the SM like Higgs boson, forming four mass degenerate 
pairs. Furthermore, each mass degenerate pair picks one field from each 
one of these doublets. We also identified the symmetry obeyed by the neutral 
mass eigenstates themselves, which is responsible for the twofold mass 
degeneracies. 

In the absence of $Z_5$ (which appears in the RIDM scalar potential) there are four mass degenerate neutral scalars 
and the symmetry of the scalar potential consists of an O$(4)$ global symmetry. 
Introducing in the potential 
the term proportional to $Z_5$, partially breaks the O$(4)$ symmetry down to an 
O(2)$\times$O(2) symmetry and the fourfold mass degeneracy is lifted, leaving a pairwise mass degeneracy. The mass degeneracy of the two charged physical fields is governed by the full O(4) symmetry.  In the case
of $Z_4 = Z_5 =0 $ there is further enhancement of the symmetry and all eight 
physical scalars contained in  $H_2$ and $H_3$ are mass 
degenerate.
 
It is instructive to examine the Higgs boson interactions with the gauge bosons 
as well as the Higgs self couplings of the RIDM, since in the RIDM the components 
of $H_2$ and $H_3$ are already states with well defined masses. We are then led to 
the conclusion that there is no experimental measurement that can physically 
distinguish the mass degenerate scalars of the RIDM on an event by event basis. 
Nevertheless, multiplicity factors due to the production of different scalar states of the same mass
do appear in physical observables and signal the existence of the mass degeneracy.

Starting with the RIDM, one can consider perturbations in which the mass degeneracies persist and yet remain natural.   By reducing the RIDM symmetries responsible for the mass degeneracies to the
smallest discrete subgroup that maintains the mass degenerate scalar states, we are led to a model that
is equivalent to a particular 3HDM that was originally proposed by Ivanov and Silva (IS) \cite{Ivanov:2015mwl}.
The IS model exhibits very special properties. The original form of the 
IS scalar potential is given in Appendix \ref{ISoriginal} and is the most general
potential respecting the symmetry given by Eq.~(\ref{Eq:X-def}). We have 
rewritten the IS potential in the notation of Eq.~(\ref{Vis}) where the symmetry is 
now given by Eq.~(\ref{newcp4}). 
In particular,
the scalar mass terms (and the corresponding mass degeneracies) are the same as in the RIDM; only the quartic couplings of the physical scalar states differ. 

One must apply the symmetry given by Eq.~(\ref{newcp4}) [denoted by CP4] four times in order to obtain the identity 
transformation.   This is to be contrasted with the conventional CP symmetry transformation (denoted by CP2) whose square is the identity.
On the other hand, if we apply the CP4 transformation while at the same 
time transforming the space coordinates from $\boldsymbol{x}$ into $-\boldsymbol{x}$,  the
end result can be identified as a generalized CP transformation. This is a very unusual type of 
CP transformation since applying it twice does not yield the identity 
transformation. However, identifying CP4 with a CP transformation
is possible because from the spacetime point of view the transformation 
remains of order two, as it should. 
Likewise, one can define a generalized time reversal operator with properties analogous to CP4 while  
transforming the time coordinate from $t$ to $-t$.  Consequently, there is no contradiction with the CPT theorem,
which remains intact.

A very interesting feature of the IS scalar potential is that the symmetry requires 
some of its coefficients to be complex (in a particular Higgs basis).  Moreover, for generic choices of the scalar potential parameters,
there is no scalar basis transformation within the family of Higgs bases,
of the form given by \eqst{hbasisu}{vtil},
that can transform the scalar potential into a new potential with only real coefficients. This is a surprising result in light of the statement that the IS potential is CP-conserving.
The IS model conserves CP independently of the existence or nonexistence of a real Higgs basis, although in the case where no real Higgs basis exists, the IS model is only invariant with respect to the generalized CP symmetry, CP4 (whereas CP2 is not a symmetry of the IS scalar potential).
Nevertheless, any CP-violating observable of the IS model must vanish.   For example,
the contributions to the CP-violating form factors of the effective $ZZZ$ 
and $ZWW$ vertices generated in the IS model must exactly cancel.  As a check of this statement, we confirmed this cancellation up to three-loop order in the IS model with no real Higgs basis.

We identified a physical quartic scalar interaction made up of an odd number of mass-degenerate neutral scalar states (e.g., $P^3 Q$ and $Q^3 P$) that is consistent with the CP4 symmetry, but would vanish if the IS scalar potential exhibits a CP2 symmetry that commutes with CP4.  This leaves open the possibility of the existence of a CP2 symmetry that does not commute with CP4.  However, we were unable to find an observable quantity of the IS model that can distinguish between the presence or absence of a noncommuting CP2 symmetry.

Finally, we stress that the possibility of a scalar potential and vacuum that is invariant with respect to a generalized CP symmetry without the existence of a real basis appears to be inexorably connected with the existence of mass-degenerate scalar states.  We strongly suspect that this connection, which has been demonstrated in this paper for the IS model, is applicable more generally to any multi-Higgs doublet model.  If true, then the existence of a generalized CP symmetry in the absence of mass degenerate scalars necessarily implies the presence of a conventional CP symmetry; i.e., the existence of a real basis of scalar fields in which the CP symmetry corresponds simply to conjugation of the scalar fields.

\section*{Acknowledgments}

We are very grateful to Igor Ivanov for many fruitful and enlightening conversations.   
We also thank Gustavo C.~Branco, Nick Mavromatos, Palash B.~Pal, Apostolos Pilaftsis, and Graham Ross with whom several aspects of this work were discussed.
H.E.H. and P.O.~acknowledge the Galileo Galilei Institute for Theoretical Physics, where this work was initiated.
H.E.H., P.O.~and M.N.R. also appreciate the hospitality of the CERN Theory group where some of this work was carried out, and M.N.R also acknowledges partial support from CERN.
H.E.H.~is supported in part by the U.S. Department of Energy grant
number DE-SC0010107, and in part by the grant H2020-MSCA-RISE-2014
No.~645722 (NonMinimalHiggs).
P.O.~is supported by the Research Council of Norway.
The work of M.N.R.~was partially supported by 
Funda\c{c}\~ao para a Ci\^encia e a Tecnologia
(FCT, Portugal) through the projects CFTP-FCT Unit 777 (UID/FIS/00777/2013),
CERN/FIS-PAR/0004/2017 and PTDC/FIS-PAR/29436/2017 which are partially funded 
through POCTI (FEDER), COMPETE, QREN and EU. H.E.H.~and M.N.R.~benefited from discussions that
took place at the University of
Warsaw during visits supported by the HARMONIA project of the National Science Centre,
Poland, under contract UMO-2015/18/M/ST2/00518 (2016--2019). M.N.R.~and P.O.~also thank the University of Bergen and
CFTP/IST/University of Lisbon, where collaboration visits took place. 

\appendix
\section{The Ivanov Silva model revisited}
\label{app:ivanov-silva}
\setcounter{equation}{0}
\renewcommand{\theequation}{A.\arabic{equation}}

Consider the most general 2HDM with a scalar potential as specified in \eq{genpot}.
Including the kinetic energy terms with SU(2)$\times$U(1) gauge covariant derivatives,
the 2HDM [after electroweak symmetry breaking under the assumption that the vacuum preserves U(1)$_{\rm EM}$]
consists of a model of two scalar doublets coupled to the gauge bosons, $W^\pm$, $Z$ and $\gamma$.
We shall ignore the couplings of the bosonic sector of the 2HDM to the fermions of the SM in the following discussion.

We now ask the following question.  Does the bosonic Lagrangian conserve CP?   For CP to be conserved, two conditions 
must be verified.  First, the scalar potential must exhibit explicit CP conservation.    Second, the vacuum must
conserve CP.   If the former is true but the latter is false, we say that CP is spontaneously broken.  
However, in this discussion, we are interested in whether both explicit and spontaneous CP violation are absent.

In the 2HDM, the answer to this question is simple.  We first transform to the Higgs basis and examine the scalar potential given
in \eq{hbasispot}.  The Higgs basis is unique up to a possible rephasing of the Higgs basis field, $H_2\to e^{i\chi}H_2$.
Then, CP is conserved if and only if there exists a choice of $\chi$ such that all Higgs basis scalar potential parameters are
real.

In the discussion above, we have not specified in detail how the scalar fields transform under a CP transformation.  
Starting from the generic $\Phi_1$--$\Phi_2$ basis employed in writing \eq{genpot}, the conventional CP transformation 
corresponds to conjugation, $\Phi_i^{\rm CP}= \Phi_i^{\star}$.   However, this is a basis-dependent statement.   Indeed, one is always free to
change the basis, $\Phi_i^\prime=U_{ij}\Phi_j$, where $U\in$~U(2).   In the new basis, $\Phi^{\prime\,CP}_i=X_{ij}\Phi^{\prime\,CP}_j$,
where $X=UU^T$ is a symmetric unitary matrix.  More generally, we can consider the \textit{generalized} CP transformation,
\beq \label{gcptrans}
\Phi_i^{\rm CP}(\boldsymbol{x},t)= X_{ij}\Phi_j^{\star}(-\boldsymbol{x},t)\,,
\eeq
where $X\in$~U(2).\footnote{Note that it is not consistent to simply define the CP transformation of a multi-Higgs doublet model without including the matrix $X$ in \eq{gcptrans}, since the form of the CP transformation depends on the choice of the scalar basis, as noted above.
Consequently, some authors prefer to call this transformation a \textit{general} CP transformation rather than \textit{generalized} CP transformation.}   
If $X$ is both unitary and symmetric, then one can find a basis in which the CP transformation is simply conjugation.\footnote{As shown in 
Appendix D of Ref.~\cite{Dreiner:2008tw} [see the Lemma below eq.~(D.3.1)], for any symmetric unitary matrix~$X$, there exists a unitary matrix $U$ such that $X=UU^T$.}
In Ref.~\cite{Ferreira:2009wh}, it is shown that
in the 2HDM there are three possible classes of generalized CP transformations (GCPs): (i) $X$ is unitary and symmetric; (ii) $X$ is unitary and antisymmetric; and (iii) $X$ is unitary but is neither symmetric nor antisymmetric.  Clearly, no basis change can convert a GCP transformation of types (ii) or (iii) into the transformation of the field into its conjugate.  Nevertheless, as shown in Ref.~\cite{Ferreira:2009wh}, any 2HDM scalar potential that is invariant under GCP transformations of types (ii) or (iii) is also separately invariant under a GCP transformation of type (i).  

Do the above results generalize to arbitrary Higgs sectors?   In particular, consider an extended Higgs sector with $N$ hypercharge-one, complex doublets (denoted henceforth as the NHDM).  To address the question of CP invariance, we transform to the so-called \textit{charged Higgs basis} defined in Ref.~\cite{Bento:2017eti}.
If the scalar fields of the charged Higgs basis are denoted by $H_i$ ($i=1,\ldots,n$), then $\vev{H_1^0}=v/\sqrt{2}$, $\vev{H_j^0}=0$ for $j=2,3,\ldots,n$, and
the fields $H_j^\pm$ (for $j=2,3,\ldots,n$) are the physical, mass-eigenstate charged Higgs fields.   Note that for $N=2$, the Higgs basis and the charged Higgs basis coincide.
For $N\geq 3$, consider first the case in which the physical charged Higgs bosons are mass non-degenerate.   In this case, the charged Higgs basis is uniquely defined up to a possible rephasing, $H_j\to e^{i\chi_j}H_j$.   In this case, CP is conserved if and only if there exist a choice of the $\chi_j$ such that all charged Higgs basis scalar potential parameters are real.   This generalizes the result of the 2HDM quoted above.

If there exist mass degeneracies among the physical charged Higgs fields, then one must re-evaluate the conditions for CP invariance.   To simplify the discussion, we focus on the case of $N=3$, in which the two physical charged Higgs bosons are mass degenerate.  In this case, the charged Higgs basis is unique up to a U(2) transformation of the charged Higgs basis fields $H_2$ and $H_3$.  Ivanov and Silva\cite{Ivanov:2015mwl} constructed a 3HDM whose scalar potential and vacuum are invariant under a generalized CP transformation such that
$({\rm GCP})^2\neq\mathds{1}$, where $\mathds{1}$ is the identity operator.
Moreover, some of the scalar potential parameters of the charged Higgs basis of the Ivanov--Silva (IS) scalar potential are complex, and no U(2) transformation of the charged Higgs basis fields $H_2$ and $H_3$ can be performed to remove all the complex phases.  Hence, the IS scalar potential is \textit{not} invariant under a separate GCP transformation that is equivalent to conjugation in another basis, in contrast to the corresponding 2HDM result.  Ivanov and Silva denote the GCP transformation of the IS scalar potential by CP4, since it has the property that $({\rm CP}4)^4=\mathds{1}$ and $({\rm CP}4)^2\neq\mathds{1}$ .  Indeed, one consequence of the CP4 symmetry of the scalar potential and the vacuum is the mass degeneracy of the 
physical charged Higgs bosons, as well as two additional mass degeneracies among pairs of neutral Higgs bosons.
In this Appendix, we consider the 3HDM scalar potential of the Ivanov and Silva model and examine some of its properties.

\subsection{The IS scalar potential}
\label{ISoriginal}

Consider the 3HDM consisting of three hypercharge-one, complex doublet fields, $\phi_i$ ($i=1,2,3$).
In the Higgs basis, the form of the scalar potential proposed initially by Ivanov and Silva (IS) in Ref.~\cite{Ivanov:2015mwl} is fixed by imposing the following generalized CP symmetry,
\begin{equation} \label{Eq:X-def}
\phi_i \to W_{ij}\phi_j^{\star}, \quad
W=\begin{pmatrix}
1 & \phm 0 & \phm 0 \\
0 & \phm 0 & \phm i \\
0 & -i & \phm 0
\end{pmatrix},
\end{equation}
which has the property that applying it four times yields the identity operator.   This is the CP4 symmetry transformation noted above.

The resulting IS scalar potential is given by 
\beq \label{vis}
V=V_0+V_1\,,
\eeq
with 
\begin{subequations} \label{Eq:IS-pot}
\begin{align} 
V_0&=-m_{11}^2(\phi_1^\dagger\phi_1)-m_{22}^2(\phi_2^\dagger\phi_2+\phi_3^\dagger\phi_3)
+\lambda_1(\phi_1^\dagger\phi_1)^2
+\lambda_2[(\phi_2^\dagger\phi_2)^2+(\phi_3^\dagger\phi_3)^2]+\lambda_3^\prime(\phi_2^\dagger\phi_2)(\phi_3^\dagger\phi_3)\nonumber \\
& \qquad
+\lambda_3(\phi_1^\dagger\phi_1)[(\phi_2^\dagger\phi_2)+(\phi_3^\dagger\phi_3)] 
+\lambda_4^\prime(\phi_2^\dagger\phi_3)(\phi_3^\dagger\phi_2)
+\lambda_4[(\phi_1^\dagger\phi_2)(\phi_2^\dagger\phi_1)+(\phi_1^\dagger\phi_3)(\phi_3^\dagger\phi_1)], 
  \\[6pt]
V_1&=\lambda_5(\phi_3^\dagger\phi_1)(\phi_2^\dagger\phi_1)
+\half\lambda_6[(\phi_2^\dagger\phi_1)^2-(\phi_1^\dagger\phi_3)^2]
+\lambda_8(\phi_2^\dagger\phi_3)^2
+\lambda_9(\phi_2^\dagger\phi_3)[(\phi_2^\dagger\phi_2)-(\phi_3^\dagger\phi_3)] +{\rm h.c.}\label{visone}
\end{align}
\end{subequations}
The hermiticity of the scalar potential implies that the coefficients of $V_0$ are real.  In contrast, the coefficients of $V_1$ are potentially complex.  However, having imposed the CP4 symmetry given by \eq{Eq:X-def}, we see that $\lambda_5$ is real.   

Under the CP4 symmetry specified in \eq{Eq:X-def}, the gauge-invariant bilinear quantities, 
$B_{ij}\equiv\phi_i^\dagger\phi_j$,
transform as follows:
\begin{subequations}
\begin{alignat}{2}
B_{11} &\to B_{11}, \\
B_{22}&\to B_{33} &\quad B_{33}&\to B_{22}, \\
B_{12}&\to i B_{31}, &\quad B_{21}&\to -i B_{13}, \\
B_{13}&\to -i B_{21}, &\quad B_{31}&\to i B_{12}, \\
B_{23}&\to -B_{23}, &\quad B_{32}&\to -B_{32}.
\end{alignat}
\end{subequations}
It follows that $V$ given by \eqst{vis}{visone}, with $\lambda_6$, $\lambda_8$ and $\lambda_9$ complex and all other scalar potential parameters real, is the most general 3HDM potential
that is invariant under the CP4 transformation given in  (\ref{Eq:X-def}).   Without loss of generality, one can furthermore assume that $\lambda_6$ is real after an appropriate rephasing of the
scalar fields $\phi_2$ and $\phi_3$.

At this stage, we have not yet found the minimum of the scalar potential and determined whether the CP4 symmetry is respected by the vacuum.
There exist a range of scalar potential parameters in which the vacuum preserves U(1)$_{\rm EM}$, in which case one can decompose the scalar doublets as,
\begin{equation} \label{Eq:phi_i}
\phi_i=\left(
\begin{array}{c}\varphi_i^+\\ (v_i+\eta_i+i \chi_i)/\sqrt{2}
\end{array}\right), \quad i=1,2,3.
\end{equation}
In particular, the vacuum conserves CP4 if the minimum of the scalar potential corresponds to $(v_1,v_2,v_3)=(v,0,0)$\cite{Ivanov:2015mwl}.
Indeed, there exists a range of scalar potential parameters for which this corresponds to the global minimum, in which case the
value of $m_{11}^2$ is fixed by the scalar potential minimum condition to be
\begin{equation}
m_{11}^2=\lambda_1 v^2.
\end{equation}
In this case, the scalar field basis employed in \eqs{vis}{Eq:IS-pot} is the Higgs basis, with the freedom to perform U(2) transformations on
$\{\phi_2,\phi_3\}$.  We shall take advantage of this freedom in the next two subsections.

It is now straightforward to determine the scalar mass spectrum of the IS model.  
Since we are in the Higgs basis, we can immediately identify the Goldstone bosons, $\varphi_1^\pm=G^\pm$ and $\chi_1=G^0$.   Moreover,
$\eta_1$ is a neutral mass-eigenstate with mass $m^2_{\eta_1}=2\lambda_1 v^2$, whose tree-level couplings to the gauge bosons and to itself are precisely those of the SM Higgs boson (corresponding to the exact alignment limit).   Indeed, this is analogous to the IDM in which $\phi_1$ is equivalent to the hypercharge-one, complex scalar doublet of the SM
and $\phi_2$ and $\phi_3$ are inert doublets.  
The two physical charged Higgs fields, $\varphi_2^\pm$ and $\varphi_3^\pm$, are mass-degenerate,
\begin{equation} \label{Eq:H_ch_mass}
m_{\varphi_2^\pm,\varphi_3^\pm}^2=\half\lambda_3 v^2 -m_{22}^2.
\end{equation}
The neutral scalar spectrum consist of the SM-like Higgs boson $\eta_1$ and a pair of mass degenerate neutral scalars made up of linear combinations of
the $\eta_{2,3}$ and $\chi_{2,3}$, with masses given by\cite{Ivanov:2015mwl},
\beq \label{Eq:IS-masses}
M^2= a+\sqrt{b^2+c^2}\,,\qquad\quad m^2= a-\sqrt{b^2+c^2}\,,
\eeq
where
\begin{equation}
a=\half(\lambda_3+\lambda_4)v^2-m_{22}^2, \qquad
b=\half\lambda_6 v^2, \qquad
c=\half\lambda_5 v^2.
\end{equation}

\subsection{A simpler form for the IS scalar potential}
\label{sect:transformations}

Given the IS scalar potential in the Higgs basis, we still have the freedom to perform a U(2) transformation on $\{\phi_2,\phi_3\}$.
It is possible to remove the $\lambda_5$ term in \eq{Eq:IS-pot} by the following basis transformation,
\beq \label{pup}
\bar{\phi}_i=U_{ij}\phi_j, 
\eeq
where
\beq \label{uuu}
U=\begin{pmatrix}
1    &  0   & 0   \\
0    & \phm \cos\theta & -\sin\theta \\
0 & \phm \sin\theta & \phm\cos\theta
\end{pmatrix}\,,
\eeq
with $0\leq\theta\leq\pi$.
With respect to the new basis, the CP4 transformation specified in \eq{Eq:X-def} is given by,
\beq
\bar\phi_i\to V_{ij}\bar\phi^{\star}_j\,, \qquad \text{where $V=UWU^T$}.
\eeq
Using the form for $U$ given in \eq{uuu}, it follows that $V=W$. Thus, in this new basis, the IS 
symmetry  takes the same form as in the original basis.

When the scalar potential is expressed in terms of the fields $\bar\phi_i$, the resulting scalar potential parameters will be denoted by $\bar{m}^2_{ii}$ and $\bar\lambda_i$.
It is straightforward to obtain expressions for $\bar{m}_{11}^2$, $\bar{m}_{22}^2$ and the $\bar\lambda_i$ in terms of the scalar potential parameters defined in \eq{Eq:IS-pot}.
In particular, $\bar{m}_{11}^2=m_{11}^2$, $\bar{m}_{22}^2=m_{22}^2$, and $\bar\lambda_i=\lambda_i$ for $i=1,3$ and 4.   Next, we note that the CP4 symmetry does not mandate that $\bar\lambda_6$ is real.   However, it is straightforward to check that
$\Im\bar\lambda_6=\Im\lambda_6$.   Having previously chosen $\lambda_6$ real (after an appropriate rephasing of $\phi_2$ and $\phi_3$),
it follows that $\bar\lambda_6$ is also real.  

The remaining transformed coefficients are given by,
\begin{subequations}
\begin{align}
\bar\lambda_2&= \lambda_2-\half\sin^2 2\theta[\lambda_2-\Re\lambda_8-\half(\lambda_3^\prime+\lambda_4^\prime)]-\sin 2\theta\cos 2\theta\Re\lambda_9\,, \\
\bar\lambda^\prime_3&=\lambda^\prime_3+\sin^2 2\theta[\lambda_2-\Re\lambda_8-\half(\lambda_3^\prime+\lambda_4^\prime)]+2\sin 2\theta\cos 2\theta\Re\lambda_9\,, \\
\bar\lambda^\prime_4&=\lambda^\prime_4+\sin^2 2\theta[\lambda_2-\Re\lambda_8-\half(\lambda_3^\prime+\lambda_4^\prime)]+2\sin 2\theta\cos 2\theta\Re\lambda_9\,, \\
\bar\lambda_5&=\lambda_5\cos 2\theta+\lambda_6\sin 2\theta\,,\\
\bar\lambda_6&=\lambda_6\cos 2\theta-\lambda_5\sin 2\theta\,,\\
\Re\bar\lambda_8&=\Re\lambda_8+\half\sin^2 2\theta[\lambda_2-\Re\lambda_8-\half(\lambda_3^\prime+\lambda_4^\prime)]+\sin 2\theta\cos 2\theta\Re\lambda_9\,,\\
\Re\bar\lambda_9&=(1-2\sin^2 2\theta)\Re\lambda_9+\sin 2\theta \cos 2\theta[\lambda_2-\Re\lambda_8-\half(\lambda_3^\prime+\lambda_4^\prime)]\,,\\
\Im\bar\lambda_8&=\cos 2\theta\Im\lambda_8+\sin 2\theta\Im\lambda_9\,,\\
\Im\bar\lambda_9&=\cos 2\theta\Im\lambda_9-\sin 2\theta\Im\lambda_8\,.
\end{align} \label{lamtrans}
\end{subequations}

One can now choose the angle $\theta$ such that $\bar\lambda_5=0$.\footnote{Note that a different choice of $\tan 2\theta$ could have been made to set either $\bar\lambda_6=0$, $\Im\bar\lambda_8=0$ or $\Im\bar\lambda_9=0$.   That is, one can always perform a change of Higgs basis to remove one degree of freedom from the coefficients of the IS scalar potential.}   This yields
$\tan 2\theta=-\lambda_5/\lambda_6$.   Then, $\sin 2\theta$ and $\cos 2\theta$ are determined up to an overall sign.  Introducing the following notation,
\beq
\lambda_{56}\equiv\sqrt{\lambda_5^2+\lambda_6^2}\,,
\eeq
we choose the angle $\theta$ such that,
\beq \label{sc}
\sin 2\theta=\frac{\lambda_5}{\lambda_{56}}\,,\qquad\quad \cos 2\theta=-\,\frac{\lambda_6}{\lambda_{56}}\,.
\eeq
Thus, the $\lambda_5$-term in \eq{Eq:IS-pot} is actually redundant.\footnote{A similar simplification was presented recently in Ref.~\cite{Ferreira:2017tvy}.}

Inserting the results of \eq{sc} back into \eq{lamtrans} yields $\bar\lambda_5=0$ and,
\begin{subequations}
\begin{align}
\bar{\lambda}_2&=\frac{\lambda _5 \left[\lambda _5 \left(\lambda^\prime _3+\lambda^\prime_4+2 \Re \lambda_8\right)+4 \lambda _6 \Re\lambda_9\right]+2 \lambda _2 \left(\lambda _5^2+2 \lambda _6^2\right)}{4 \lambda^2_{56}},\\
\bar{\lambda}^\prime_3&=\frac{\lambda _5^2 \left(\lambda^\prime_3-\lambda^\prime_4-2 \Re\lambda_8\right)+2 \lambda _6^2 \lambda^\prime_3-4 \lambda _5 \lambda _6 \Re\lambda_9+2 \lambda _2 \lambda _5^2}{2\lambda^2_{56}},\\
\bar{\lambda}^\prime_4&=\frac{\lambda _5^2 \left(-\lambda^\prime_3+\lambda^\prime_4-2 \Re\lambda_8\right)+2 \lambda _6^2 \lambda^\prime_4-4 \lambda _5 \lambda _6 \Re\lambda_9+2 \lambda _2 \lambda _5^2}{2 \lambda^2_{56}},\\
\bar{\lambda}_6&=-\lambda_{56},\\
\Re \bar{\lambda}_8&=\frac{-\lambda _5^2 \left(\lambda^\prime_3+\lambda^\prime_4-2 \Re\lambda_8\right)-4 \lambda _5 \lambda _6 \Re\lambda_9+4 \lambda _6^2 \Re\lambda_8+2 \lambda _2 \lambda _5^2}{4 \lambda^2_{56}},\\
\Re \bar{\lambda}_9&=\frac{\lambda _5\lambda _6 \left(\lambda^\prime_3+\lambda^\prime_4+2 \Re\lambda_8\right)-2 \lambda _5^2 \Re\lambda_9+2 \lambda _6^2 \Re\lambda_9-2 \lambda _2 \lambda _5 \lambda _6}{2 \lambda^2_{56}},\\
\Im \bar{\lambda}_8&=\frac{\lambda _5 \Im\lambda_9-\lambda _6 \Im\lambda_8}{\lambda_{56}},\\
\Im \bar{\lambda}_9&=\frac{-\lambda _5 \Im\lambda_8-\lambda _6 \Im\lambda_9}{\lambda_{56}}.
\end{align}
\end{subequations}

An additional feature of the IS scalar potential with $\bar\lambda_5=0$ is that the real and imaginary parts of the neutral fields $\bar{\phi}_2^0$ and $\bar{\phi}_3^0$ are mass eigenstates.  That is, the neutral squared-mass matrices are already diagonal in the $\{\bar\phi_1,\bar\phi_2,\bar\phi_3\}$ basis.
In particular, the lightest of the two mass-degenerate states lives in the imaginary part of $\bar\phi_2$ and in the real part of $\bar\phi_3$.  The heaviest of the two mass-degenerate neutral states lives in the real part of $\bar\phi_2$ and in the imaginary part of $\bar\phi_3$. 

It is convenient to make an additional field redefinition, $\bar\phi_3\to i\bar\phi_3$.   
The effect of this modification is to modify $\bar{V}_1$ by flipping the sign of 
$(\bar\phi^\dagger_1\bar\phi_3)^2$ in the term proportional to $\bar\lambda_6$ and to transform $\bar\lambda_8\to -\bar\lambda_8$ and $\bar\lambda_9\to -i\bar\lambda_9$.
To make contact with the $H23$-basis employed in \eq{Vis}, 
we define,
\beq \label{hphi}
H_1=\bar\phi_1\,,\qquad H_2=\bar\phi_2\,,\qquad 	H_3=i\bar\phi_3\,,
\eeq
corresponding to a basis change, $H_i\to \widetilde{U}_{ij}\bar\phi_j$, with $\widetilde{U}={\rm diag}(1\,,\,1\,,\,i)$.
Note that
the heaviest mass degenerate neutral fields now reside in the real part of the neutral components of $H_2$ and $H_3$, and the lightest mass degenerate neutral fields reside in the imaginary part of the neutral components of $H_2$ and $H_3$.
When expressed in the $H23$-basis, the IS scalar potential is given by,
\beq \label{app:vishh}
\mathcal{V}_{\rm IS}= \mathcal{V}_{\rm RIDM}+Z_3^\prime (H_2^\dagger H_2)(H_3^\dagger H_3)+Z_4^\prime(H_2^\dagger H_3)(H_3^\dagger H_2) +\bigl[Z_8 (H_2^\dagger H_3)^2+Z_9(H_2^\dagger H_3)(H_2^\dagger H_2-H_3^\dagger H_3)+{\rm h.c.}\bigr]\,,
\eeq
where $\mathcal{V}_{\rm RIDM}$ is given by \eq{ridmpot},
with $Z_8$ and $Z_9$ potentially complex and all other scalar potential parameters real. 
\Eq{app:vishh} is the version of the IS scalar potential employed in section~\ref{sec:beyondridm}.  To make contact with the previous notation used above, we note that
\beqa
Y_1&=& -m_{11}^2\,,\qquad Y_2=-m_{22}^2\,,\qquad
Z_1=2\lambda_1\,,\qquad Z_2=2\bar\lambda_2\,,\qquad Z_3=\lambda_3\,,\qquad Z_4=\lambda_4 \nonumber  \\
Z_3^\prime&=&\bar\lambda_3^\prime-2\bar\lambda_2\,,\qquad Z_4^\prime=\bar\lambda_4^\prime\,,\qquad Z_5=\bar\lambda_6\,,\qquad
Z_8=-\bar\lambda_8\,,\qquad Z_9=-i\bar\lambda_9\,.
\eeqa
The corresponding CP4 symmetry transformation now takes the form
\beq \label{cp4preferred}
H_i\to X_{ij}H_j^{\star}\,,\qquad \text{where $X=\widetilde{U}W\widetilde{U}^T=\begin{pmatrix} 1 & \phm 0 & \phm 0 \\ 0 & \phm 0 & -1 \\ 0 & \phm 1 & \phm 0\end{pmatrix}$}\,,
\eeq
as indicated in \eq{newcp4}.

\subsection{Non-existence of a real Higgs basis}
\label{app:real}

Consider the IS scalar potential [cf.~\eq{app:vishh} with $\mathcal{V}_{\rm RIDM}$ given by \eq{ridmpot}] expressed in terms of the Higgs basis of scalar doublet fields, $\{H_1,H_2,H_3\}$, where $\vev{H_1^0} \neq 0$ and the vevs of the other two doublet fields vanish.     
The coefficients  $Z_8$ and $Z_9$ are potentially complex and all other scalar potential parameters are real.  Recall that the Higgs basis is unique only up to an arbitrary U(2) transformation of $\{H_2, H_3\}$. 
Is it possible to transform to a new Higgs basis in which all the IS scalar potential parameters are real?  Such a Higgs basis, if it exists, is called a real Higgs basis.

The most general basis transformation that preserves the general class of Higgs bases is given  
(in block diagonal form) by,
\beq \label{hbasisu}
\begin{pmatrix} \bar{H}_1 \\ \bar{H}_{23}\end{pmatrix}=\begin{pmatrix} 1 & \,\,0 \\ 0 &\,\, \widetilde{V}\end{pmatrix} \begin{pmatrix} H_1 \\ H_{23}\end{pmatrix}\,,
\eeq
where
\beq
H_{23}\equiv\begin{pmatrix} H_2\\ H_3\end{pmatrix}\,,\qquad \quad
\bar{H}_{23}\equiv\begin{pmatrix} \bar{H}_2\\ \bar{H}_3\end{pmatrix}\,,
\eeq
and $\widetilde{V}$ is the most general U(2) matrix,
\beq \label{vtil}
\widetilde{V}=e^{i\psi/2}\begin{pmatrix} e^{i\alpha}\cos\phi & -e^{-i\beta}\sin\phi \\ e^{i\beta}\sin\phi & \phm e^{-i\alpha}\cos\phi\end{pmatrix}\,,
\eeq
where $0\leq\phi<\pi$, $-\pi<\psi\leq\pi$, $0\leq\alpha\leq\pi$ and  $0\leq\beta\leq\pi$.  Applying \eq{hbasisu} to the IS scalar potential given in \eq{app:vishh} yields
\beqa 
\mathcal{V}_{\rm IS}&=&Y_1 \bar{H}_1^\dagger  \bar{H}_1+ Y_2 \left( \bar{H}_2^\dagger  \bar{H}_2 + \bar{H}_3^\dagger  \bar{H}_3\right)
+\half Z_1( \bar{H}_1^\dagger  \bar{H}_1)^2+\half \bar{Z}_2( \bar{H}_2^\dagger  \bar{H}_2+ \bar{H}_3^\dagger  \bar{H}_3)^2 \label{visbar} \\
&&
+Z_3(\bar{H}_1^\dagger  \bar{H}_1)( \bar{H}_2^\dagger  \bar{H}_2+ \bar{H}_3^\dagger \bar{H}_3)
+Z_4\bigl[(  \bar{H}_1^\dagger  \bar{H}_2)( \bar{H}_2^\dagger  \bar{H}_1)+(  \bar{H}_1^\dagger  \bar{H}_3)( \bar{H}_3^\dagger  \bar{H}_1)\bigr]
\nonumber\\
&&
+\bar{Z}_3^\prime ( \bar{H}_2^\dagger  \bar{H}_2)( \bar{H}_3^\dagger  \bar{H}_3)+\bar{Z}_4^\prime( \bar{H}_2^\dagger  \bar{H}_3)( \bar{H}_3^\dagger  \bar{H}_2) 
 +i\bar{Z}_5^\prime\bigl[e^{i\psi} (\bar{H}_3^\dagger  \bar{H}_1)( \bar{H}_2^\dagger  \bar{H}_1)-e^{-i\psi}( \bar{H}_1^\dagger \bar{H}_2)( \bar{H}_1^\dagger  \bar{H}_3)\bigr]\,,
\nonumber \\
&&
+\bigl\{\half \bar{Z}_5\bigl[e^{i\psi} (\bar{H}_2^\dagger  \bar{H}_1)^2 +e^{-i\psi}( \bar{H}_1^\dagger  \bar{H}_3)^2\bigr]+\bar{Z}_8 ( \bar{H}_2^\dagger  \bar{H}_3)^2+\bar{Z}_9( \bar{H}_2^\dagger  \bar{H}_3)(\bar{H}_2^\dagger  \bar{H}_2- \bar{H}_3^\dagger  \bar{H}_3)+{\rm h.c.} \bigr\}\,.\nn
\eeqa

\noindent
The coefficients $Y_1$, $Y_2$, $Z_1$, $Z_3$ and $Z_4$ are unmodified, 
whereas,
\beqa
&& \bar{Z}_2=Z_2+\half\sin^2 2\phi\bigl(Z_3^\prime+Z_4^\prime+Z_8 e^{2i\xi}+Z^*_8 e^{-2i\xi}\bigr) 
 -\sin 2\phi\cos 2\phi\bigl(Z_9e^{i\xi}+Z^*_9e^{-i\xi}\bigr),\label{zbtwo} \\
&&\bar{Z}_3^\prime=Z_3^\prime-\sin^2 2\phi\bigl(Z_3^\prime+Z_4^\prime+Z_8 e^{2i\xi}+Z^*_8 e^{-2i\xi}\bigr) 
+2\sin 2\phi\cos 2\phi\bigl(Z_9e^{i\xi}+Z^*_9e^{-i\xi}\bigr), \\
&&\bar{Z}_4^\prime=Z_4^\prime-\half\sin^2 2\phi\bigl(Z_3^\prime+Z_4^\prime+Z_8 e^{2i\xi}+Z^*_8 e^{-2i\xi}\bigr) 
+\sin 2\phi\cos 2\phi\bigl(Z_9e^{i\xi}+Z^*_9e^{-i\xi}\bigr), \\
&&\bar{Z}_5^\prime=Z_5\sin 2\phi \sin\xi\,,\label{zbfivep} \\
&&\bar{Z}_5=e^{i\chi}Z_5\bigl(e^{i\xi}\cos^2\phi+e^{-i\xi}\sin^2\phi\bigr)\,,\label{zbfive} \\
&&\bar{Z}_8= e^{2i\chi}\bigl\{-\tfrac14 \sin^2 2\phi\bigl(Z_3^\prime+Z_4^\prime\bigr)+e^{2i\xi}\cos^4\phi\,Z_8+e^{-2i\xi}\sin^4\phi\, Z_8^* \nonumber \\
&& \qquad\qquad\qquad +\sin 2\phi\bigl[e^{i\xi}\cos^2\phi\,Z_9-e^{-i\xi}\sin^2\phi\, Z_9^*\bigr]\bigr\}\,, \label{zbeight} \\
&&\bar{Z}_9=e^{i\chi}\bigl\{-\half \sin 2\phi\cos 2\phi(Z_3^\prime+Z_4^\prime) -\sin 2\phi\bigl[e^{2i\xi}\cos^2\phi\,Z_8-
e^{-2i\xi} \sin^2\phi\,Z_8^*\bigr] \nonumber \\
&& \qquad\qquad\qquad +\half e^{i\xi}(\cos 4\phi+\cos 2\phi)Z_9+\half e^{-i\xi}(\cos 4\phi-\cos 2\phi)Z_9^*\bigr\}\,,\label{zbnine}
\eeqa
where
\beq \label{sumdiff}
\xi\equiv \alpha+\beta\,,\qquad\quad \chi\equiv\alpha-\beta\,.
\eeq
By definition of the $H23$-basis, $Z_5$ is real and $Z_5^\prime=0$ [the latter is a consequence of the absence of a term in \eq{app:vishh} that involves $(H_3^\dagger H_1)(H_2^\dagger H_1)$ and its hermitian conjugate].   After employing a generic U(2) basis change [\eq{vtil}], a nonzero $\bar{Z}_5^\prime$ and a complex $\bar{Z}_5$ are generated [cf.~\eqs{zbfivep}{zbfive}], such that
\beq \label{z55}
Z_5^2=|\bar{Z}_5|^2+\bar{Z}_5^{\prime\,2}\,.
\eeq

It is instructive to examine the form of the CP4 transformation in the $\{\bar{H}_1$, $\bar{H}_2$, $\bar{H_3}\}$ basis.  Starting from \eq{cp4preferred} and transforming
$H_i\to  \bar{H}_i= \widetilde{V}_{ij}H_j$ ($i,j=2,3$), it follows that the CP4 transformation of the barred fields is given by,
\beq
\bar{H}_i\to \bar{X}_{ij}\bar{H}_j^{\star}\,,\qquad \text{where $\bar{X}=VW{V}^T$}\,,
\eeq
where the $3\times 3$ matrices $\bar{X}$, ${V}$  and $W$ in block form are given by
\beq
\bar{X}=\begin{pmatrix} 1 & 0 \\ 0 & \widetilde{X}\end{pmatrix}\,,\qquad
{V}=\begin{pmatrix} 1 & 0 \\ 0 & \widetilde{V}\end{pmatrix}\,,\qquad W=\begin{pmatrix} 1 & 0 \\ 0 & \epsilon\end{pmatrix}\,,
\eeq
and $\epsilon\equiv\left(\begin{smallmatrix} 0 & -1 \\ 1 & \phm 0\end{smallmatrix}\right)$.  For any $\widetilde{V}\in$~U(2), we have
\beq \label{XV}
\widetilde{X}=\widetilde{V}\epsilon\widetilde{V}^T=e^{i\psi}\epsilon\,,
\eeq
after taking the determinant of \eq{XV} and noting that $\det \widetilde V=e^{i\psi}$.  
Indeed, if we impose invariance of the scalar potential under CP4 in the $\{\bar{H}_1$, $\bar{H}_2$, $\bar{H_3}\}$ basis, then the IS scalar potential must have the form given by \eq{visbar}, with $\bar{Z}_8$ and $\bar{Z}_9$ potentially complex and all other scalar potential coefficients [excluding factors of $i$ or $e^{\pm i\psi}$ that explicitly appear in \eq{visbar}] real.

It is possible to choose a basis in which all but one of the scalar potential parameters are real.  This can be achieved by choosing $\psi=\alpha=\beta=0$ in \eq{vtil}.  In this case,
\eqs{zbeight}{zbnine} yield $\bar{Z}_5^\prime=0$, $\bar{Z}_5=Z_5$ and 

\beqa
\Im\bar Z_8&=&\cos 2\phi\Im Z_8+\sin 2\phi\Im Z_9\,,\nn \\
\Im\bar Z_9&=&\cos 2\phi\Im Z_9-\sin 2\phi\Im Z_8\,.\label{zee89}
\eeqa
Indeed, there is a choice of $\phi$ in \eq{zee89} such that $\Im \bar Z_9=0$ (and another choice of $\phi$ such that $\Im\bar{Z}_8=0$). Thus, by a series of basis changes, we have reduced the number of independent parameters in the IS scalar potential from 14 to 12.

\subsubsection{Transforming to a Higgs basis where $Z_8$ and $Z_9$ are real}
We now examine whether a choice of $\psi$, $\chi$, $\xi$ and $\phi$ exists such that $i\bar{Z}_5^\prime e^{\pm i\psi}$, $\bar{Z}_5 e^{\pm i\psi}$, $\bar{Z}_8$ and $\bar{Z}_9$ are all real.  To begin, we first show that a Higgs basis exists in which $\bar{Z}_8$ and $\bar{Z}_9$ are both real.   Here, we follow the analysis given in Appendix C of Ref.~\cite{Gunion:2005ja}.   By assumption, $Z_9$ is real and $Z_8=|Z_8|e^{i\theta_8}$ (where $\theta_8$ is not an integer multiple of $\pi$ so that $\Im Z_8\neq 0$).  Setting $\Im \bar{Z_8}=\Im\bar{Z_9}=0$ in \eqs{zbeight}{zbnine} yields,
\beq \label{abcd}
\Im\bar{Z}_8=f_a\cos 2\chi-f_b\sin 2\chi=0\,,\qquad\quad  
\Im\bar{Z}_9=f_c\cos \chi-f_d\sin \chi=0\,,\
\eeq
where
\beqa
f_a&=& |Z_8|\cos 2\phi\sin(2\xi+\theta_8)+Z_9\sin 2\phi\sin\xi\,, \label{fadef}\\
f_b&=&\tfrac14(Z_3^\prime+Z_4^\prime)\sin^2 2\phi-|Z_8|(1-\half\sin^2 2\phi)\cos(2\xi+\theta_8)-Z_9\sin 2\phi\cos 2\phi \cos\xi\,,\label{fbdef}\\
f_c&=&-|Z_8|\sin 2\phi\sin(2\xi+\theta_8)+Z_9\cos 2\phi\sin\xi\,,\label{fcdef}\\
f_d&=&\half(Z_3^\prime+Z_4^\prime) \sin 2\phi\cos 2\phi+|Z_8|\sin 2\phi \cos 2\phi\cos(2\xi+\theta_8)-Z_9\cos 4\phi\cos\xi\,.\label{fddef}
\eeqa
Assuming that $f_a\neq 0$ and $f_c\neq 0$, 
\eq{abcd} implies that
\beq \label{chivalue}
\cot\chi=\frac{f_d}{f_c}\,,\qquad\quad \cot 2\chi=\frac{f_b}{f_a}\,.
\eeq
Employing the trigonometric identity, $\cot 2\chi=(\cot^2\chi-1)/(2\cot\chi)$, we end up with,
\beq \label{Gdef}
G(\phi,\xi)\equiv f_a(f_d^2-f_c^2)-2f_b f_c  f_d=0\,.
\eeq
Note that the above condition is independent of the angle $\psi$.
Inserting the results of \eqst{fadef}{fddef} into \eq{Gdef} leads to a very complicated expression.  However, it is quite easy to check that
\beq \label{plusminus}
G(0,\xi)=-G(\half\pi,\xi)=Z_9^2\Im Z_8\,.
\eeq
As a consequence of \eq{plusminus}, for any choice of $\xi$, there must exist a
value of $\phi$ between 0 and $\half\pi$ such that $G(\phi,\xi)=0$.  
Plugging these values of $\phi$ and $\xi$ back into \eqst{fadef}{fddef}, we can then use \eq{chivalue} to determine $\chi$.   Thus, we have shown that for any choice of $\xi$ and $\psi$, there must exist a corresponding $\phi$ and $\chi$ (whose values depend on the choice of $\xi$) such that $\Im\bar{Z}_8=\Im\bar{Z}_9=0$.\footnote{Although we have reached this conclusion under the assumption that $f_c$ and $f_a$ are nonzero, it is straightforward to modify the analysis if either $f_a=0$ and/or $f_c=0$.  If $f_a=f_b=f_c=f_d=0$, then \eq{abcd} immediately yields $\Im\bar{Z}_8=\Im\bar{Z}_9=0$.  If at least one of the quantities $f_a$, $f_b$, $f_c$ and $f_d$ is nonzero, then $\chi$ can be determined from one of the two expressions in \eq{chivalue}.}

\subsubsection{Does a Higgs basis exist were all scalar potential parameters are real?}
\label{app:rbasis}

Having found a Higgs basis with real $\bar{Z}_8$ and $\bar{Z}_9$ for an arbitrary choice of $\xi$ and $\psi$ (where the parameters $\phi$ and $\chi$ have been determined), we now examine whether it is also possible to choose particular values of $\xi$ and $\psi$ such that $i\bar{Z}_5^\prime e^{\pm i\psi}$ and $\bar{Z}_5 e^{\pm i\psi}$ are both real.  If this were possible, then one would have succeeded in finding a U(2) transformation, $\bar{H}_i = \widetilde{V}_{ij} H_j$ ($i,j=2,3$) such that all the coefficients of the IS scalar potential are real.  
For example, if $Z_5=0$, then it follows from \eqs{zbfivep}{zbfive} that $\bar{Z}_5^\prime=\bar{Z}_5=0$, in which case all the coefficients of the IS scalar potential, when expressed in terms of the barred scalar doublet fields, are real.   Thus a real Higgs basis exists when $Z_5=0$.   

It therefore follows that if $Z_5=0$, then the IS scalar potential must possess a CP2 symmetry of the form, $H_i\to Y_{ij}H_j^{\star}$, where $Y$ is a symmetric unitary matrix, which in block diagonal form [cf.~\eq{hbasisu}] is given by,
\beq \label{Ydef}
Y=\begin{pmatrix} 1 & \,\,\,0 \\ 0 &\,\,\, \widetilde{Y}\end{pmatrix}\,,\qquad\quad \text{with $\widetilde{Y}\equiv (\tilde V^T\tilde V)^*$},
\eeq
and $\tilde{V}$ [given by \eq{vtil}] is the unitary matrix that transforms the $H23$ basis into a real Higgs basis.  Suppose one performs a CP4 transformation [\eq{cp4preferred}] followed by a CP2 transformation, $H_i\to Y_{ij}H_j^{\star}$, and compares the result obtained by performing these two transformations in the opposite order.   
The results of applying CP4 followed by CP2 as compared to CP2 followed by CP4 are equivalent
to the Higgs family transformations, $YX^*$ and $XY^*$, respectively~\cite{Ivanov:2018ime}.  Using \eqss{cp4preferred}{vtil}{Ydef}, it follows that
\beq \label{xystar}
XY^*=e^{2i\psi}YX^*\,.
\eeq
That is, the CP2 and CP4 transformations commute if and only if $\det Y= e^{2i\psi}=1$.

For example, the $H23$ basis is a real Higgs basis in the trivial case where $\Im Z_8=\Im Z_9=0$, independently of the value of $Z_5$.   In this case, the corresponding CP2 symmetry, $H_i\to Y_{ij}H_j^{\star}$, with $Y=\mathds{1}$, commutes with the CP4 symmetry of the IS potential.   If $Z_5=0$, the real Higgs basis obtained above is independent of $\psi$.  In this case, it is convenient to choose $\psi=0$ in defining the CP2 transformation.  It then follows from \eq{xystar} that the CP2 and CP4 transformations commute
if $Z_5=0$ and either $\Im Z_8$ and/or $\Im Z_9$ is nonzero. 

Two other special cases, first pointed out in Appendix B of Ref.~\cite{Ivanov:2018ime}, are noteworthy.   First, suppose that $Z_9=0$.   In this case, the choice of $\psi=\chi=\half\pi$, $\xi=0$ and $\phi=\tfrac14\pi$ inserted into \eqst{zbfivep}{zbnine} will yield a real Higgs basis, with $\bar{Z}_5^\prime=0$, $e^{\pm i\psi}\bar{Z}_5=\mp Z_5$, $\bar{Z_8}=\tfrac14(Z_3^\prime+Z_4^\prime-2\Re Z_8)$ and $\bar{Z}_9=\Im Z_8$.  In light of \eqs{vtil}{sumdiff}, the barred and unbarred scalar fields are related by
\beq
\bar{H}_2=\frac{i}{\sqrt{2}}(H_2-H_3)\,,\qquad\quad \bar{H}_3=\frac{1}{\sqrt{2}}(H_2+H_3)\,.
\eeq  
In the $H23$ basis, we can identify the corresponding CP2 transformation as $H_i\to Y_{ij}H_j^{\star}$, with
\beq
Y=\begin{pmatrix} 1 & \,\,\, 0 &\,\,\, 0 \\ 0 &\,\,\, 0 &\,\,\, 1 \\ 0 &\,\,\, 1 &\,\,\, 0\end{pmatrix}\,.
\eeq
Since $XY^*\neq YX^*$, it follows that the CP2 and CP4 transformations do not commute.

Second, suppose that $\Im Z_8=0$, $\Re Z_9=0$ and $\Im Z_9\neq 0$.  In this case, we simply choose $\bar{H}_2=H_2$ and $\bar{H_3}=iH_3$, corresponding to 
$\psi=\half\pi$, $\chi=\xi=-\tfrac14\pi$ and $\phi=0$ in \eq{vtil}.
A real Higgs basis is then achieved with $\bar{Z}_5^\prime=0$, $e^{\pm i\psi}\bar{Z}_5=\pm Z_5$,
$\bar{Z}_8=-\Re Z_8$, and $\bar{Z}_9=\Im Z_9$.
In the $H23$ basis, we can identify the corresponding CP2 transformation as $H_i\to Y_{ij}H_j^{\star}$, with
\beq
Y=\begin{pmatrix} 1 & \phm 0 &\phm 0 \\ 0 & \phm 1 &\phm 0 \\ 0 & \phm 0 & -1\end{pmatrix}\,.
\eeq
Once again, the CP2 and CP4 transformations do not commute.  
It should be noted that the last two cases are related by a simple basis transformation.  Namely, starting from an $H23$ basis with $\Im Z_8=0$ and $\Re Z_9=0$ and employing $\chi=\psi=\xi=0$ and $\phi=\tfrac14\pi$ in \eqst{zbfivep}{zbnine}  yields $\bar{Z}_5^\prime=0$,
$\bar{Z}_5=Z_5$, $\Im\bar{Z}_8\neq 0$ and $\bar{Z}_9=0$, thereby reducing to the previous case above.

We now consider the IS scalar potential with generic parameters (excluding the special cases considered above) and investigate whether a real Higgs basis exists.  In particular, consider the $H23$ basis with $Z_5$, $Z_9\neq 0$.  
As noted above, we can assume without loss of generality that $Z_5$ and $Z_9$ are real.\footnote{The case of $\Im Z_8=0$ and $\Re Z_9=0$ is thus eliminated from consideration, since
it is related by a scalar basis transformation to the case of $Z_9=0$ as noted above.}   
We examine two different cases:
\begin{itemize}
	\item Case 1: $\psi\neq\pm\half\pi$
	\item Case 2: $\psi=\pm\half\pi$
\end{itemize}

  In case 1, a real Higgs basis would require $\bar{Z}_5^\prime=0$.   Since $\sin 2\phi\neq 0$ [in light of \eq{plusminus}], it follows that $\sin\xi=0$, in which case $e^{\pm i\psi}\bar{Z}_5=\pm e^{i(\chi\pm\psi)}Z_5$ is real if and only if \hbox{$\sin(\chi\pm\psi)=0$}.  
This equation must be satisfied for both sign choices, which yields $\sin\chi\cos\psi=\cos\chi\sin\psi=0$.   For generic values of the parameters, \eqst{fadef}{chivalue} imply that
$\sin\chi\neq 0$ and $\cos\chi\neq 0$.   Thus, in general no value of $\psi$ exists such that $\sin(\chi\pm\psi)=0$ holds for both sign choices.  That is, case 1 cannot yield a real Higgs basis for a generic choice of the IS scalar potential parameters.

In case 2, $i\bar{Z}_5^\prime e^{\pm i\psi}$ is real for all choices of $\xi$ and one must check whether there exists a $\xi$ that yields a real value of $i\bar{Z}_5=ie^{i\chi}Z_5(e^{i\xi}\cos^2\phi+e^{-i\xi}\sin^2\phi$).  The condition that $i\bar{Z}_5$ is real is equivalent to
\beq 
\Re\bigl[e^{i(\chi+\xi)}\cos^2\phi+e^{i(\chi-\xi)}\sin^2\phi\bigr]=0\,,
\eeq
which can be simplified to the condition,
\beq \label{realcond}
\cot\chi=\cos 2\phi\tan\xi\,.
\eeq
It follows that either $\chi\pm\xi$ are both half odd integer multiples of $\half\pi$ or $\cos 2\phi=\cot\chi\cot\xi$.  
If $\chi\pm\xi$ are both half odd integer multiples of $\half\pi$, then either $\chi$ is a half odd integer of $\half\pi$ and $\xi$ is an integer multiple of $\pi$ or vice versa.
If $\chi$ is a half odd integer multiple of $\half\pi$ and $\xi$ is an integer multiple of~$\pi$, then \eq{abcd} yields $f_a=f_d=0$.  However, these latter two equations cannot be simultaneously satisfied if $Z_9\neq 0$.  Similarly, if
$\chi$ is an integer multiple of $\pi$ and $\xi$ is a half odd integer multiple of~$\half\pi$, then \eq{abcd} yields $f_a=f_c=0$ which cannot be simultaneously satisfied if $\Im Z_8$ and $Z_9$ are nonzero.
Thus, if $\chi\pm\xi$ are both half odd integer multiples of $\half\pi$, then no real Higgs basis exists for generic values of the IS scalar potential parameters.

Finally, we examine the possibility that $i\bar{Z}_5$ is real due to $\cos 2\phi=\cot\chi\cot\xi$.  We can also assume that $\xi$ is not an integer multiple of $\half\pi$, as this case was already treated above.
In order that $\Im\bar{Z}_8=\Im\bar{Z}_9=0$, one must satisfy $\cot\chi=f_d/f_c$ and $G(\phi,\xi)=0$, under the assumption of $f_c\neq 0$.
In this case we can satisfy $\Im\bar{Z}_9=0$ if $\phi=\phi_\xi$, where   
\beq \label{phixi}
\cos 2\phi_\xi=\left(\frac{f_d}{f_c}\right)_{\phi=\phi_\xi}\cot\xi\,.
\eeq
Using \eqs{fcdef}{fddef}, one can employ \eq{phixi} to obtain a quadratic equation for $\cot 2\phi_\xi$, whose solution is given by
\beqa
\cot2\phi_\xi&=&\frac{1}{2Z_9}
\biggl[\half(Z_3^\prime+Z_4^\prime)\cos\xi+|Z_8|\cos(\xi+\theta_8) \nonumber \\
&&\qquad\qquad\quad \pm\sqrt{[\half(Z_3^\prime+Z_4^\prime)\cos\xi+|Z_8|\cos(\xi+\theta_8)]^2+4Z_9^2\cos^2\xi}
\biggr]\,.\label{phixisol}
\eeqa
\clearpage

\noindent
\Eq{phixisol} determines $\sin 2\phi_\xi$ up to an overall sign.  It is convenient to choose this sign to be positive.

It is sufficient to demonstrate one example of the IS scalar potential parameters in which no real Higgs basis exists.   Thus, consider an example where
\beq \label{z}
Z_8=-\half(Z_3^\prime+Z_4^\prime)+2iZ_9\,.
\eeq
Then, 
$\half(Z_3^\prime+Z_4^\prime)\cos\xi+|Z_8|(\cos(\xi+\theta_8) =-2Z_9\sin\xi$.
It follows that
\beq
\cot 2\phi_\xi=\pm 1-\sin\xi\,.
\eeq

We now investigate whether a value of $\xi\neq \half n\pi$ (where $n$ is an integer) exists such that $G(\phi_\xi,\xi)=0$.   We introduce the notation, $\tilde f\equiv f(\phi_\xi,\xi)$, where $\phi_\xi$ has been determined from \eq{phixisol}.  Then, \eq{phixi} implies that $\tilde{f}_d=\tilde f_c\cos 2\phi_\xi\tan\xi$.  Inserting this result into \eq{Gdef} yields,
\beq
G(\phi_\xi,\xi)= \tilde{f}_c^2\bigl[\tilde{f}_a(\cos^2 2\phi_\xi\tan^2\xi-1)-2\tilde{f}_b \cos 2\phi_\xi\tan\xi\bigr]\,.
\eeq
An explicit calculation yields
\beq
\tilde{f}_c=\sin 2\phi_\xi\bigl[-\Re Z_8\sin 2\xi+(3\sin\xi\mp 2)(\sin\xi\pm 1)Z_9\bigr]\,,
\eeq
and
\beq
\tilde{f}_a(\cos^2 2\phi_\xi\tan^2\xi-1)-2\tilde{f}_b \cos 2\phi_\xi\tan\xi =2Z_9\frac{\sin 2\phi_\xi}{\cos^2\xi}(\sin\xi\mp 1)\,.
\eeq
Hence, we end up with\footnote{We have made use of the identity, $1+\cot^2 2\phi_\xi=1/\sin^2 2\phi_\xi$.  As noted below \eq{phixisol} , we have assumed that $\sin 2\phi_\xi$ is positive.}
\beq \label{Gspecial}
G(\phi_\xi,\xi)=\frac{2Z_9}{\cos^2\xi}\bigl[1+(1\mp\sin\xi)^2\bigr]^{-3/2}(\sin\xi\mp 1)\bigl[\Re Z_8\sin 2\xi-(3\sin\xi\mp 2)(\sin\xi\pm 1)Z_9\bigr]^2\,.
\eeq
Since the above analysis has assumed that $\tilde{f}_c\neq 0$ and $\xi\neq \half n\pi$ (for integer $n$) it follows that $G(\phi_\xi,\xi)$ is strictly nonzero, which implies that $\Im\bar{Z}_8\neq 0$.\footnote{By expanding to squared expression in \eq{Gspecial}, one sees that the factor of $\cos^2\xi$ in the denominator is canceled by terms in the numerator.  Hence, there is no singularity in the limit of $\cos\xi\to 0$.}  

\subsubsection{Special cases}
Cases where $f_c=0$ need to be treated separately.  First, we
examine the case of $f_c=0$ and $f_d\neq 0$.  Inserting \eq{z} into \eq{fcdef}, one can solve for $\cot 2\phi_\xi$,
\beq \label{tantwo}
\cot 2\phi_\xi=\frac{\Re Z_8\sin 2\xi+2Z_9\cos 2\xi}{Z_9\sin \xi}\,.
\eeq  
We next impose $\Im\bar{Z}_9=0$.  Then \eq{abcd} implies that $\sin\chi=0$, in which case \eq{realcond} yields $\cos\xi=0$.  Inserting the latter result back into \eq{tantwo} yields $\cot 2\phi_\xi=\pm 2$.
If one now attempts to impose $\Im\bar{Z}_8=0$ using \eq{abcd} with $\sin\chi=0$, then one would conclude that $f_a=0$.  However,
one can explicitly show that $f_a\neq 0$ by inserting \eq{z} into \eq{fcdef} and employing $\cot 2\phi_\xi=\pm 2$ and $\cos\xi=0$.  

Finally, we briefly consider the case of $f_c=f_d=0$.  Inserting \eq{z} into \eqs{fcdef}{fddef} yields equations for 
$\tan 2\phi$ and $\tan 4\phi$ respectively.  The compatibility of these two equations then fixes the value of $\xi$.
In this case, $\Im\bar{Z}_9=0$ is automatic, and  $\Im\bar{Z}_8=0$ implies via \eq{abcd} that $\cot 2\chi=f_b/f_a$.  At this stage, $\chi$, $\xi$ and $\phi$ are all determined prior to imposing \eq{realcond}.  The latter is an independent condition; hence for generic values of $\Re Z_8$ and $Z_9$, it is not possible to perform a basis change such that $i\bar{Z}_5$, $\bar{Z}_8$ and $\bar{Z}_9$ are simultaneously real.  This completes all the subcases of case 2.   

We conclude that if the IS scalar potential possesses at least one non-real coefficient (for generic choices of the scalar potential parameters), then no real Higgs basis exists and it is not possible to perform a U(2) transformation of the Higgs basis fields $\{H_2,H_3\}$ such that all coefficients of the scalar potential are real.  

\subsection{Basis-invariant polynomial functions of the IS scalar potential parameters}
\label{app:inv}

In our analysis of the IS model, we have advocated the choice of a particular class of Higgs bases in which $\bar{Z}_5^\prime=0$.   Nevertheless, it is instructive to show that physical observables that depend on the parameters of the IS scalar potential are independent of the choice of the scalar basis.   In this appendix, we introduce a number of basis invariant quantities and evaluate them in the $H23$-basis.

Consider a generic basis of scalar fields, $\{\Phi_a\}$, where $a=1,2,3$ labels hypercharge-one, doublet fields of the 3HDM.  Basis transformations
that leave invariant the form of the canonical kinetic energy terms correspond to global U(3) transformations,
$\Phi_a\to U_{a\bbar}\Phi_b$ [and $\Phi_\abar^\dagger\to\Phi_\bbar^\dagger
U^\dagger_{b\abar}$], where the $3\times 3$ unitary matrix $U$ satisfies
$U^\dagger_{b\abar}U_{a\cbar}=\delta_{b\cbar}$.  Here, we follow the index
conventions introduced in Ref.~\cite{Davidson:2005cw}, in which replacing an unbarred index with a barred index is
equivalent to hermitian conjugation.   We only allow sums over 
barred--unbarred index pairs, which are
performed by employing
the U(3)-invariant tensor $\delta_{a\bbar}$.    
In this notation, the 3HDM scalar potential in a generic $\Phi_a$-basis is given by,
\beq \label{genericpot}
\mathcal{V}=Y_{a\bbar}\Phi_\abar^\dagger\Phi_b
+\half Z_{a\bbar c\dbar}(\Phi_\abar^\dagger\Phi_b)
(\Phi_\cbar^\dagger\Phi_d)\,,
\eeq
where $Z_{a\bbar c\dbar}=Z_{c\dbar a\bbar}$.
Hermiticity of $\mathcal{V}$ implies that
$Y_{a \bbar}= (Y_{b \abar})^\ast$ and $Z_{a\bbar c\dbar}= (Z_{b\abar d\cbar})^\ast$.  Minimizing the scalar potential, under the assumption that the vacuum preserves U(1)$_{\rm EM}$,  yields the neutral Higgs vacuum expectation values, $\vev{\Phi^0_a}=v \widehat{v}_a/\sqrt{2}$, where $v=246$~GeV and $\widehat v_a$ is a vector of unit norm.   It is convenient to define the hermitian matrix\cite{Branco:1999fs}
\beq \label{vabdef}
V_{a\bbar}\equiv \widehat v_a \,\widehat v_\bbar^\ast\,.
\eeq

One can now construct basis-invariant quantities that depend on knowledge of the scalar potential minimum by forming products of $V_{a\bbar}$ and $Z_{a\bbar c\dbar}$ such that all barred--unbarred index pairs are summed over.  We define six invariant quantities below, 
\clearpage

\beqa
J_1&=&V_{a\cbar}V_{b\dbar}Z_{c\abar d\bbar},\\
J_2&=&V_{a\bbar}Z_{b\abar c\cbar},\\
J_3&=&V_{a\bbar}Z_{b\cbar c\abar},\\
J_4&=&V_{a\bbar}Z_{b\dbar c\ebar}Z_{d\abar e\cbar},\\
J_5&=&V_{a\bbar}Z_{b\dbar c\ebar}Z_{d\fbar e\gbar}Z_{f\abar g\cbar},\\
J_6&=&V_{a\bbar}Z_{b\dbar c\ebar}Z_{d\fbar e\gbar}Z_{f\hbarr g\kbar}Z_{h\abar k\cbar}.
\eeqa
The invariants above can be evaluated in any basis.   In particular, in the $H23$-basis, the only nonzero component of $V_{a\bbar}$ is $V_{11}=1$.
We thus obtain,
\beqa
J_1&=&Z_1,\\
J_2&=&Z_1+2Z_3,\\
J_3&=&Z_1+2Z_4,\\
J_4&=&Z_1^2+2 Z_3^2+2 Z_4^2+2 Z_5^2,\\
J_5&=&Z_1^3+4 Z_5^2 Z_1+2 Z_3^3+6 Z_3 Z_4^2+2 Z_2 Z_5^2+4 Z_5^2 \Re Z_8 ,\\
J_6&=&Z_1^4+2 Z_3^4+2 Z_4^4+12 Z_3^2 Z_4^2+4 Z_5^4+2Z_5^2(3Z_1^2+2Z_1 Z_2+Z_2^2)\nonumber\\
&&+8 Z_5^2\bigl[|Z_8|^2+(Z_1+Z_2)\Re Z_8+ \left(\Im Z_9\right)^2\bigr].
\eeqa

Using the first four invariant quantities above, one can show that $Z_5$ can be expressed in terms of an invariant quantity.\footnote{In the $H23$-basis (where $Z_5^\prime=0$), one expects that $Z^2_5$ can be expressed in terms of an invariant quantity in light of the mass relation,
$M_P^2-M_Q^2=Z_5 v^2$, which implies that $Z_5^2$ is a physical parameter.  In a general class of Higgs bases, the corresponding invariant quantity is $|Z_5|^2+|Z_5^\prime|^2$ [cf.~\eq{z55}].}    
In particular,
\beq \label{zeefiveinv}
Z_5^2=-J_1^2+\half J_1 \left(J_2+J_3\right) -\tfrac14 (J_2^2+J_3^2)+\half J_4\,.
\eeq

Finally, we have discovered a remarkable invariant quantity,
\beqa
\mathcal{N}&=&32Z_5^2 J_6
-16J_5^2+8J_5(3J_{21}J_{31}^2+K) -J_{31}^4(9J_{21}^2+4Z_5^2) -6KJ_{21}J_{31}^2  
-24Z_5^2 J_{21}^2 J_{31}^2  \nonumber \\[5pt]
&& -J_{21}^6 - 4Z_5^2J_{21}^4-8J_1(J_1^2+2Z_5^2)J_{21}^3 -16J_1^6 
-96 Z_5^2 J_1^4  -192 Z_5^4 J_1^2  -128Z_5^6\,, \label{Ninv}
\eeqa
where $J_{ij}\equiv J_i-J_j$, the invariant quantity $Z^2_5$ is given by \eq{zeefiveinv} and
\beq
K\equiv 4J_1^3+8Z_5^2 J_1+J_{21}^3\,.
\eeq

Plugging in the expressions for $J_1\,,\ldots\,,J_6$ given above, we find
\beq \label{N}
\mathcal{N}=256 Z_5^4\bigl[(\Im Z_8)^2 + (\Im Z_9)^2\bigr]\,.
\eeq
It follows that if $Z_5\neq 0$ then there exists a ratio of invariant quantities, which when evaluated in the $H23$-basis, is equal to $(\Im Z_8)^2 + (\Im Z_9)^2$.   
In contrast, if $Z_5=0$, then there is no invariant quantity that reduces in the $H23$-basis to $(\Im Z_8)^2 + (\Im Z_9)^2$.  Nevertheless, the invariant condition, $Z_5=0$, signals the presence of four mass-degenerate neutral scalars.

The significance of the invariant $\mathcal{N}$ is as follows.  The CP4-conserving IS model possesses a CP2 symmetry that commutes with CP4 if and only if $\mathcal{N}=0$.  Note that the nonvanishing of $\mathcal{N}$ does not exclude the possibility of a CP2 symmetry that does not commute with CP4.  Two explicit examples of this phenomenon were presented in Appendix~\ref{app:rbasis}: (i) $Z_9=0$ and $\Im Z_8\neq 0$; and (ii) $\Im Z_8=\Re Z_9=0$ and $\Im Z_9\neq 0$.  In both these cases, a real Higgs basis exists, and the corresponding CP2 transformation does not commute with CP4.\footnote{One cannot employ \eq{N} to compute $\mathcal{N}$ in the real Higgs basis in cases (i) and (ii), since in both cases, the real Higgs basis lies outside the set of $H23$ bases.  Nevertheless, we have checked that evaluating $\mathcal{N}$ directly in the real Higgs basis in cases (i) and (ii)  reproduces the corresponding results obtained in the $H23$ basis via \eq{N}.}   
We also noted in Section~\ref{sect:Z-to-QQQQ} that under the assumption that $M_P\neq M_Q$ (or equivalently for $Z_5\neq 0$ in the $H23$ basis), the decay rate for $Z\to QQQQ^*$, $QQ^* Q^* Q^*$, if kinematically allowed, is nonzero if and only if $\mathcal{N}\neq 0$.

Note that the invariant quantity $\mathcal{N}$ constructed above has been expressed in terms of Higgs basis parameters.   This means that this invariant quantity depends on the knowledge of the vacuum, i.e. the minimum of the scalar potential (which is needed to formally define the Higgs basis).    Given an explicitly CP4-invariant scalar potential, one could ask a slightly different question: is there an invariant quantity that can differentiate between scalar potentials that explicitly preserve or violate the CP2 symmetry, independently of the vacuum.  This question has been recently addressed and answered in Ref.~\cite{Ivanov:2018ime}.  However, it is not clear that such an invariant quantity can be directly related in practice to a physical observable.

\subsection{An alternative Higgs basis}
\label{app:basis}

In this paper, we first defined the $H23$-basis by employing the scalar doublet fields $\{H_1,H_2,H_3\}$, which was one particular choice among possible Higgs bases.  An arbitrary Higgs basis can be obtained by performing the U(2) basis transformation given by \eqs{hbasisu}{vtil}.  The corresponding IS scalar potential is given by \eq{visbar}, where the barred coefficients in terms of the unbarred coefficients are given in \eqst{zbtwo}{zbnine}.

In section~\ref{alt-basis}, we explored another Higgs basis choice, called the $RS$-basis, which employs the scalar doublet fields,
$\{H_1,\mathcal{R},\mathcal{S}\}$.   The relations between the $H23$-basis and $RS$-basis are given by,
\beq
\mathcal{R}\equiv \frac{1}{\sqrt{2}}\bigl(H_2+iH_3\bigr)=\begin{pmatrix} R^\dagger \\ \frac{1}{\sqrt{2}}\bigl(P+iQ^\dagger\bigr)\end{pmatrix}\,,\quad\qquad
\mathcal{S}\equiv \frac{1}{\sqrt{2}}\bigl(H_2-iH_3\bigr)=\begin{pmatrix} S^\dagger \\ \frac{1}{\sqrt{2}}\bigl(P^\dagger+iQ\bigr)\end{pmatrix}\,.
\eeq
Note that the form of the CP4 transformation in this basis is
\beq
\begin{pmatrix} H_1 \\ \mathcal{R} \\ \mathcal{S}\end{pmatrix}\longrightarrow \begin{pmatrix} 1 & \phm 0 & \phm 0 \\ 0 & \phm 0 & \phm i \\ 0 & -i & \phm 0\end{pmatrix}\begin{pmatrix} H_1^\dagger \\ \mathcal{R}^\dagger \\ \mathcal{S}^\dagger\end{pmatrix}\,.
\eeq
The change of basis from $\{H_1,H_2,H_3\}$ to $\{H_1,R,S\}$ corresponds to choosing $\alpha=\beta=\phi=\tfrac14\pi$ and $\psi=-\half\pi$.
Inserting these results into \eqst{zbtwo}{zbnine} yields,
\enlargethispage{1.25\baselineskip}
\beqa
\bar{Z}_2&=&Z_2+\half(Z_3^\prime+Z_4^\prime-2\Re Z_8)\,,\\
\bar{Z}_3^\prime&=&-Z_4^\prime+2\Re Z_8\,,\\
\bar{Z}_4^\prime&=&\half(Z_4^\prime-Z_3^\prime+2\Re Z_8)\,,\\
\bar{Z}_5^\prime&=&Z_5\,,\\
\bar{Z}_5&=&0\,,\\
\bar{Z}_8&=&-\tfrac14(Z_3^\prime+Z_4^\prime+2\Re Z_8)+i\Re Z_9\,,\\
\bar{Z}_9&=& \Im Z_9+i\Im Z_8\,.\label{app:z89}
\eeqa

Note that in the $RS$-basis, the invariant $\mathcal{N}$ defined in \eq{Ninv} is 
\beq
\mathcal{N}=256\bar{Z}_5^{\prime\,4}|\bar{Z}_9|^2\,.
\eeq
In particular, the absence [or presence]  of the $Z\to QQQQ^*$, $QQ^*Q^*Q^*$ decay discussed in Appendix~\ref{app:QQQQ} depends on 
the [non-]vanishing of $\bar{Z}_9$.  These decays are governed by \eq{lin}, which when expressed in the $RS$-basis is given by,
\beq \label{linalt}
 \delta\mathscr{L}_{4h}\ni \half i(PQ-P^\dagger Q^\dagger)\bigl[\bar{Z}_9(P^{\dagger\,2}+Q^2)+\bar{Z}_9^*(P^2+Q^{\dagger\,2})\bigr]\,.
 \eeq

\section{$\boldsymbol{Z}$ decay into four inert neutral scalars}
\label{app:QQQQ}
\setcounter{equation}{0}
\renewcommand{\theequation}{B.\arabic{equation}}

Consider a universe (not ours) in which the electroweak theory of elementary particles at the electroweak scale consists of the IS model, with
$
M_{H^\pm,h^\pm}<M_Q<\tfrac14 m_Z\ll M_P\,.
$
In this case, the decay of the $Z$ to four neutral inert scalars would be consistent with a CP4-symmetric IS scalar potential that does not possess a real scalar basis.  Experimentally, the final state would be detected via the decay $Q\to (H^\pm, h^\pm)+W^{*\,\mp}$, with the virtual $W^{*\,\mp}$ decaying to quark or lepton pairs.
In this universe, the $H^\pm, h^\pm$ are the lightest particles of the inert scalar sector and hence stable.  Although this is not our universe, this example provides a proof in principle of the existence of an experimental distinction between the CP4-conserving/CP2-nonconserving case and  the CP4/CP2-conserving case.   

\begin{figure}[t!]
\refstepcounter{figure}
\label{Fig:cpc-br-ratios-low}
\addtocounter{figure}{-1}
\begin{center}
\includegraphics[scale=0.7]{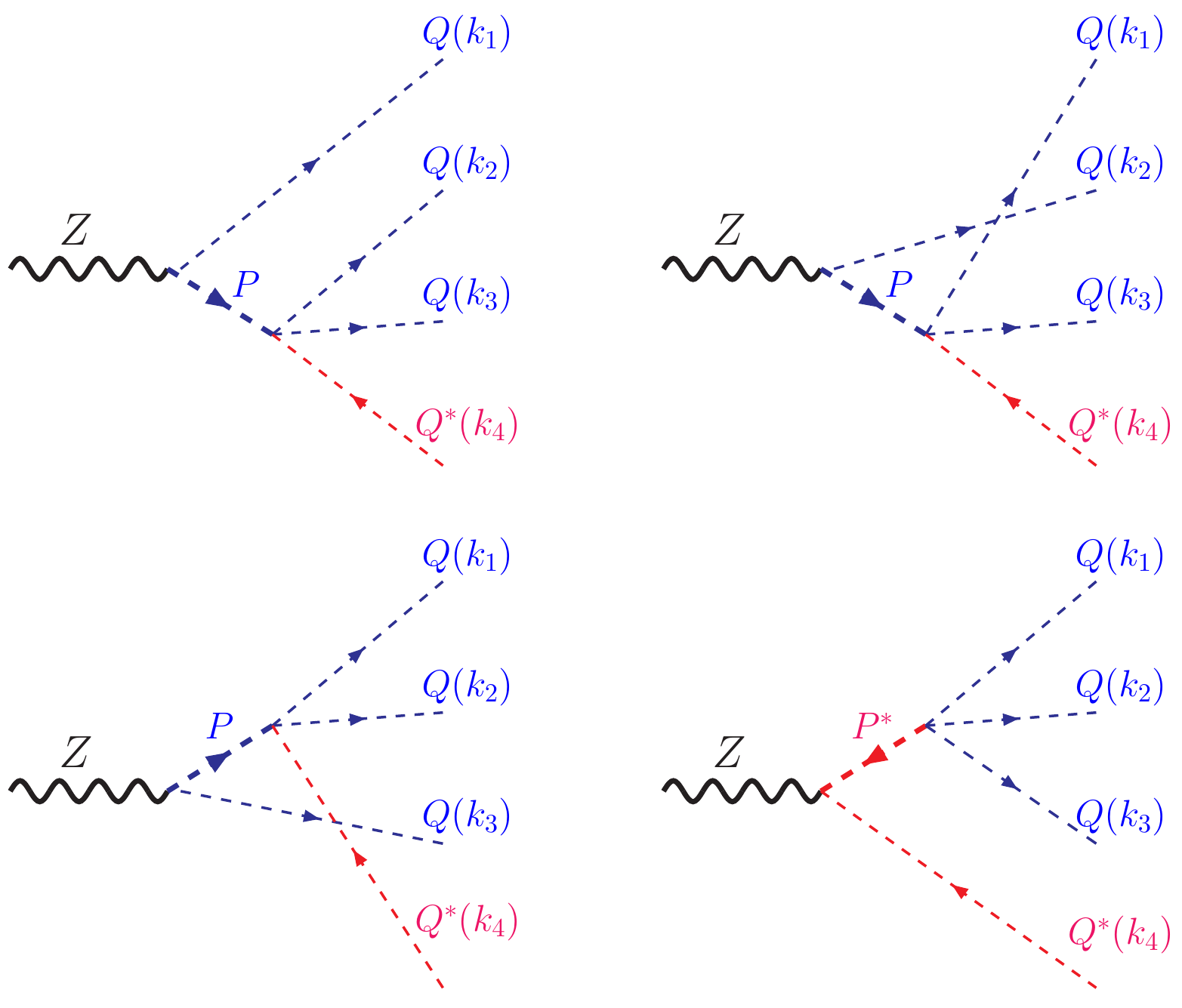}
\end{center}
\vspace*{-4mm}
\caption{\small Feynman diagrams for $Z\to QQQQ^\ast$}
\end{figure}

In light of \eq{lin}, there are four contributing tree-level Feynman diagrams to the decay amplitude $Z\to QQQQ^\ast$, which are shown in Fig.~\ref{Fig:cpc-br-ratios-low}.
Employing the Feynman rules obtained from \eq{lin} (and including the appropriate symmetry factors in obtaining the rules for the four-scalar vertex), the invariant matrix element is given by,
\beqa
i\mathcal{M}&=&\frac{g}{2\cos\theta_W}\bigl(\Im Z_8 +i\Im Z_9\bigr)\,\varepsilon_\lambda(p)\cdot\left[\frac{p-2k_1}{(p-k_1)^2-M_P^2}+\frac{p-2k_2}{(p-k_2)^2-M_P^2}\right. \nn \\[6pt]
&&\qquad\qquad\qquad
\left.+\frac{p-2k_3}{(p-k_3)^2-M_P^2}-\frac{3(p-2k_4)}{(p-k_4)^2-M_P^2}\right]\,,\nn
\eeqa
where $p$ is the four-momentum of the $Z$ and the $k_i$ are the final state momenta (with $k_4$ the momentum of $Q^*$).  We then square the matrix element and average over the initial state spins, using
\beq
|\mathcal{M}|^2_{\rm ave}\equiv \tfrac13\sum_\lambda |X\cdot \varepsilon_\lambda(p)|^2=-\tfrac13 X_\mu X^*_\nu\left( g^{\mu\nu}-\frac{p^\mu p^\nu}{m_Z^2}\right)\,.
\eeq
where $X$ is the four vector dotted into the polarization vector in the expression for $i\mathcal{M}$.

We shall work in the approximation that $M_P\gg m_Z$ and $M_Q=0$.   In this case,
\beq
X=\frac{g}{M_P^2\cos\theta_W}\bigl(\Im Z_8 +i\Im Z_9\bigr)(p-4k_4)\,,
\eeq
where we have used conservation of momentum, $p=k_1+k_2+k_3+k_4$.
It then follows after some simplification (with $p^2=m_Z^2$) that,
\beq
|\mathcal{M}|^2_{\rm ave}=\frac{16g^2}{3m_Z^2 M_P^4\cos^2\theta_W}\bigl[(\Im Z_8)^2+ (\Im Z_9)^2\bigr] (p\cdot k_4)^2\,.
\eeq

The four body decay width for $Z\to QQQQ^*$ is given by
\beq
\Gamma=\frac16\,\frac{(2\pi)^{-8}}{2m_Z}\int\left(\prod_{i=1}^4 \frac{d^3 k_i}{2E_i}\right)\delta^4(p-k_1-k_2-k_3-k_4)|\mathcal{M}|^2_{\rm ave}\,,
\eeq
where the factor of $1/6$ is due to the three identical Qs in the final state (which means we overcount by a factor of 3! by integrating over the full phase space).

Using the above results, we obtain,
\beq \label{e1}
\Gamma=\frac{4g^2(2\pi)^{-8}}{9m_Z^3 M_P^4\cos^2\theta_W}\bigl[(\Im Z_8)^2+ (\Im Z_9)^2\bigr]
\int\left(\prod_{i=1}^4 \frac{d^3 k_i}{2E_i}\right)\delta^4(p-k_1-k_2-k_3-k_4) \,(p\cdot k_1)^2
\eeq
after changing integration variables $k_1\longleftrightarrow k_4$.

To perform the phase space integration, we follow Ref.~\cite{Haber:1993jr}.
   To integrate over $d^3k_1d^3k_2$ we use,
  \beq
  \int\frac{d^3k_1}{2E_1}\frac{d^3k_2}{E_2}
\delta^{4}(N-k_1-k_2)\times\begin{cases}
1& \text{$=\frac12\pi$,}\\
k_{1\mu}&\text{$=\frac14\pi N_\mu$,}\\
k_{1\mu}k_{1\nu}&\text{$=-\frac{1}{24}\pi(N^2g_{\mu\nu}-4N_\mu N_\nu)$,}\\
k_{1\mu}k_{2\nu}&\text{$=\frac{1}{24}\pi(N^2g_{\mu\nu}+2N_\mu N_\nu)$,}\end{cases}
\eeq
where $N$ is an arbitrary four-vector.  In the present application, $N=p-k_3-k_4$.  
After performing this integration, we have two further integrations to do over $k_3$ and $k_4$.
It is convenient to
work in the $Z$ rest frame:
\beq \label{kinematics}
p=(m_Z;0,0,0);\ \  k_3=E_3(1;0,0,1);\ \
k_4=E_4(1;\sin\theta,0,\cos\theta)\,.
\eeq
 We introduce
the following scaled kinematic variables
\beq
w\equiv\frac{1-\cos\theta}{2},\ \ \
y\equiv\frac{2E_3}{m_Z},\ \ \ z\equiv\frac{2E_4}{m_Z}\,.
\eeq
Then, 
\beq \label{e3e4}
\int\frac{d^3 k_3}{2E_3}\,\frac{d^3k_4}{2E_4}=\frac{\pi^2 m_Z^4}{4} 
\int_0^{1}\!z\,dz\! \left\{\int_0^{1-z}y\,dy\int_0^1 dw+\!
\int_{1-z}^1 y\,dy\! \int_{(y+z-1)/yz}^1 \!dw\right\}\,.
\eeq

We now evaluate the integral in \eq{e1}.  Using the above results,
\beqa
\int\frac{d^3k_1}{2E_1}\frac{d^3k_2}{E_2}
\delta^{4}(N-k_1-k_2)\,(p\cdot k_1)^2&=&-\frac{\pi}{24}\biggl\{m_Z^2(p-k_3-k_4)^2-4\bigl[p\cdot(p-k_3-k_4)\bigr]^2\biggr\} \nn \\[6pt]
&=&\frac{\pi m_Z^4}{24}\bigl[(2-y-z)^2-1+y+z-yzw\bigr]\,,
\eeqa
where $N\equiv p-k_3-k_4$.  In obtaining the above result, we have used [cf.~\eq{kinematics}],
\beqa
(p-k_3-k_4)^2&=&m_Z^2-2p\cdot (k_3+k_4)+2k_3\cdot k_4=m_Z^2(1-y-z+yzw)\,, \\[5pt]
p\cdot (p-k_3-k_4) &=&m^2_Z\bigl[1-\half(y+z)\bigr]\,.
\eeqa
Hence, after employing \eq{e3e4}, we end up with
\beqa
&& \int\left(\prod_{i=1}^4 \frac{d^3 k_i}{2E_i}\right)\delta^4(p-k_1-k_2-k_3-k_4) \,(p\cdot k_1)^2
\nn \\[10pt]
&& \qquad =\frac{\pi^3 m_Z^8}{96}\int_0^{1}\!z\,dz\! \left\{\int_0^{1-z} y\,dy\int_0^1 dw+\!
\int_{1-z}^1 y\,dy\! \int_{(y+z-1)/yz}^1 \!dw\right\}\bigl[(2-y-z)^2-1+y+z-yzw\bigr]
\nn \\[6pt] 
&&\qquad\,\, =\frac{\pi^3 m_Z^8}{1280}\,. 
\eeqa
Collecting all our results, we end up with
\beq
\Gamma=\frac{g^2 m_Z^5\bigl[(\Im Z_8)^2+(\Im Z_9)^2\bigr]}{3^2\cdot  5\cdot2^{14}\, \pi^5 M_P^4\cos^2\theta_W}\,.
\eeq
Relative to the decay rate of $Z$ into a neutrino pair, $\Gamma(Z\to \nu\bar{\nu})= g^2 m_Z/(96\pi\cos^2\theta_W)$, 
\beq \label{app:zqqqq}
\frac{\Gamma(Z\to QQQQ^*)}{\Gamma(Z\to \nu\bar{\nu})}=\frac{(\Im Z_8)^2+(\Im Z_9)^2}{3\cdot  5\cdot 2^{9}\, \pi^4 }\left(\frac{m_Z}{M_P}\right)^4\,.
\eeq

Finally, we note that the decay rate for $Z\to QQ^*Q^*Q^*$ is identical to the one given above.  Since $Q$ and $Q^*$ are mass degenerate, the experimentally observable width would be a factor of~2 times the one given in \eq{app:zqqqq}, as quoted in \eq{zqqqq}.

\bigskip



\begin{thebibliography}{99}
    
\bibitem{Aad:2012tfa}
  G.~Aad {\it et al.} [ATLAS Collaboration],
  ``Observation of a new particle in the search for the Standard Model Higgs boson with the ATLAS detector at the LHC,''
  Phys.\ Lett.\ B {\bf 716} (2012) 1
  [arXiv:1207.7214 [hep-ex]].

\bibitem{Chatrchyan:2012xdj}
  S.~Chatrchyan {\it et al.} [CMS Collaboration],
  ``Observation of a new boson at a mass of 125 GeV with the CMS experiment at the LHC,''
  Phys.\ Lett.\ B {\bf 716} (2012) 30
  [arXiv:1207.7235 [hep-ex]].
   
\bibitem{Gunion:2012gc}
  J.F.~Gunion, Y.~Jiang and S.~Kraml,
  ``Could two NMSSM Higgs bosons be present near 125 GeV?,''
  Phys.\ Rev.\ D {\bf 86} (2012) 071702
   [arXiv:1207.1545 [hep-ph]].

\bibitem{Gunion:2012he}
  J.F.~Gunion, Y.~Jiang and S.~Kraml,
  ``Diagnosing Degenerate Higgs Bosons at 125 GeV,''
  Phys.\ Rev.\ Lett.\  {\bf 110} (2013) 051801
   [arXiv:1208.1817 [hep-ph]].

\bibitem{Ferreira:2012nv}
  P.M.~Ferreira, R.~Santos, H.E.~Haber and J.P.~Silva,
  ``Mass-degenerate Higgs bosons at 125 GeV in the two-Higgs-doublet model,''
  Phys.\ Rev.\ D {\bf 87} (2013) 055009
  [arXiv:1211.3131 [hep-ph]].

\bibitem{Drozd:2012vf}
  A.~Drozd, B.~Grzadkowski, J.F.~Gunion and Y.~Jiang,
  ``Two-Higgs-Doublet Models and Enhanced Rates for a 125 GeV Higgs,''
  JHEP {\bf 1305} (2013) 072
  [arXiv:1211.3580 [hep-ph]].

\bibitem{Grossman:2013pt}
  Y.~Grossman, Z.~Surujon and J.~Zupan,
  ``How to test for mass degenerate Higgs resonances,''
  JHEP {\bf 1303} (2013) 176
  [arXiv:1301.0328 [hep-ph]].
  
\bibitem{Munir:2013wka}
  S.~Munir, L.~Roszkowski and S.~Trojanowski,
  ``Simultaneous enhancement in $\gamma \gamma, b\bar{b}$ and $\tau^{+} \tau^{-}$ rates in the NMSSM with nearly degenerate scalar and pseudoscalar Higgs bosons,''
  Phys.\ Rev.\ D {\bf 88} (2013) 055017
  [arXiv:1305.0591 [hep-ph]].

\bibitem{Efrati:2013ini}
  A.~Efrati, D.~Grossman and Y.~Hochberg,
  ``A tale of two Higgs,''
  JHEP {\bf 1309} (2013) 118
   [arXiv:1302.7215 [hep-ph]].

\bibitem{David:2014jla}
  A.~David, J.~Heikkil\"a and G.~Petrucciani,
  ``Searching for degenerate Higgs bosons: a profile likelihood ratio method to test for mass-degenerate states in the presence of incomplete data and uncertainties,''
  Eur.\ Phys.\ J.\ C {\bf 75} (2015) 49
  [arXiv:1409.6132 [hep-ph]].

\bibitem{Han:2015pwa}
  X.F.~Han, L.~Wang and J.M.~Yang,
  ``Higgs pair signal enhanced in the 2HDM with two degenerate 125 GeV Higgs bosons,''
  Mod.\ Phys.\ Lett.\ A {\bf 31} (2016) 1650178
  [arXiv:1509.02453 [hep-ph]].

\bibitem{Khachatryan:2016vau}
  G.~Aad {\it et al.} [ATLAS and CMS Collaborations],
  ``Measurements of the Higgs boson production and decay rates and constraints on its couplings from a combined ATLAS and CMS analysis of the LHC pp collision data at $ \sqrt{s}=7 $ and 8 TeV,''
  JHEP {\bf 1608} (2016) 045
  [arXiv:1606.02266 [hep-ex]].

\bibitem{Sirunyan:2018koj}
  A.M.~Sirunyan {\it et al.} [CMS Collaboration],
``Combined measurements of Higgs boson couplings in proton-proton collisions at $\sqrt{s}=$ 13 TeV,''
  [arXiv:1809.10733 [hep-ex]].
  
  \bibitem{ATLAS:2018doi}
  ATLAS collaboration,
  ``Combined measurements of Higgs boson production and decay using up to 80 fb$^{-1}$ of proton--proton collision data at $\sqrt{s}=$ 13 TeV collected with the ATLAS experiment,''
  ATLAS-CONF-2018-031 (July, 2018).
  
\bibitem{Bian:2017gxg}
  L.~Bian, N.~Chen, W.~Su, Y.~Wu and Y.~Zhang,
  ``Future prospects of mass-degenerate Higgs bosons in the CP -conserving two-Higgs-doublet model,''
  Phys.\ Rev.\ D {\bf 97} (2018) 115007
  [arXiv:1712.01299 [hep-ph]].

\bibitem{Donoghue:1978cj}
  J.F.~Donoghue and L.F.~Li,
  ``Properties of Charged Higgs Bosons,''
  Phys.\ Rev.\ D {\bf 19} (1979) 945.

\bibitem{Georgi:1978ri}
  H.~Georgi and D.~V.~Nanopoulos,
  ``Suppression of Flavor Changing Effects From Neutral Spinless Meson Exchange in Gauge Theories,''
  Phys.\ Lett.\  {\bf 82B} (1979) 95.

\bibitem{Botella:1994cs}
  F.J.~Botella and J.P.~Silva,
  ``Jarlskog--like invariants for theories with scalars and fermions,''
  Phys.\ Rev.\ D {\bf 51} (1995) 3870
  [hep-ph/9411288].

\bibitem{Branco:1999fs}
  G.C.~Branco, L.~Lavoura and J.P.~Silva,
 \textit{CP Violation} (Oxford University Press, Oxford, UK, 1999).
  
\bibitem{Davidson:2005cw}
  S.~Davidson and H.E.~Haber,
  ``Basis-independent methods for the two-Higgs-doublet model,''
  Phys.\ Rev.\ D {\bf 72} (2005) 035004
   [Erratum: Phys.\ Rev.\ D {\bf 72} (2005) 099902]
   [hep-ph/0504050].


\bibitem{Ogreid:2017alh}
  O.M.~Ogreid, P.~Osland and M.N.~Rebelo,
  ``A Simple Method to detect spontaneous CP Violation in multi-Higgs models,''
  JHEP {\bf 1708} (2017) 005
  [arXiv:1701.04768 [hep-ph]].

\bibitem{Gunion:1989we}
  J.F.~Gunion, H.E.~Haber, G.L.~Kane and S.~Dawson,
  \textit{The Higgs Hunter's Guide}
  (Westview Press, Boulder, CO, 2000).


\bibitem{Branco:2011iw}
  G.C.~Branco, P.M.~Ferreira, L.~Lavoura, M.N.~Rebelo, M.~Sher and J.P.~Silva,
  ``Theory and phenomenology of two-Higgs-doublet models,''
  Phys.\ Rept.\  {\bf 516} (2012) 1
   [arXiv:1106.0034 [hep-ph]].
 
\bibitem{Gerard:2007kn}
  J.-M.~Gerard and M.~Herquet,
  ``A Twisted custodial symmetry in the two-Higgs-doublet model,''
  Phys.\ Rev.\ Lett.\  {\bf 98} (2007) 251802
  [hep-ph/0703051 [hep-ph]].
 
\bibitem{Haber:2010bw}
  H.E.~Haber and D.~O'Neil,
  ``Basis-independent methods for the two-Higgs-doublet model III: The CP-conserving limit, custodial symmetry, and the oblique parameters $S$, $T$, $U$,''
  Phys.\ Rev.\ D {\bf 83} (2011) 055017
  [arXiv:1011.6188 [hep-ph]].
 
\bibitem{Grzadkowski:2010dj} 
  B.~Grzadkowski, M.~Maniatis and J.~Wudka,
  ``The bilinear formalism and the custodial symmetry in the two-Higgs-doublet model,''
  JHEP {\bf 1111} (2011) 030
    [arXiv:1011.5228 [hep-ph]].
  
\bibitem{Nishi:2011gc} 
  C.C.~Nishi,
  ``Custodial SO(4) symmetry and CP violation in N-Higgs-doublet potentials,''
  Phys.\ Rev.\ D {\bf 83}  (2011) 095005
  [arXiv:1103.0252 [hep-ph]].
 
\bibitem{Ma:2006km}
  E.~Ma,
  ``Verifiable radiative seesaw mechanism of neutrino mass and dark matter,''
  Phys.\ Rev.\ D {\bf 73} (2006) 077301
   [hep-ph/0601225].
 
\bibitem{Barbieri:2006dq}
  R.~Barbieri, L.J.~Hall and V.S.~Rychkov,
  ``Improved naturalness with a heavy Higgs: An Alternative road to LHC physics,''
  Phys.\ Rev.\ D {\bf 74} (2006) 015007
   [hep-ph/0603188].

\bibitem{Ivanov:2015mwl}
  I.P.~Ivanov and J.P.~Silva,
  ``CP-conserving multi-Higgs model with irremovable complex coefficients,''
  Phys.\ Rev.\ D {\bf 93} (2016) 095014
  [arXiv:1512.09276 [hep-ph]].

\bibitem{Kopke:2018vyw}
  M.~K\" opke,
  ``Investigation of the GCP Structure of Three-Higgs-Doublet Models and a 
General Method to Derive Boundedness Constraints for Multi-Higgs Potentials,''
Master Thesis, Karlsruhe Instititue of Theoretical Physics (KIT), 14 February 2018.

\bibitem{Ivanov:2018ime}
  I.P.~Ivanov, C.C.~Nishi, J.P.~Silva and A.~Trautner,
  ``Basis-invariant conditions for CP symmetry of order 4,''
  arXiv:1810.13396 [hep-ph].

\bibitem{Craig:2013hca}
  N.~Craig, J.~Galloway and S.~Thomas,
  ``Searching for Signs of the Second Higgs Doublet,''
  arXiv:1305.2424 [hep-ph].

\bibitem{Haber:2013mia}
  H.E.~Haber,
  ``The Higgs data and the Decoupling Limit,'' in
Proceedings of the 1st Toyama International Workshop on Higgs as a Probe of
  New Physics 2013 (HPNP2013), Toyama, Japan, February 13--16, 2013,
  arXiv:1401.0152 [hep-ph].

\bibitem{Asner:2013psa}
  D.M.~Asner {\it et al.},
  ``ILC Higgs White Paper,''
  arXiv:1310.0763 [hep-ph].

\bibitem{Carena:2013ooa}
  M.~Carena, I.~Low, N.R.~Shah and C.E.M.~Wagner,
  ``Impersonating the Standard Model Higgs Boson: Alignment without Decoupling,''
  JHEP {\bf 1404} (2014) 015
  [arXiv:1310.2248 [hep-ph]].

\bibitem{Carena:2014nza}
  M.~Carena, H.E.~Haber, I.~Low, N.R.~Shah and C.E.M.~Wagner,
  ``Complementarity between Nonstandard Higgs Boson Searches and Precision Higgs Boson Measurements in the MSSM,''
  Phys.\ Rev.\ D {\bf 91} (2015)  035003
  [arXiv:1410.4969 [hep-ph]].

\bibitem{Dev:2014yca}
  P.S.~Bhupal Dev and A.~Pilaftsis,
  ``Maximally Symmetric Two Higgs Doublet Model with Natural Standard Model Alignment,''
  JHEP {\bf 1412} (2014) 024
   [Erratum: JHEP {\bf 1511} (2015) 147]
  [arXiv:1408.3405 [hep-ph]].

\bibitem{Bernon:2015qea}
  J.~Bernon, J.F.~Gunion, H.E.~Haber, Y.~Jiang and S.~Kraml,
  ``Scrutinizing the alignment limit in two-Higgs-doublet models: $m_h=125$~GeV,''
  Phys.\ Rev.\ D {\bf 92} (2015) 075004
  [arXiv:1507.00933 [hep-ph]].

\bibitem{Bernon:2015wef}
  J.~Bernon, J.F.~Gunion, H.E.~Haber, Y.~Jiang and S.~Kraml,
  ``Scrutinizing the alignment limit in two-Higgs-doublet models. II. $m_H=125$~GeV,''
  Phys.\ Rev.\ D {\bf 93} (2016) 035027
  [arXiv:1511.03682 [hep-ph]].
  
  \bibitem{Pilaftsis:2016erj}
  A.~Pilaftsis,
  ``Symmetries for standard model alignment in multi-Higgs doublet models,''
  Phys.\ Rev.\ D {\bf 93} (2016) 075012
  [arXiv:1602.02017 [hep-ph]].

\bibitem{Lavoura:1994fv}
  L.~Lavoura and J.P.~Silva,
  ``Fundamental CP violating quantities in a SU(2)$\times$U(1) model with many Higgs doublets,''
  Phys.\ Rev.\ D {\bf 50} (1994) 4619
   [hep-ph/9404276].
  
\bibitem{Lee:1973iz}
  T.D.~Lee,
  ``A Theory of Spontaneous T Violation,''
  Phys.\ Rev.\ D {\bf 8} (1973) 1226.

\bibitem{Branco:1985aq}
  G.C.~Branco and M.N.~Rebelo,
  ``The Higgs Mass in a Model With Two Scalar Doublets and Spontaneous {CP} Violation,''
  Phys.\ Lett.\  {\bf 160B} (1985) 117.

\bibitem{Branco:2005em}
  G.C.~Branco, M.N.~Rebelo and J.I.~Silva-Marcos,
  ``CP-odd invariants in models with several Higgs doublets,''
  Phys.\ Lett.\ B {\bf 614} (2005) 187
  [hep-ph/0502118].

\bibitem{Gunion:2005ja} 
  J.F.~Gunion and H.E.~Haber,
  ``Conditions for CP-violation in the general two-Higgs-doublet model,''
  Phys.\ Rev.\ D {\bf 72}, 095002 (2005)
  [hep-ph/0506227].

\bibitem{Grzadkowski:2016szj}
  B.~Grzadkowski, O.M.~Ogreid and P.~Osland,
  ``Spontaneous CP violation in the 2HDM: physical conditions and   
  the alignment limit,''
  Phys.\ Rev.\ D {\bf 94} (2016) 115002
  [arXiv:1609.04764 [hep-ph]].

  
\bibitem{Mendez:1991gp}
  A.~Mendez and A.~Pomarol,
  ``Signals of CP violation in the Higgs sector,''
  Phys.\ Lett.\ B {\bf 272} (1991) 313.

\bibitem{Haber:2006ue}
  H.E.~Haber and D.~O'Neil,
  ``Basis-independent methods for the two-Higgs-doublet model. II. The Significance of $\tan\beta$,''
  Phys.\ Rev.\ D {\bf 74} (2006) 015018
   [Erratum: Phys.\ Rev.\ D {\bf 74} (2006) 059905]
  [hep-ph/0602242].

\bibitem{Grzadkowski:2014ada}
  B.~Grzadkowski, O.M.~Ogreid and P.~Osland,
  ``Measuring CP violation in Two-Higgs-Doublet models in light of the LHC Higgs data,''
  JHEP {\bf 1411} (2014) 084
   [arXiv:1409.7265 [hep-ph]].


\bibitem{Branco:2015gna}
  G.C.~Branco, I.~de Medeiros Varzielas and S.F.~King,
  ``Invariant approach to CP in unbroken $\Delta(27)$,''
  Nucl.\ Phys.\ B {\bf 899} (2015) 14
  [arXiv:1505.06165 [hep-ph]].

\bibitem{Varzielas:2016zjc}
  I.~de Medeiros Varzielas, S.F.~King, C.~Luhn and T.~Neder,
  ``CP-odd invariants for multi-Higgs models: applications with discrete symmetry,''
  Phys.\ Rev.\ D {\bf 94} (2016)  056007
  [arXiv:1603.06942 [hep-ph]].

\bibitem{deMedeirosVarzielas:2017ote}
  I.~de Medeiros Varzielas, S.F.~King, C.~Luhn and T.~Neder,
  ``Spontaneous CP violation in multi-Higgs potentials with triplets of $\Delta(3n^2)$ and $\Delta(6n^2)$,''
  JHEP {\bf 1711} (2017) 136
  [arXiv:1706.07606 [hep-ph]].
  
\bibitem{Deshpande:1977rw}
  N.G.~Deshpande and E.~Ma,
  ``Pattern of Symmetry Breaking with Two Higgs Doublets,''
  Phys.\ Rev.\ D {\bf 18} (1978) 2574.

\bibitem{Ivanov:2007de} 
  I.P.~Ivanov,
  ``Minkowski space structure of the Higgs potential in 2HDM. II. Minima, symmetries, and topology,''
  Phys.\ Rev.\ D {\bf 77} (2008) 015017 
  [arXiv:0710.3490 [hep-ph]].

\bibitem{Ferreira:2009wh}
  P.M.~Ferreira, H.E.~Haber and J.~P.~Silva,
  ``Generalized CP symmetries and special regions of parameter space in the two-Higgs-doublet model,''
  Phys.\ Rev.\ D {\bf 79} (2009) 116004
  [arXiv:0902.1537 [hep-ph]].

\bibitem{Ferreira:2010yh}
  P.M.~Ferreira, H.E.~Haber, M.~Maniatis, O.~Nachtmann and J.P.~Silva,
  ``Geometric picture of generalized-CP and Higgs-family transformations in the two-Higgs-doublet model,''
  Int.\ J.\ Mod.\ Phys.\ A {\bf 26} (2011) 769
  [arXiv:1010.0935 [hep-ph]].

\bibitem{Battye:2011jj}
  R.A.~Battye, G.D.~Brawn and A.~Pilaftsis,
  ``Vacuum Topology of the Two Higgs Doublet Model,''
  JHEP {\bf 1108} (2011) 020
   [arXiv:1106.3482 [hep-ph]].
  
 \bibitem{Pilaftsis:2011ed} 
  A.~Pilaftsis,
  ``On the Classification of Accidental Symmetries of the Two Higgs Doublet Model Potential,''
  Phys.\ Lett.\ B {\bf 706} (2012) 465
  [arXiv:1109.3787 [hep-ph]].

 \bibitem{Peccei:1977ur}
  R.D.~Peccei and H.R.~Quinn,
  ``Constraints Imposed by CP Conservation in the Presence of Instantons,''
  Phys.\ Rev.\ D {\bf 16} (1977) 1791.
  
  \bibitem{Weinberg:1977ma}
  S.~Weinberg,
  ``A New Light Boson?,''
  Phys.\ Rev.\ Lett.\  {\bf 40} (1978) 223.

\bibitem{Wilczek:1977pj}
  F.~Wilczek,
``Problem of Strong  $P$  and  $T$  Invariance in the Presence of Instantons,''
  Phys.\ Rev.\ Lett.\  {\bf 40} (1978) 279.

\bibitem{Haber:2015pua}
  H.E.~Haber and O.~St\r{a}l,
  ``New LHC benchmarks for the CP-conserving two-Higgs-doublet model,''
  Eur.\ Phys.\ J.\ C {\bf 75} (2015) 491
   [Erratum: Eur.\ Phys.\ J.\ C {\bf 76} (2016)  312]
  [arXiv:1507.04281 [hep-ph]].

\bibitem{Hall:1981bc}
  L.J.~Hall and M.~B.~Wise,
  ``Flavor Changing Higgs  Boson Couplings,''
  Nucl.\ Phys.\ B {\bf 187} (1981) 397.

\bibitem{Olaussen:2010aq}
  K.~Olaussen, P.~Osland and M.A.~Solberg,
  ``Symmetry and Mass Degeneration in Multi-Higgs-Doublet Models,''
  JHEP {\bf 1107} (2011) 020
 [arXiv:1007.1424 [hep-ph]].

\bibitem{Aranda:2016qmp} 
  A.~Aranda, I.P.~Ivanov and E.~Jim\'{e}nez,
  ``When the C in CP does not matter: anatomy of order-4 CP eigenstates and their Yukawa interactions,''
  Phys.\ Rev.\ D {\bf 95} (2017) 055010 
   [arXiv:1608.08922 [hep-ph]].

\bibitem{Ishimori:2012zz}
  H.~Ishimori, T.~Kobayashi, H.~Ohki, H.~Okada, Y.~Shimizu and M.~Tanimoto,
``An introduction to non-Abelian discrete symmetries for particle physicists,''
  Lect.\ Notes Phys.\  {\bf 858} (2012) 1.

\bibitem{Ivanov:2012ry}
  I.P.~Ivanov and E.~Vdovin,
  ``Discrete symmetries in the three-Higgs-doublet model,''
  Phys.\ Rev.\ D {\bf 86} (2012) 095030
    [arXiv:1206.7108 [hep-ph]].
  
  \bibitem{Ivanov:2012fp}
  I.P.~Ivanov and E.~Vdovin,
  ``Classification of finite reparametrization symmetry groups in the three-Higgs-doublet model,''
  Eur.\ Phys.\ J.\ C {\bf 73} (2013)  2309
   [arXiv:1210.6553 [hep-ph]].

\bibitem{Ferreira:2017tvy}
  P.~M.~Ferreira, I.P.~Ivanov, E.~Jim\'enez, R.~Pasechnik and H.~Ser\^{o}dio,
  ``CP4 miracle: shaping Yukawa sector with CP symmetry of order four,''
  JHEP {\bf 1801} (2018) 065
  [arXiv:1711.02042 [hep-ph]].
  
\bibitem{Ivanov:2017bdx}
  I.P.~Ivanov,
  ``Radiative neutrino masses from order-4 CP symmetry,''
  JHEP {\bf 1802} (2018) 025
   [arXiv:1712.02101 [hep-ph]].
  
 \bibitem{Ivanov:2018qni}
  I.P.~Ivanov and M.~Laletin,
  ``Multi-Higgs models with CP-symmetries of increasingly high order,''
  Phys.\ Rev.\ D {\bf 98} (2018)  015021
  [arXiv:1804.03083 [hep-ph]].

\bibitem{Grzadkowski:2016lpv}
  B.~Grzadkowski, O.M.~Ogreid and P.~Osland,
  ``CP-Violation in the $ZZZ$ and $ZWW$ vertices at $e^+e^-$ colliders in Two-Higgs-Doublet Models,''
  JHEP {\bf 1605} (2016) 025
   [Erratum: JHEP {\bf 1711} (2017) 002]
  [arXiv:1603.01388 [hep-ph]].

\bibitem{Hagiwara:1986vm}
  K.~Hagiwara, R.D.~Peccei, D.~Zeppenfeld and K.~Hikasa,
  ``Probing the Weak Boson Sector in $e^+ e^- \to W^+ W^-$,''
  Nucl.\ Phys.\ B {\bf 282} (1987) 253.
  
\bibitem{Nieves:1996ff}
  J.F.~Nieves and P.B.~Pal,
  ``Electromagnetic properties of neutral and charged spin 1 particles,''
  Phys.\ Rev.\ D {\bf 55} (1997) 3118
  [hep-ph/9611431].

\bibitem{Gounaris:1999kf}
  G.J.~Gounaris, J.~Layssac and F.M.~Renard,
  ``Signatures of the anomalous $Z\gamma$ and $Z Z$ production at the lepton and hadron colliders,''
  Phys.\ Rev.\ D {\bf 61} (2000) 073013
   [hep-ph/9910395].

\bibitem{Gounaris:2000dn}
  G.J.~Gounaris, J.~Layssac and F.M.~Renard,
  ``Off-shell structure of the anomalous $Z$ and $\gamma$ self-couplings,''
  Phys.\ Rev.\ D {\bf 62} (2000) 073012
  [Addendum: Phys.\ Rev.\ D {\bf 65} (2002) 017302]
   [hep-ph/0005269].
 
\bibitem{Baur:2000ae}
  U.~Baur and D.L.~Rainwater,
  ``Probing neutral gauge boson self-interactions in $ZZ$ production at hadron colliders,''
  Phys.\ Rev.\ D {\bf 62} (2000) 113011
  [hep-ph/0008063].
 

\bibitem{Hahn:2000kx}
  T.~Hahn,
  ``Generating Feynman diagrams and amplitudes with FeynArts 3,''
  Comput.\ Phys.\ Commun.\  {\bf 140} (2001) 418
  [hep-ph/0012260].
   
\bibitem{Dreiner:2008tw}
  H.K.~Dreiner, H.E.~Haber and S.P.~Martin,
  ``Two-component spinor techniques and Feynman rules for quantum field theory and supersymmetry,''
  Phys.\ Rept.\  {\bf 494} (2010) 1
   [arXiv:0812.1594 [hep-ph]].
  
\bibitem{Bento:2017eti}
  M.P.~Bento, H.E.~Haber, J.C.~Rom\~{a}o and J.P.~Silva,
  ``Multi-Higgs doublet models: physical parametrization, sum rules and unitarity bounds,''
  JHEP {\bf 1711} (2017) 095
   [arXiv:1708.09408 [hep-ph]].
  
\bibitem{Haber:1993jr}
  H.E.~Haber and Y.~Nir,
  ``$Z \to A^0 A^0\nu\bar{\nu}$ and $e^+ e^- \to A^0 A^0 Z$ in two Higgs doublet models,''
  Phys.\ Lett.\ B {\bf 306} (1993) 327
   [hep-ph/9302228].


\end{thebibliography}
\end{document}